\documentclass[review]{elsarticle}

\usepackage{amssymb}
\usepackage{mathrsfs}
\usepackage{amsmath}
\usepackage{amsthm}
\usepackage{graphicx}
\usepackage{subfig}
\usepackage{epstopdf}
\usepackage{amsmath,bm} 
\usepackage{dsfont}
\usepackage{array}
\usepackage{multirow}
 \usepackage{flushend}
 \usepackage{booktabs}
 \usepackage{threeparttable}

 \usepackage{color}

\newcommand{\ychf}{\textcolor[rgb]{0.00,0.00,1.00}}
\usepackage{fancyhdr,lipsum}
\journal{Journal of The Franklin Institute}

\newcommand{\ycf}{\textcolor[rgb]{0.00,0.00,0.00}}

\begin{document}
\begin{frontmatter}
\title{Development of a General Momentum Exchange Devices Fault Model for Spacecraft Fault-Tolerant Control System Design}

\author[mymainaddress]{Chengfei Yue}
\ead{chengfei\_yue@u.nus.edu}
\author[mysecondaryaddress]{Qiang Shen}
\ead{qiang.shen@hotmail.com}
\author[mymainaddress]{Xibin Cao}
\ead{xbcao@hit.edu.cn}
\author[mymainaddress]{Feng Wang \corref{mycorrespondingauthor}}
\ead{wfhitsat@hit.edu.cn}
\cortext[mycorrespondingauthor]{Corresponding author}
\author[mythirdaddress]{Cher Hiang Goh}
\ead{elegch@nus.edu.sg}
\author[mythirdaddress]{Tong Heng Lee}
\ead{eleleeth@nus.edu.sg}

\address[mymainaddress]{Research Center of Satellite Technology, Harbin Institute of Technology}
\address[mysecondaryaddress]{School for Engineering of Matter, Transport
and Energy, Arizona State University}
\address[mythirdaddress]{Department of Electrical and Computer Engineering, National University of Singapore}

\begin{abstract} This paper investigates the mechanism of various faults of momentum exchange devices. These devices are modeled as a cascade electric motor (EM) - variable speed drive(VSD) system. Considering the mechanical part of  the EM and the VSD system, the potential faults are reviewed and summarized. Thus with a clear understanding of these potential faults, a general fault model in a cascade multiplicative structure is established for momentum exchange devices. Based on this general model, various fault scenarios can be simulated, and the possible output can be appropriately visualized.
In this paper, six types of working condition are identified and the corresponding fault models are constructed. Using this fault model, the control responses using reaction wheels and single gimbal control moment gyros under various fault conditions are demonstrated. The simulation results show the severities of the faults and  demonstrate that the additive fault is more serious than the multiplicative fault from the viewpoint of control accuracy. Finally, existing fault-tolerant control strategies are briefly summarized and potential approaches including both passive and active ones to accommodate gimbal fault of single gimbal control moment gyro \ycf{are} demonstrated.
\end{abstract}
\end{frontmatter}

\section{Introduction}
\ychf{Momentum exchange devices(MEDs) have significant advantages of cleanliness, without the expulsion of gases, over thrusters.} In addition, these devices came always with small volume and light weight. Thus they have been widely employed in spacecraft attitude determination and control system (ADCS) \cite{tsiotras2001satellite,shen2017robust,wie2002rapid,ahmed2002adaptive}.
Among all momentum exchange devices, the reaction wheel (RW) is the primary attitude control actuator due to its mechanical simplicity and low cost. However, most RWs only provide less than 1 N$\cdot$m maximum torque that is much smaller than control moment gyros (CMGs) with 100 $-$ 5000 N$\cdot$m maximum torque \cite{wie2001singularity}. Thus RWs are replaced by CMGs in the agile spacecraft for rapid maneuver, such as Pleaides \cite{gleyzes2012pleiades} and Wordview-2 \cite{poli2010radiometric}. However, failures of these momentum exchange devices occur occasionally in practical missions. For instance, four RWs failed on the Far Ultraviolet Spectroscopy Explorer (FUSE) spacecraft in 2001, 2002, 2004, and 2007, respectively; and two each on Hayabusa in July and October 2005, Dawn in 2010 and 2012, and Kepler in 2012 and 2013 \cite{markley2014fundamentals}. The control moment gyros failed and prevented the spacecraft WorldView-4 from pointing accurately in 2019. Thus the development of fault-tolerant control system for the MEDs actuated spacecraft is urgent.

In existing fault-tolerant researches, most of the work such as \cite{fekih2014fault, jiang2012fault, murugesan1987fault, shen2015finite, noumi2013fault, zhang2017fault} focus on the control problem itself,
 except for \cite{murugesan1987fault} and \cite{shen2015finite}, where the four potential faults (recoverable) and/or failures (irrecoverable) of RWs are briefly introduced. Why the fault are modeled in additive way and multiplicative way is not clear. For the CMG actuated spacecraft, the fault model of the CMG is unclear and the fault-tolerant result is rare.  In \cite{noumi2013fault}, the skew angle of CMG configuration is analyzed and a genetic algorithm is adopted to simultaneously tune the skew angle and controller gains to achieve fault tolerant. In \cite{zhang2017fault}, Zhang et al. considered the gimbal fault and employed sliding mode control theory to change the CMG gimbal rate directly to void the singularity and achieve fault tolerance. \ycf{Comparing with these existing works, this paper gives insight into the fault in general momentum exchange devices (MEDs). We explain clearly why the fault of the reaction wheel can be modeled in additive and multiplicative way, then generalize  the fault model of the reaction wheel to a wide range of MEDs, especially the single gimbal control moment gyros. The contribution of this paper paves the way of developing fault-tolerant control strategies for the CMG actuated spacecraft. Since this work focuses on the fault modeling instead of the fault-tolerant controller design, the description of the fault-tolerant controller design is just to describe the potential implementation of our fault model.}

For MED fault modeling, MEDs can be regarded as a cascade combination of an electric motor (EM) and its variable speed drive (VSD) system from a systemic point of view. \ychf{More specifically, a RW is a flywheel mounted to an electric brushless DC motor (BLDC) \cite{bialke98high,rahimi2017fault} and the torque is generated through wheel's acceleration or deceleration. For CMGs, a momentum wheel is mounted on one or two gimbals containing two kinds of motors: stepper motor and BLDC motor \cite{choi2015fault}. The stepper motor provides precision gimbal control of CMGs while the BLDC motor provides an efficient way of driving the momentum wheel to store the angular momentum.} Thus the RW is a one-EM-VSD-loop system and the CMG can essentially be regarded as a cascade combination of two (SGCMG) or three (DGCMG) EM-VSD loops. All potential faults of RW and CMGs would lie in the mechanical part of the EM, sensors and actuators of VSD, or the electrical part of these components.

\begin{figure}[h]
  \centering
    \includegraphics[scale=0.62]{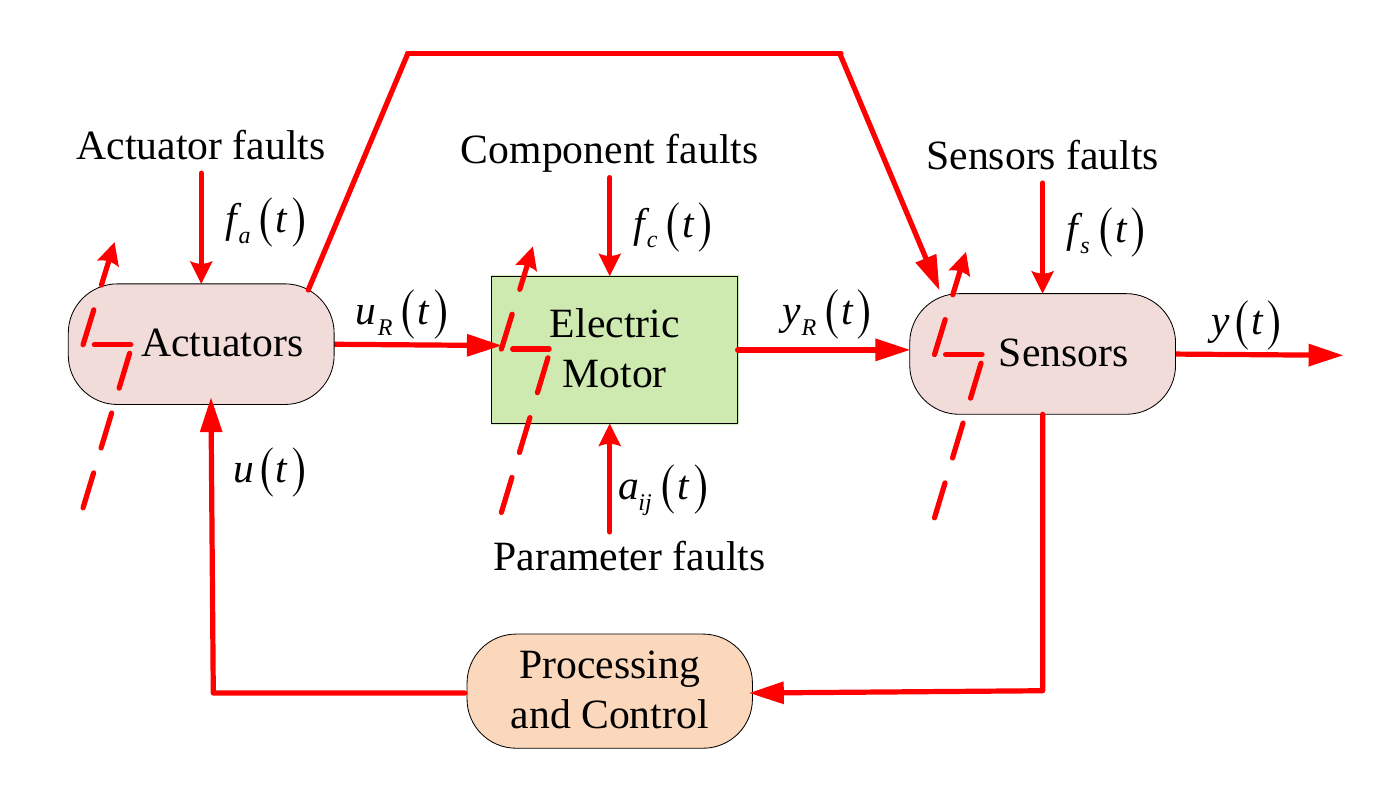}
  \caption{Fault model of EM-VSD system}
\label{fig1}
\end{figure}

 The details of potential faults in EM-VSD are analyzed in \cite{choi2015fault,book:931874, kastha1994investigation,nandi2005condition,campos2008fault,rajagopalan2006detection,moseler2000application} and \cite{rajagopalan2006thesis}.
\ychf{For the EM, potential faults are categorized into: stator faults, rotor faults, eccentricity-related faults and bearing or gear faults.}
These faults belong to multiplicative faults. \ychf{In the VSD, the faults can be categorized into sensor faults and actuator faults}. These faults are considered as  additive faults \cite{campos2008fault,chen2012robust}. The schematic diagram of EM-VSD system containing the potential fault is shown in Fig. \ref{fig1}, where $f_a(t)$, $f_c(t)/a_{ij}(t)$ and $f_s(t)$  represent the parameter errors of actuators, electric motors and sensors caused by fault and the detailed explanations can be found in Section 2 to Section 4.

Motivated by the aforementioned observations, this paper investigates the potential faults in the EM-VSD system, which are categorized into multiplicative or additive fault  through analyzing an EM-VSD model. The momentum exchange devices are considered as being in a cascade mechanical structure, in which an EM-VSD system governs one degree of control freedom and works independently. Based on this model, and considering potential faults in the EM-VSD system, a general fault model for momentum exchange devices is then established. \ycf{To the best of the authors' knowledge, this is the first attempt to propose a generalized fault model to a wide range of momentum exchange devices with a clear understanding of mechanical mechanism.With the utilization of this model, the contribution in our work makes it possible to describe the potential fault scenarios efficiently and effectively, and removes the existing barrier lacking of a unified fault model in developing the necessary fault-tolerant controllers for CMG-actuated spacecraft system.}
 Simulations of the control responses of the RW and SGCMG actuated spacecraft under various fault scenarios \ycf{visualize} the severity of the multiplicative and additive faults and show that the additive fault is more serious than the multiplicative fault from the viewpoint of control accuracy. Finally, existing fault-tolerant control strategies are briefly summarized and potential fault-tolerant strategies based on the proposed fault model to accommodate the gimbal fault of SGCMGs are demonstrated to describe the potential implementation of our proposed model.

The rest of this paper is organized as follows. Section 2 presents the EM system, the potential faults of the EM and its fault model. Section 3 addresses the potential faults of sensors and actuators, and the corresponding models are given. Combined with the model of the EM, the overall structure of the EM-VSD system and the fault model is established in Section 4. Based on the fault model of the EM-VSD, a general fault frame of momentum exchange devices is established in Section 5. Various fault scenarios in different momentum exchange devices are then modeled through choosing different model parameters. In Section 6, simulations are conducted to demonstrate system performance in the presence of different faults, illustrating the appropriate suitability and applicability of this general fault frame. \ychf{In Section 7, the fault-tolerant control strategies are summarized and potential approaches to hanlde the gimbal faults of the SGCMGs are given.} Finally, conclusions are noted in Section 8.

\section{EM Fault Modeling}\label{EM_sec}
\subsection{Mathematical Model of EM}

\begin{figure}[!b]
  \centering
  \subfloat[BLDC drive system]{\label{fig2a}
    \includegraphics[scale=0.55]{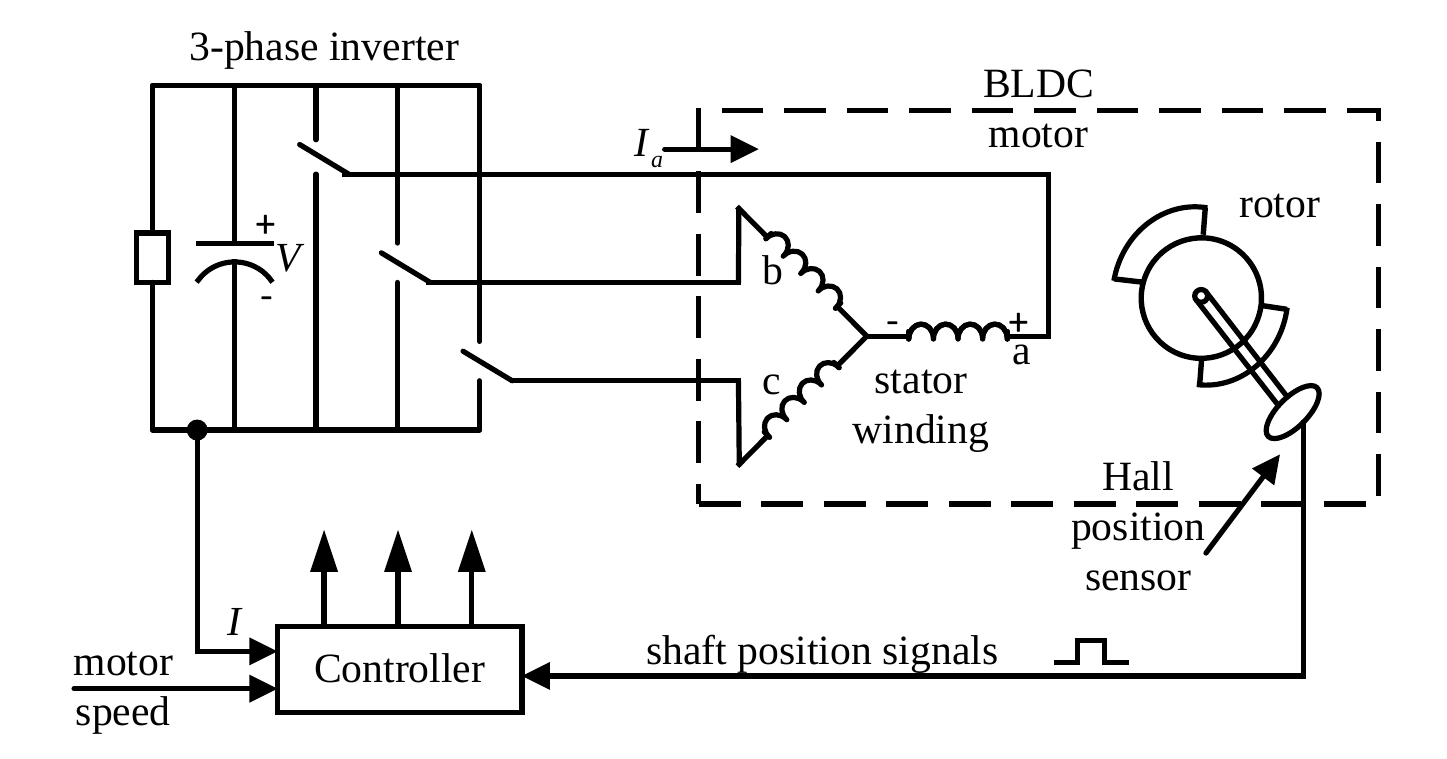}}
   \subfloat[Equivalent circuit]{\label{fig2b}
    \includegraphics[scale=0.45]{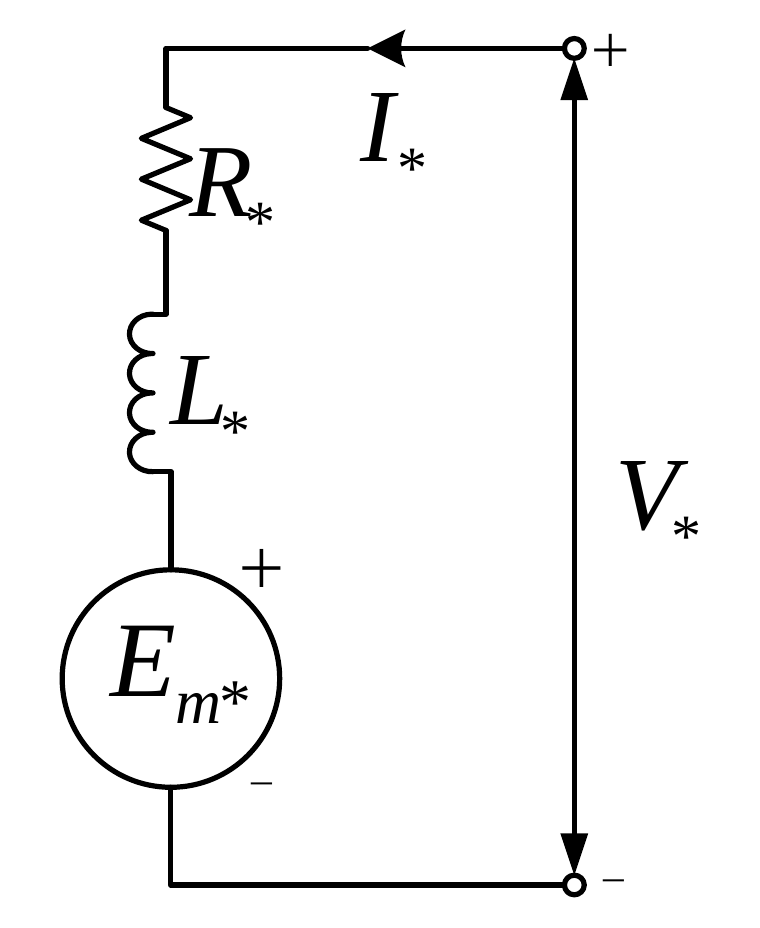}}
  \caption{\ychf{BLDC motor}}
  \label{fig2}
\end{figure}

\ychf{For the momentum exchange devices, the permanent magnet (PM) brushless DC motors (BLDC) are widely employed due to their advantages of high power density, high efficiency, long operating life, noiseless operation, high speed ranges and etc. The BLDC motors come in single-phase, two-phase, and three phase configurations. Among these, the three-phase motor is the most popular and widely used.} The 3-phase electronically commutated BLDC motor drive system is shown in Fig. \ref{fig2a} and the one phase equivalent circuit of BLDC motor is illustrated in Fig. \ref{fig2b}\cite{joksimovic2000detection}. Without loss of generality and considering the phase A,  the control input is denoted as $V_a$ and the phase current is denoted as $I_a$. The electrical equivalent of the armature coil can be described by a resistance $R_a$, a self-inductance $L_a$ and an induced voltage referring to the back electromotive force (emf) $E_{ma}$ which opposes the voltage source. The relationship can be modeled as:
\begin{subequations}\label{1}
 \begin{align}\label{1a}
  V_a&={{E}_{ma}}+R_a I_a+L_a\frac{d {I_a}}{dt}, \\ \label{1b}
{{E}_{ma}}&={{K}_{Ea}}\omega_r,
\end{align}
\end{subequations}
where $K_{Ea}$ is the back-emf constant and $\omega_r$ is the angular velocity of the rotor.

Using the similar method as in \cite{moseler2000application} and taking the mean value, the electrical subsystem of the BLDC motor can be described as:
\begin{equation}\label{R1}
{V}=k_E \omega_r+R {I}+L\frac{d {{I}}}{dt}.
\end{equation}

Performing an energy balance on the system, the sum of the torques of the motor must equal zero. Therefore, we have
\begin{equation}\label{2}
J_m \dot{\omega}_r+\sigma\omega_r =T_e-{{T}_{l}},
\end{equation}
where $J_m$ is the inertia of the rotor and the equivalent mechanical load, $\sigma$ is the viscous friction coefficient, and $T_l$ is the load torque. $T_e$ is the electromagnetic torque and expressed as:
\begin{equation}\label{3}
T_e = K_t I ,
\end{equation}
with $K_t$ being  the torque constant depending on the flux density of  the fixed magnets, the reluctance of the iron core, and the number of turns in the armature winding.

Denoting $x={[ \begin{matrix}
   I & \omega_r   \\
\end{matrix} ]^{T}}$ as the state variable and $V$ as the control input, the state space  model of a BLDC motor is obtained as:
\begin{equation}\label{4}
\left\{ \begin{matrix}
   \left[ \begin{matrix}
   {\dot{I}}  \\
   {\dot{\omega }_r}  \\
\end{matrix} \right]=\left[ \begin{matrix}
   -\frac{R}{L} & -\frac{{{K}_{E}}}{L}  \\
   \frac{{{K}_{t}}}{J_m} & -\frac{\sigma}{J_m}  \\
\end{matrix} \right]\left[ \begin{matrix}
   I  \\
   \omega_r   \\
\end{matrix} \right]+\left[ \begin{matrix}
   \frac{1}{L}  \\
   0  \\
\end{matrix} \right]V\text{+}\left[ \begin{matrix}
   0  \\
   -\frac{1}{J_m}  \\
\end{matrix} \right]{{T}_{l}}  \\
   y=C\left[ \begin{matrix}
  I  \\
   \omega_r   \\
\end{matrix} \right] \\
\end{matrix}, \right.
\end{equation}
with $y$ being the measurements. \ycf{$C$ is the output matrix that decides the measurement output. For example, the matrix $C$ could be the identity matrix to measure both current $I$ and the angular velocity of the rotor $\omega_r$, or it could be matrix $[0, 1]$ to measure the angular velocity of the rotor $\omega_r$ only. For EM-VSD system, $\omega_r$ is more important, hence we choose matrix $C$ to be $[0, 1]$}.

Denoting
\[A=\left[ \begin{matrix}
   -\frac{R}{L} & -\frac{{{K}_{E}}}{L}  \\
   \frac{{{K}_{t}}}{J_m} & -\frac{\sigma}{J_m}  \\
\end{matrix} \right], \ B=\left[ \begin{matrix}
   \frac{1}{L}  \\
   0  \\
\end{matrix} \right], D=\left[ \begin{matrix}
   0  \\
   -\frac{1}{J_m}  \\
\end{matrix} \right],\]
and $u= V$, equation \eqref{4} can be further written into a compact form as:
\begin{equation}\label{5}
\left\{ \begin{matrix}
   \dot{x}=Ax+Bu\text{+}D{{T}_{l}}  \\
   y=Cx  \\
\end{matrix} \right. .
\end{equation}
Other motors such as the stepper motors have the similar compact form as  \eqref{5} \cite{betin1999closed,morar2003stepper}.

\subsection{Potential Fault of EM}
In an EM, there may exist mechanical and electrical faults or failures, or a combination of these mechanical and electrical faults.
\ychf{Specifically, the potential faults may be \cite{choi2015fault,nandi2005condition,rajagopalan2006thesis,collacott2012mechanical}:}
\begin{itemize}
\item \ychf{Stator faults. ``The most frequently occurring stator fault is the breakdown of the winding insulation in the region where the end windings enter the stator slots. It may be caused by large electrical voltage stresses, electro-dynamic forces produced by winding currents, thermal aging from multiple heating and cooling cycles, and mechanical vibrations from internal and external sources. This winding insulation breakdown can result in turn-to-turn faults that eventually lead to short circuits to ground. \cite{rajagopalan2006thesis}"}

\item \ychf{Rotor faults. The rotor of the BLDC are PM and the major fault of rotor is the damaged rotor magnet \cite{rajagopalan2006thesis}. Some permanent magnets corrode \cite{pal1991direct} and cracks formed during manufacturing \cite{pal1991design} can lead to disintegration. The partial demagnetization of the magnets may also influence the magnetic flux density distribution \cite{fisher1991design}. Other faults may be the broken rotor bar or cracked rotor end-rings \cite{choi2015fault}. These faults are mainly caused by excess stresses}.

\item Eccentricity-related faults can be categorized as static and/or dynamic air-gap irregularities. The dynamic eccentricity is the character describing the displacement between the center of the rotor and the center of the rotation. The possible reasons are bent rotor shaft, bearing wear or misalignment, and mechanical resonance. Static eccentricity may be caused by the ovality of the stator core or incorrect positioning of the rotor or stator. When the eccentricity becomes large, it can result in damage of stator and rotor.

  \item Bearing and gearbox faults or failures. \ychf{Bearing failures account for the vast majority of the recorded motor failure \cite{motor1985report}.
   The bearing failures are caused by continued stress, inherent eccentricity,  fatigue and other external causes, such as unbalanced load, improper installation, contamination and corrosion, and improper lubrication. This kind of fault may lead to excessive noises and vibrations.}

\end{itemize}

\subsection{Fault Model of EM}
\begin{figure}[!h]
  \centering
    \includegraphics[scale=0.62]{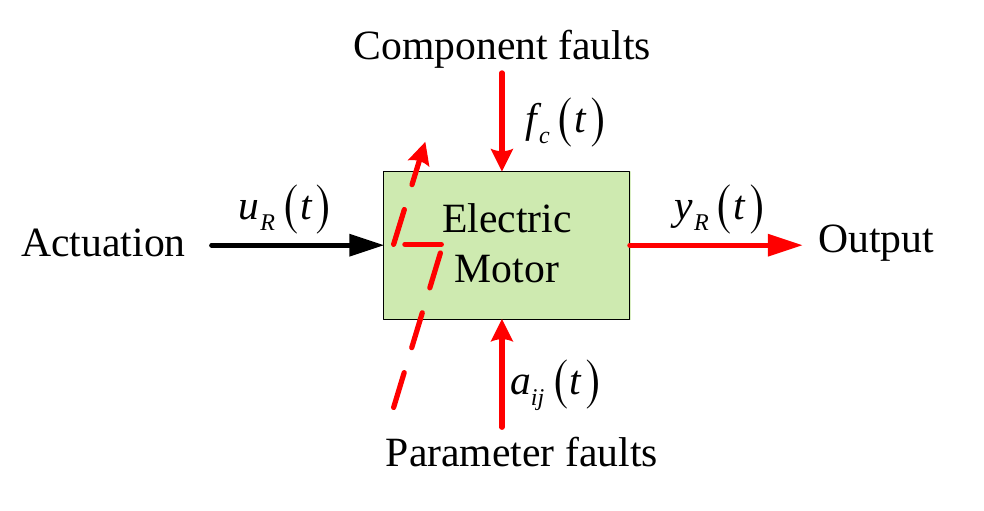}
  \caption{Fault model of electric motor}
\label{fig3}
\end{figure}

As addressed in \cite{chen2012robust,isermann2011fault}, the fault of EM can be modeled as component faults $f_c(t)$ or parameter faults $a_{ij}(t)$ as shown in Fig. \ref{fig3}.

The component fault occurs when some condition changes in the system. In some other cases, the faults can be expressed  as a change in the system parameter. Then the mathematical fault model can be constructed as:
 \begin{equation}\label{6}
\dot x = Ax + Bu{\rm{ + }}D{T_l} + {f_c}\left( t \right)
 \end{equation}
 or
\begin{eqnarray}\label{7} \notag
\dot x &=& Ax + Bu{\rm{ + }}D{T_l} + \sum {\left( {\sum {{a_{ij}}} {x_j}} \right)} {e_i}\\
& =& \left( {A + \Delta A} \right)x + Bu{\rm{ + }}D{T_l},
\end{eqnarray}
where $e_i$ is the $i$th basis vector, and $\Delta A$ is the discrepancy of the state-transition matrix caused by parameter faults. Indexes $i$, $j$ are related to the EM system and can be determined by fault diagnosis. Fault models \eqref{6} and \eqref{7} are equivalent in describing the component fault and the system matrix $A$ is influenced by the component faults.

Solving the foregoing equation \eqref{7} from the time instant $k$ to $k+1$ with a constant control input $u(k)$, we obtain
\begin{equation}\label{8}
{{x}_{R}}\left( k+1 \right)={{e}^{\left( A+\Delta A \right)h}}x_R\left( k \right)+{{e}^{\left( A+\Delta A \right)h}}\left\{ \left. \int_{0}^{h}{{{e}^{-\left( A+\Delta A \right)t }}}dt  \right\} \right.\left( Bu(k)+D{{T}_{l}} \right),
\end{equation}
where $h$ is the step size, $x_R(k+1)$ and $x_R(k)$ are real sates at the time instants $k+1$ and $k$, respectively and the subscript ``R" represents real value rather than the measured one.
Integrating $\int_{0}^{h}{{{e}^{-\left( A+\Delta A \right)t }}}dt $ and using the first order approximation ${{e}^{-\left( A+\Delta A \right)h}}\approx{I}-\left( A+\Delta A \right)h$, we have:
\begin{equation}\label{9}
\int_{0}^{h}{{{e}^{-\left( A+\Delta A \right)t }}}dt\approx-{{\left( A+\Delta A \right)}^{-1}}\left[ {{e}^{-\left( A+\Delta A \right)h}}-I \right]\approx h.
\end{equation}

Moreover, substituting \eqref{9} into \eqref{8}, it follows that
\begin{equation}\label{10}
  {{x}_{R}}\left( k+1 \right)\approx{{e}^{\Delta Ah}}\left\{ \left. {{e}^{Ah}}\left[ x_R\left( k \right)+\left( Bu(k)+D{{T}_{l}} \right)h \right] \right\} \right. .
\end{equation}
To compensate the load torque and cancel the influence of the system state $x_R(k)$, the control torque $u(k)$ can be designed as:
\begin{equation}\label{11}
u(k) = \frac{1}{h}u_0(k) - \frac{1}{h} B^+(x_R(k)+DT_lh),
\end{equation}
where $B^+= B^T(BB^T+\varsigma I)^{-1})$ is the pseudo-inverse of the matrix $B$ and $\varsigma$ is a small positive number.

Then, substituting \eqref{11} into \eqref{10}, we obtain
\begin{equation}\label{12}
  {{x}_{R}}\left( k+1 \right)  \approx{{e}^{\Delta Ah}}{e}^{Ah}Bu_0(k) .
\end{equation}
 When the motor is assumed to be a $1$-dimensional speed or torque control system, we obtain:
 \begin{subequations}\label{13}
 \begin{equation}\label{13a}{{x}_{R}}\left( k+1 \right)\approx{{e}^{\Delta ah}}\left\{ \left. {{e}^{ah}}b \right\} \right.{{u}_{0}}\left( k \right)=\eta g{{u}_{0}}\left( k \right)\end{equation}
  \begin{equation}\label{13b}{{y}_{R}}\left( k+1 \right)=c{{x}_{R}}\left( k+1 \right)\approx\eta cg{{u}_{0}}\left( k \right)=\eta {{u}^{+}}\left( k \right)\end{equation}
 \end{subequations}
with $g={{e}^{ah}}b$ being the transfer function from input $u_0(k)$ to state $x_R(k+1)$ without component faults, and $u^+(k)=cgu_0(k)$ being the equivalent input. In the following section, the control input $u(k)$ will refer to this equivalent input $u^+(k)$. The term ${{e}^{\Delta ah}}$ is denoted as the effectiveness factor $\eta$ describing the component faults, and it is constrained in the interval $\eta \in [0,1]$ practically.
Different values of the $\eta$ correspond to different scenarios, for example, a) $\eta =1$, the EM works normally and no fault occurs; b) $0 < \eta <1$ refers to malfunctions, in which the EM partially loses effectiveness, but not fail totally; c) $\eta = 0$ denotes a complete failure.

It is clearly observed from \eqref{13b} that the component fault of the EM can be represented by a multiplicative effectiveness factor $\eta$. The result in this section establishes the mathematical foundation to describe the component fault in a multiplicative way.

\section{VSD System Fault Modeling}\label{VSD_sec}
\subsection{Introduction of VSD and Potential Faults}
To effectively and  precisely drive the EM, VSD system is essential. It consists of the EM as the plant, sensors to measure the system state and output signal, processing and control component to generate control command, and actuators to drive the EM.

In the VSD system, sensors are adopted to measure the motor velocity, position and the output voltage of the inverter. Sensors can be mechanical or electrical, and the reliability of mechanical sensor is lower than the electrical one due to mechanical complexity \cite{campos2008fault}. \ychf{For the BLDC, the sensors may be the Hall position sensor and the electrical tachometer. The loss of a Hall sensor results in torque pulsations when the rotor is moving.}

\ychf{Actuators in VSD are inverters which convert DC electricity to AC electricity since almost all PM motors are inverter-fed. Various faults can occur in the inverter, such as the loss of one or more switches of a phase, the short circuit of a switch, and the opening of one of the lines to the machine \cite{choi2015fault}.}
\subsection{Fault Model of Sensors and Actuators in VSD}
\begin{figure}[!ht]
  \centering
  \subfloat[Fault model of VSD sensors]{
    \label{fig3a}
    \includegraphics[scale=0.5]{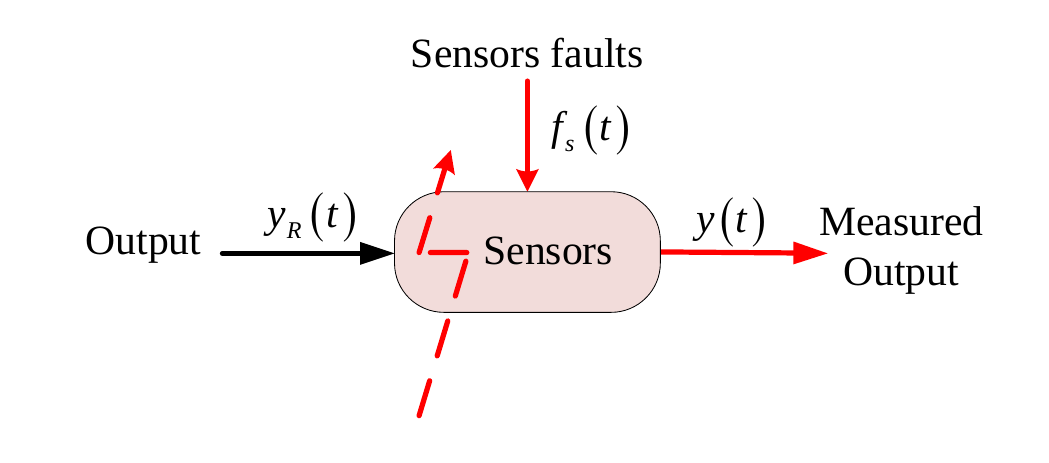}}
  \subfloat[Fault model of VSD actuators]{
    \label{fig3b}
    \includegraphics[scale=0.5]{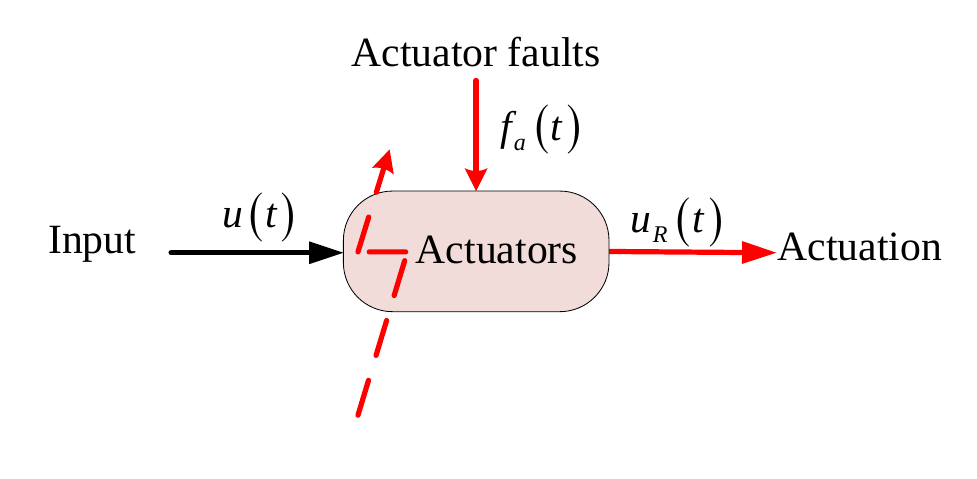}}
  \caption{Fault model of sensors and actuators}
  \label{fig4}
\end{figure}

Fault of sensors and actuators are often modeled as the additive fault as shown in Fig. \ref{fig4}. As described in Fig, \ref{fig3a}, the parameter $f_s(t)$ denotes the sensors faults. Then the output of sensor under fault is obtained as:
\begin{equation}\label{14}
y(t) = y_R(t) + f_s(t).
\end{equation}
All the sensors' faults can be described by choosing a proper $f_s(t)$ vector. For example, when the output stucks at a particular value $a$, $f_s(t)$ can be chosen as $a - y_R(t)$.

Similar to the sensors fault, the real actuation $u_R(t)$ and the actuator command $u(t)$ are connected via the actuator fault vector $f_a(t)$:
\begin{equation}\label{15}
u_R(t) = u(t) +f_a(t).
\end{equation}

It is observed that actuator will generate an additional control command to feed the plant after fault occurs. When the sensor is faulty, the output will have a direct bias affecting nominal measurement. Thus, the fault in actuators and sensors belongs to the additive category.
This section establishes the foundation of modeling the sensor and actuator fault in an additive way.


\section{Fault Model of EM-VSD System}
Considering all potential faults in EM, sensors and actuators, the schematic diagram of the overall EM-VSD system is shown in Fig. \ref{fig1}.
Based on the analysis in the previous two sections, the EM suffers the multiplicative fault represented by $\Delta A$, and the sensors and actuators in VSD system may occur the additive fault $f_s(t)$ and $f_a(t)$.
Then the state space model of the EM-VSD system in the presence of faults/failures can be constructed as:
\begin{equation}\label{16}
\left\{ \begin{matrix}
   \dot{x}=\left( A+\Delta A \right)x+B\left[ u+{{f}_{a}}\left( t \right) \right]\text{+}D{{T}_{l}}  \\
   y=Cx+{{f}_{s}}\left( t \right)  \\
\end{matrix} \right. .
\end{equation}

In view of the output relationship \eqref{13}, and substituting the potential fault model \eqref{14} and \eqref{15} into \eqref{16},
we get the output of the overall system as:
\begin{equation}\label{17}
y=\eta \left( u^++f_a(t)\right)+f_s(t) .
\end{equation}
Denoting the offset as $ y_o = \eta  f_a(t)+f_s(t)$, which is on the basis of precise output $y$ under the nominal control signal $u^+$, equation \eqref{17} is further written as:
\begin{equation}\label{18}
y=\eta u^+ + y_o .
\end{equation}

Now, we have developed the fault model of an EM-VSD system as presented in \eqref{18}.
Considering the fact that the EM-VSD system is working either in the speed control mode or in the torque control mode. Then we can get the general fault frame of an EM-VSD system in speed control mode and torque control mode as:
\begin{equation}\label{19}
\left\{ \begin{matrix}
   \omega = \eta_{\omega} {{\omega }_{c}} +{{\omega }_{o}},  \\
   T= \eta_T {{T}_{c}} +{{T}_{o}}, \\
\end{matrix} \right.\quad \begin{matrix}
   \text{Speed control mode}  \\
   \hspace{0.2cm} \text{Torque control mode}  \\
\end{matrix}
\end{equation}
where $\omega_c$ and $T_c$ are the angular velocity control command and the torque control command corresponding to the speed control mode and the torque control mode, respectively, $\eta_{\omega}$ and $\eta_{T}$ are the effectiveness factors representing the component fault of the EM in speed control mode and torque control mode, $\omega_o$ amd $T_o$ are the additive measurement offset as a combination of the actuator fault and the sensor fault in VSD system, and $\omega$ and $T$ are the measured angular velocity and output torque.

\section{General Fault Model of Momentum Exchange Devices}
The most commonly used momentum exchange devices in aerospace mission are RWs and CMGs. CMGs can be further divided into SGCMG, DGCMG and VSCMG. The differences between these categories are the number of degree of control freedom and the working mode of flywheels. Specifically, the RW and SGCMG are single degree-of-freedom devices, while DGCMG and VSCMG have two and three degree-of-freedom, respectively.
From the working mode of flywheels perspective, SGCMGs and DGCMGs hold their flywheel speed at a constant value \cite{markley2014fundamentals}, whereas the the rotor speed of RWs and VSCMG are time-varying.


Although RWs and CMGs have different working principles, these momentum exchange devices can be modeled by a series of cascade EM-VSD systems to drive the wheel and the gimbal frame separately. As stated in \cite{leve2015spacecraft}, the dynamics of CMG gimbal is independent of rotor momentum for the case of a very stiff gimbal. In parallel, we assume that dynamics of gimbal are working independent. In each control loop, the potential fault and the corresponding model are given in Section \ref{EM_sec} and \ref{VSD_sec}. When they are constructed as a cascade instrument, there will be a high dependence among each system and the overall fault model will be in a multiplicative form. Consequently, the general fault model of a momentum exchange device is given by:
 \begin{equation}\label{20}
y=\sum\limits_{j=1}^{m}\left\{ \left. \underset{{{i}_{j}}=1}{\overset{{{n}_{j}}}{\mathop{\Pi }}}\,\left[ \eta^{i_j} {{u}_{c}}^{{{i}_{j}}}+{{y}_{o}}^{{{i}_{j}}}\right] \right\} \right. ,
 \end{equation}
where the superscript $i_j$ means the $i$th element in series of the $j$th parallel term, $m$ and $n_j$ represents the total number of parallel terms and the total number of term in series of the $j$th parallel term, $u_c^{i_j}$ is the nominal control command, $y_o^{i_j}$ is the offset caused by fault, failure and malfunctions, and $\eta^{i_j} \in [0, 1]$ is the control effectiveness factor. In the following, by using the general fault model of the momentum exchange devices proposed in \eqref{20}, we give the specific fault model of the RW, SGCMG, DGCMG and VSCMG, respectively.

\subsection{RW Fault Model}
\ychf{RW contains a rotating flywheel and internal BLDC motor as well as associated electronics \cite{markley2014fundamentals}}. This device can be regarded as a single loop EM-VSD system.
It is always fed by control commands in order to generate desired control torque via acceleration or deceleration. Therefore, the whole loop is considered as the acceleration control loop. When the RW works in the torque control mode, the fault model can be given by:
\begin{equation}\label{22}
\tau_{rw} =\eta_{rw} {{\tau }_{c}}+{{\tau }_{o}}.
\end{equation}
where $\tau_c$ is the command torque, $\eta_{rw}$ is the effectiveness of the RW, $\tau_o$ is the output offset due to the fault and $\tau_{rw}$ is the real output.
This equation is consistent with the fault model of RW-actuated spacecraft system in existing literature, such as \cite{shen2015finite}, \cite{jin2008fault} and \cite{xiao2012adaptive}.

\subsection{SGCMG Fault Model} \label{sec_SGCMG_F}
SGCMG contains a spinning rotor mounted on a gimbal. In nominal condition, the rotor holds a constant speed \ychf{using a BLDC motor} while the gimbal is manipulated to change the direction of angular momentum \ychf{by a stepper motor}. Then a gyroscopic reaction torque orthogonal to both the rotor spin and gimbal axes is generated. With a small input of the gimbal, a much larger control torque is produced to act on the spacecraft, which is the so-called torque amplification characteristic. More specifically, the torque is proportional to both the angular momentum and gimbal angular rate calculated as:
\begin{equation}\label{23}
 \tau =-{{h}_{0}}\dot{\delta }\hat{t},
\end{equation}
where $h_0$ is the constant angular momentum of the spinning rotor determined by the moment inertia of the wheel $J_{sc}$ and motor speed $\omega_{sc}$ as $h_0 = J_{sc} \omega_{sc}$, $\delta$ is the gimbal angle, and $\hat{t}$ is a unit vector in the direction of output torque. The minus symbol `` - '' in  \eqref{23} means the output torque lies in the opposite direction of $\hat{t}$.

The SGCMG is considered as a combination of two EM-VSD systems.
The potential fault of SGCMG may exist in the rotor control loop and the gimbal frame control loop. We denote the rotor's EM-VSD control loop as the first degree-of-freedom and this loop is marked with a superscript ``r'', and the control loop of gimbal frame as the second degree-of-freedom with a superscript ``g''. According to the independent assumption, the dynamics of rotor and gimbal will not influence each other in their own loop control.
Then for the rotor, the angular momentum is the product of moment of inertia and the spindle speed, i.e. $h_0 = J_{sc} \omega_{sc}$, and the motor speed is controlled by the VSD system. Considering the possible fault in the rotor control loop, the angular momentum can be computed as:
\begin{equation}\label{24}
  {{h}_{0}}= \eta^r J_{sc}{{\omega }_{c}}+{{h}_{o}},
\end{equation}
where $h_o$ is the output offset. When the flywheel works normally, the output $h_0$ equals to the command angular momentum $h_c=J_{sc}  {{\omega }_{c}}$.
Considering the control loop of gimbal, its fault model is expressed as:
\begin{equation}\label{25}
\dot{\delta }= {{\eta }^{g}}{\dot{\delta }_{c}}+{\dot{\delta }_{o}}
\end{equation}
with $\dot{\delta}_c$ and $\dot{\delta}_o$ being the commanded gimbal rate and the gimbal rate offset caused by fault.
Substituting equations \eqref{24} and  \eqref{25} into \eqref{23}, we obtain the fault model of a SGCMG as
\begin{equation}\label{26}
  \tau =-\left[\eta^r J_{sc}{{\omega }_{c}}+{{h}_{o}} \right]\left[ {{\eta }^{g}}{\dot{\delta }_{c}}+{\dot{\delta }_{o}} \right]\hat{t}.
\end{equation}

\renewcommand{\arraystretch}{0.6}
\begin{table}[!t]
\centering
\caption{Work condition and fault model of SGCMG}\label{tab1}
\begin{threeparttable}
\begin{tabular}{ccc}
\toprule
Rotor&Gimbal&Model\\
\midrule
\multirow{6}*{$N,F_d$}&$N$&$-{{h}_{*}} {{{\dot{\delta }}}_{c}}\hat{t}$\\
&$F_a$& $-{{h}_{*}}\left[\eta^g {{{\dot{\delta }}}_{c}}\right] \hat{t}$\\
&$F_b$& $0$\\
&$F_c$& $-{{h}_{*}}\left[  {{\eta }^{g}}{{{\dot{\delta }}}_{c}} +{{{\dot{\delta }}}_{o}} \right]\hat{t}$\\
&$F_d$&$-{{h}_{*}}{{\dot{\delta }}_{o}}\hat{t}$ \\
&$F_e$&$-{{h}_{*}}\left[  {{{\dot{\delta }}}_{c}} +{{{\dot{\delta }}}_{o}} \right]\hat{t}$\\
\midrule
\multirow{6}*{$F_a$}&$N$&$-\left[\eta^r J_{sc} {{\omega }_{c}}\right]  {{{\dot{\delta }}}_{c}}\hat{t}$\\
&$F_a$& $-\left[\eta^r J_{sc} {{\omega }_{c}}\right]\left[ \eta^g {{{\dot{\delta }}}_{c}} \right]\hat{t}$\\
&$F_b$& $0$\\
&$F_c$& $-\left[\eta^r J_{sc} {{\omega }_{c}}\right]\left[ \eta^g {{{\dot{\delta }}}_{c}} +{{{\dot{\delta }}}_{o}} \right]\hat{t}$\\
&$F_d$& $-\left[\eta^r J_{sc} {{\omega }_{c}}\right]{{\dot{\delta }}_{o}}\hat{t}$\\
&$F_e$& $-\left[\eta^r J_{sc} {{\omega }_{c}}\right]\left[  {{{\dot{\delta }}}_{c}} +{{{\dot{\delta }}}_{o}} \right]\hat{t}$\\
\midrule
\multirow{2}*{$F_b$}&$N$,$ F_a$,$ F_b$& \multirow{2}*{$0$}\\
&$ F_c$,$ F_d$,$ F_d$\\
\midrule
\multirow{6}*{$F_c$}&$N$& $-\left[\eta^r J_{sc} {{\omega }_{c}}+{{h}_{o}} \right]{{{\dot{\delta }}}_{c}}\hat{t}$\\
&$F_a$& $-\left[ J_{sc}sa{{t}_{{{\omega }_{m}}}}\left( \eta^r {{\omega }_{c}} \right)+{{h}_{o}} \right]\left[\eta^g {{{\dot{\delta }}}_{c}} \right]\hat{t}$\\
&$F_b$& $0$\\
&$F_c$& $-\left[\eta^r J_{sc} {{\omega }_{c}}+{{h}_{o}} \right]\left[ {{\eta }^{g}}{\dot{\delta }_{c}} +{\dot{\delta }_{o}} \right]\hat{t}$\\
&$F_d$&$-\left[\eta^r J_{sc} {{\omega }_{c}}+{{h}_{o}} \right]{{\dot{\delta }}_{o}}\hat{t}$\\
&$F_e$& $-\left[\eta^r J_{sc} {{\omega }_{c}}+{{h}_{o}} \right]\left[ {\dot{\delta }_{c}} +{\dot{\delta }_{o}} \right]\hat{t}$\\
\midrule
\multirow{6}*{$F_e$}&$N$&$-\left[ J_{sc} {{\omega }_{c}}+{{h}_{o}} \right] {{{\dot{\delta }}}_{c}}\hat{t}$\\
&$F_a$&$-\left[ J_{sc} {{\omega }_{c}}+{{h}_{o}} \right]\left[ \eta^g {{{\dot{\delta }}}_{c}} \right]\hat{t}$\\
&$F_b$&$0$\\
&$F_c$& $-\left[ J_{sc} {{\omega }_{c}}+{{h}_{o}} \right]\left[\eta^g {{{\dot{\delta }}}_{c}}+{{{\dot{\delta }}}_{o}} \right]\hat{t}$\\
&$F_d$& $-\left[J_{sc} {{\omega }_{c}}+{{h}_{o}} \right]{{\dot{\delta }}_{o}}\hat{t}$\\
&$F_e$& $-\left[ J_{sc} {{\omega }_{c}}+{{h}_{o}} \right]\left[  {{{\dot{\delta }}}_{c}} +{{{\dot{\delta }}}_{o}} \right]\hat{t}$\\
\bottomrule
\end{tabular}
\small
Note: $\eta^r, \eta^g \in (0,1)$ in this table, $h_{*} = h_c$, command angular momentum for working condition N, and $h_{*} = h_o$, pure angular momentum offset for working condition $F_d$.
\end{threeparttable}
\end{table}

Noting that the saturation is not treated as fault since it is a kind of constraints.
Based on \eqref{26}, it is clear that different combinations of $\eta^r,\eta^g, h_o$ and $\dot{\delta}_o$ represent the different fault scenarios in a SGCMG.
To illustrate all potential faults and fault-free situation of a SGCMG, the potential fault cases in each single control loop (either rotor control loop or gimbal frame control loop) are defined as:
\begin{itemize}
\item[$N$]: Nominal working condition;
\item[$F_a$]: Partially lose effect, without offset;
 \item[$F_b$]: Totally fail, without offset;
\item[$F_c$]: Partially lose effect and have offset;
\item[$F_d$]: Totally fail and have offset;
\item[$F_e$]: Pure offset without losing effect.
\end{itemize}

Combining these different work conditions, the fault model list of SGCMG is obtained as shown in Table \ref{tab1}. This model is consistent with that in \cite{zhang2017fault}, where only the gimbal fault is considered. To the best knowledge of authors, this is the first attempt to establish a systematic fault model of SGCMG.

\subsection{DGCMG Fault Model}
Double gimbal control moment gyro is a flywheel with a constant angular speed mounted on an orthogonal installed gimbal frame. Different from the SGCMG, a DGCMG has two gimbals including the inner gimbal $\hat{g}_i$ and outer gimbal $\hat{g}_o$. All these gimbals together with the direction of angular momentum $\hat{h}$ form the CMG frame, denoted as $\mathbb{G}=\left\{ \left. \hat{h},{{{\hat{g}}}_{i}},{{{\hat{g}}}_{o}} \right\} \right.$.
Therefore, the DGCMG can be modeled by three EM-VSD systems working independently.
Similar to the mechanism of SGCMG, the output torque is generated by rotating gimbals \cite{leve2015spacecraft}. Based on the independence assumption, the torques generated by inner gimbal and outer gimbal are independent. Then the nominal overall control torque of a DGCMG can be expressed as \cite{sasaki2016fault}:
\begin{equation}\label{27}
 \tau = \tau_i + \tau_o = -{{h}_{0}}{{\dot{\delta }}_{i}}{{\hat{t}}_{i}}-{{h}_{0}}{{\dot{\delta }}_{o}}{{\hat{t}}_{o}},
\end{equation}
where
$\tau_i$ and $\tau_o$ are the torques generated by the inner gimbal loop and outer gimbal loop,
$\dot{\delta}_i$ is the gimbal rate of the inner gimbal, $\dot{\delta}_o$ is the gimbal rate of the outer gimbal, and $\hat{t}_i$ and $\hat{t}_o$ are the directions of the inner and outer output torque.
Then for the inner gimbal and outer gimbal, the fault models are:
\begin{equation}\label{28}
 {{\tau }_{i}}=-\left[ \eta^{r}J_{dc} {{\omega }_{c}}+{{h}_{o}} \right]\left[ {{\eta }^{g_i}}{{{\dot{\delta }}}_{{{c}_{i}}}}+{{{\dot{\delta }}}_{{{o}_{i}}}} \right]{{\hat{t}}_{i}}
\end{equation}
and
\begin{equation}\label{29}
{{\tau }_{o}}=-\left[ \eta^{r}J_{dc} {{\omega }_{c}}+{{h}_{o}} \right]\left[ {{\eta }^{g_o}}{{{\dot{\delta }}}_{{{c}_{o}}}} +{{{\dot{\delta }}}_{{{o}_{o}}}} \right]{{\hat{t}}_{o}},
\end{equation}
respectively. The subscript ``$dc$" represents DGCMG, ``$i$'' represents inner gimbal and ``$o$'' represents outer gimbal.


\subsection{VSCMG Fault Model}
VSCMG can be considered as a combination of RW and  CMGs. It has two major categories: single gimbal VSCMG (SGVSCMG) and double gimbal VSCMG (DGVSCMG). The fault model of these two types of VSCMG are addressed in the following.

\emph{a). SGVSCMG}

  This SGVSCMG is a combination of a RW and a SGCMG, so it has two working modes, i.e., RW working mode and SGCMG working mode. The output torque is generated by the acceleration in the direction of angular momentum working as the RW and the gimbal rotation working as the SGCMG \cite{schaub2000singularity, yoon2004singularity}. The nominal output torque of a SGVSCMG can be expressed as:
  \begin{equation}\label{30}
  \tau = \tau_r + \tau_c =-J_{svs}\dot{\omega}_{svs}{{\hat{t}}_{r}}-h\left( \omega_{svs}  \right)\dot{\delta }{{\hat{t}}_{g}},
  \end{equation}
  where
  $\tau_r$ and $\tau_c$ are output torque generated in the RW working mode and the SGCMG working mode, and the subscript ``$svs$'' represents SGVSCMG.
  Considering the potential fault in RW, the fault model of the RW working mode is:
\begin{equation}\label{31}
{{\tau }_{r}}=-\left[ {{\eta }^{{{r}_{svs}}}}{{J}_{svs}}{{{\dot{\omega }}}_{{{c}_{svs}}}} +{{\tau }_{{{o}_{{{svs}}}}}} \right]{{\hat{t}}_{r}}.
  \end{equation}
When the SGVSCMG is in the SGCMG working mode, only the measurement of angular momentum, or the angular velocity equivalently, is used to design the gimbal rate command. Thus we just need to consider the additive fault caused by the sensors in the rotor control loop when SGVSCMG is in SGCMG working mode. Then the fault model of the SGCMG working mode can be established as:
\begin{equation}\label{32}
{{\tau }_{c}}=-\left[ J_{svs} {{\omega }_{c_{svs}}}+{{h}_{o_{svs}}} \right]\left[ {{\eta }^{g}}{{{\dot{\delta }}}_{c}} +{{{\dot{\delta }}}_{o}} \right]{{\hat{t}}_{g}}.
\end{equation}


\emph{b). DGSGCMG}

The DGSGCMG can be regarded as a combination of the RW and the DGCMG. The output torque contains three parts: 1) $\tau_r$ caused by rotor acceleration; 2) $\tau_{g_i}$ generated by by inner gimbal rotation; and 3) ${\tau}_{g_o}$ caused by outer gimbal rotation. The output torque can be expressed as\cite{sasaki2016fault,cui2013steering}:
\begin{equation}\label{33}
\tau =\tau_r + \tau_{g_i} + \tau_{g_o}=-J_{dvs}\dot{\omega }_{dvs}{{\hat{t}}_{r}}-h\left( \omega_{dvs}  \right){{\dot{\delta }}_{{{g}_{i}}}}{{\hat{t}}_{{{g}_{i}}}}-h\left( \omega_{dvs}  \right){{\dot{\delta }}_{{{g}_{o}}}}{{\hat{t}}_{{{g}_{o}}}},
\end{equation}
where the subscript ``$dvs$'' represents DGVSCMG.
The fault model of RW working mode is same as equation \eqref{31}. The fault relating to theDGCMG is described as
\begin{equation}\label{34}
{{\tau }_{{{g}_{i}}}}=-\left[ J_{dvs} {{\omega }_{c_{dvs}}}+{{h}_{o_{dvs}}} \right]\left[ {{\eta }^{{{g}_{i}}}}{{{\dot{\delta }}}_{{{i}_{c}}}} +{{{\dot{\delta }}}_{{{i}_{o}}}} \right]{{\hat{t}}_{{{g}_{i}}}}
\end{equation}
and
\begin{equation}\label{35}
{{\tau }_{{{g}_{o}}}}=-\left[ J_{dvs} {{\omega }_{c_{dvs}}}+{{h}_{o_{dvs}}} \right]\left[  {{\eta }^{{{g}_{o}}}}{{{\dot{\delta }}}_{{{o}_{c}}}}+{{{\dot{\delta }}}_{{{o}_{o}}}} \right]{{\hat{t}}_{{{g}_{o}}}}.
\end{equation}


\section{Simulation Demonstration}
This Section demonstrates attitude control results of RWs or SGCMGs actuated spacecraft under different fault scenarios. The severities of the faults are qualitatively analyzed to give a guideline for the fault-tolerant control system design.

\subsection{Fault Effects in EM-VSD}
\begin{figure}[h]
  \centering
    \includegraphics[scale=0.72]{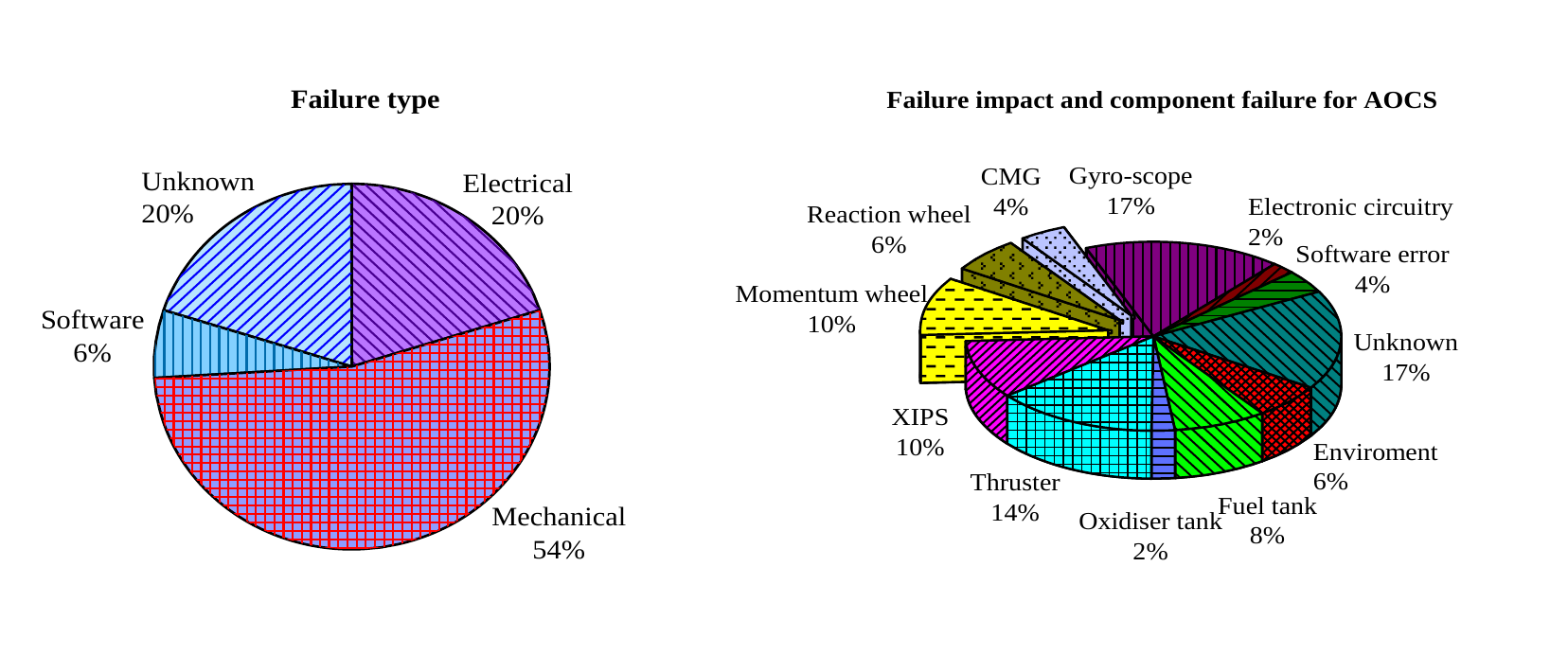}
  \caption{\ychf{AOCS failure distributions \cite{tafazoli2009study}}}
\label{figadd}
\end{figure}

\ychf{According to the failure analysis reported in \cite{tafazoli2009study}, 32\% of the failures comes from the Attitude and Orbit Control System (AOCS), which is a combination of the Attitude Determination and Control System (ADCS) and Guidance, Navigation and Control (GNC) subsystem. Among the failures of AOCS, 44\% of them are related to actuators including Thrusters, RWs, CMG and XIPS. For the failure type, more than half of the failures (54\%) are mechanical, and a relative small portion (20\%) are electrical. The detailed analysis are shown in Fig. \ref{figadd}. }

\begin{figure}[!hb]
  \centering
  \subfloat[Condition $N$ and $F_a$]{
    \label{fig5a}
    \includegraphics[scale=0.42]{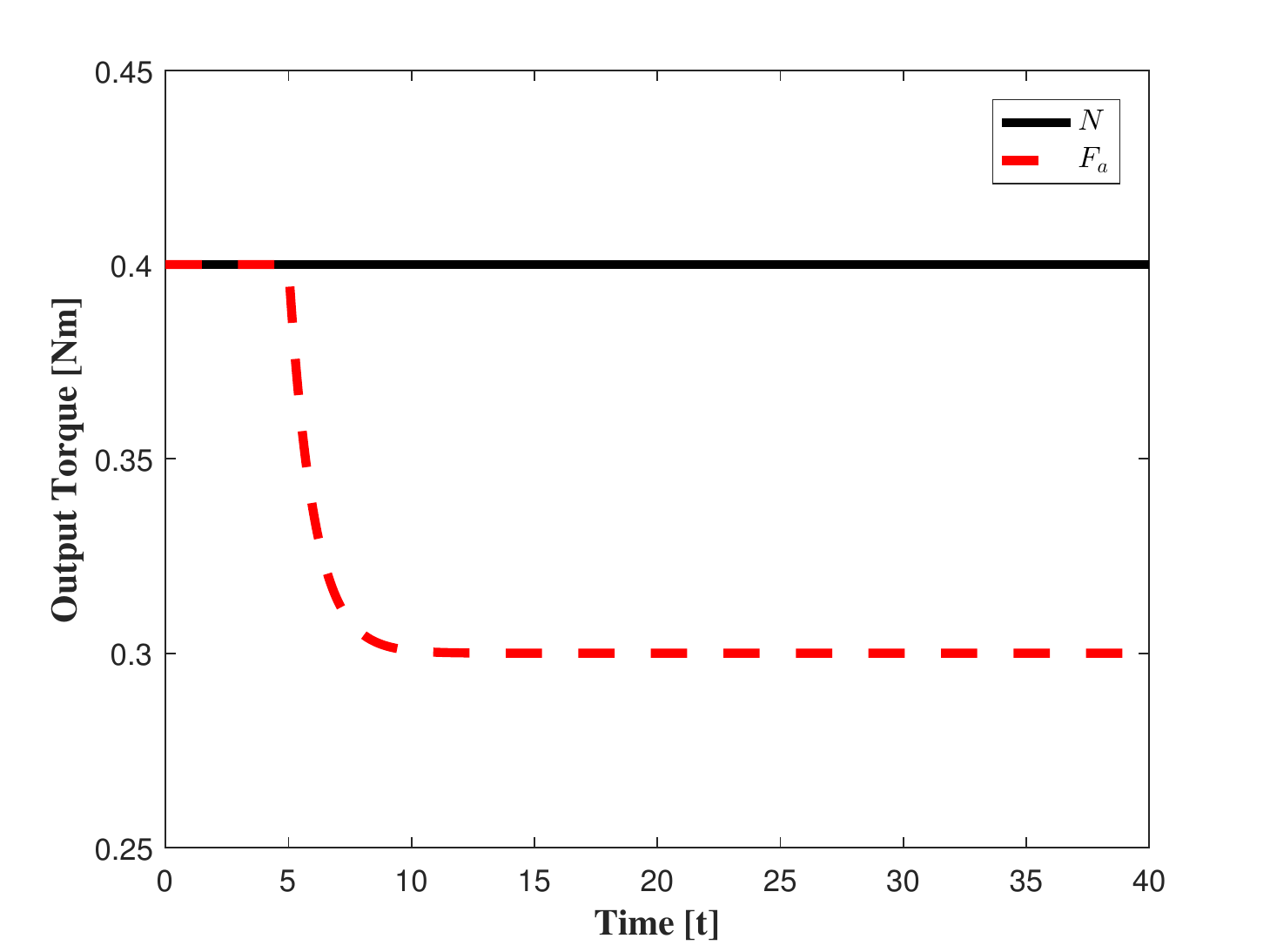}}
  \subfloat[Condition $N$ and $F_b$]{
    \label{fig5b}
    \includegraphics[scale=0.42]{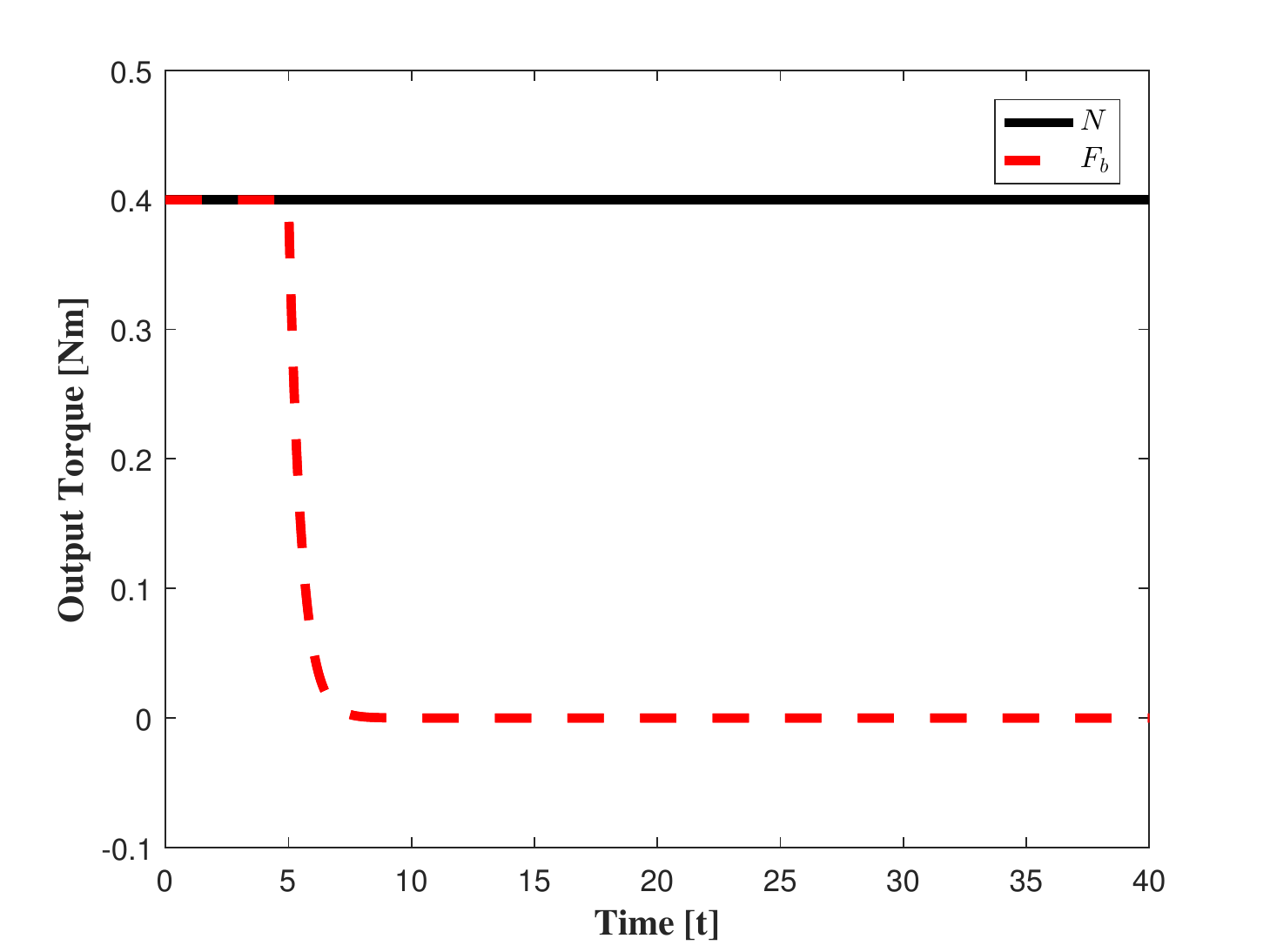}}

 \subfloat[Condition $N$ and $F_c$]{
    \label{fig5c}
    \includegraphics[scale=0.42]{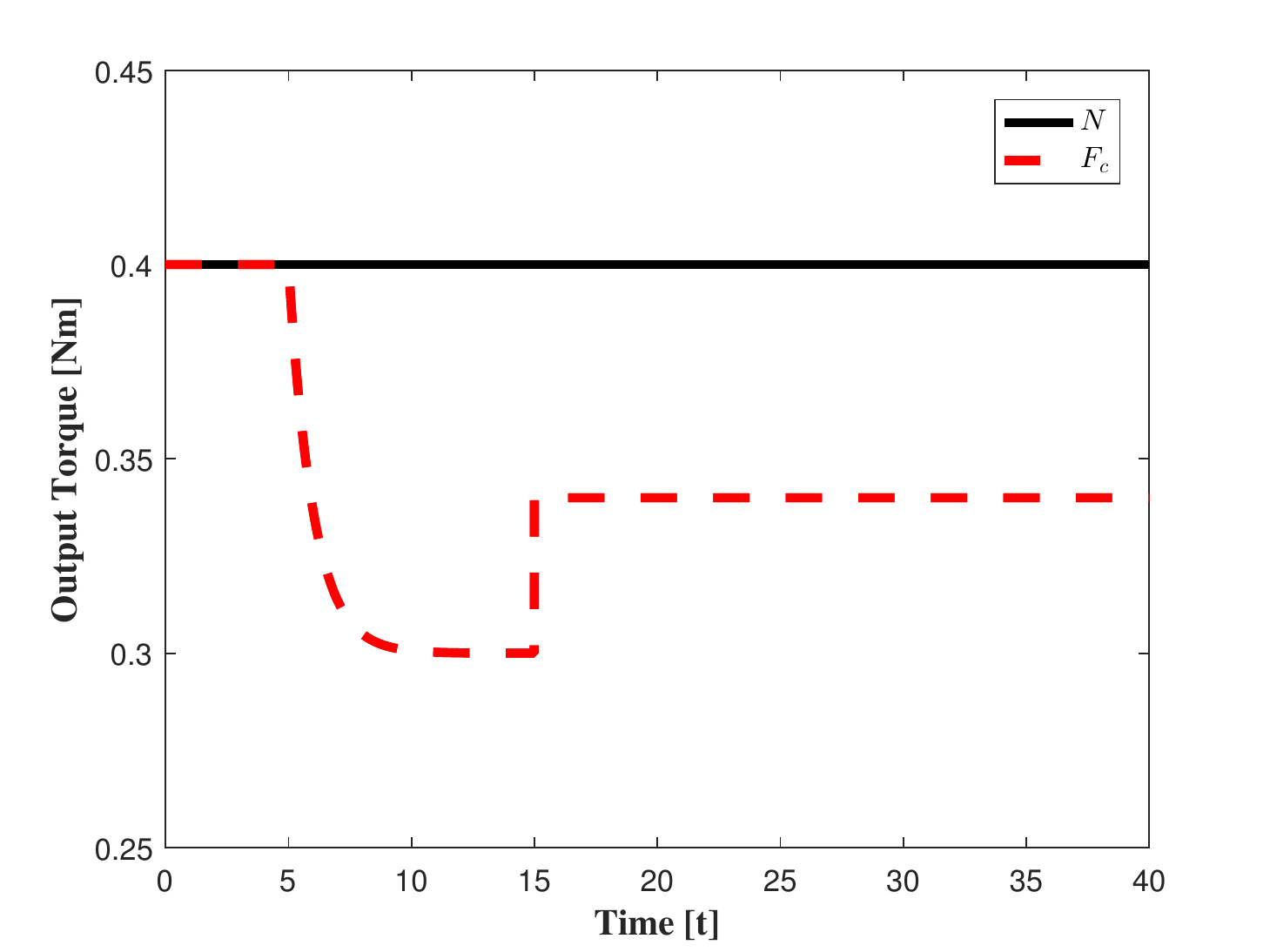}}
  \subfloat[Condition $N$ and $F_d$]{
    \label{fig5d}
    \includegraphics[scale=0.42]{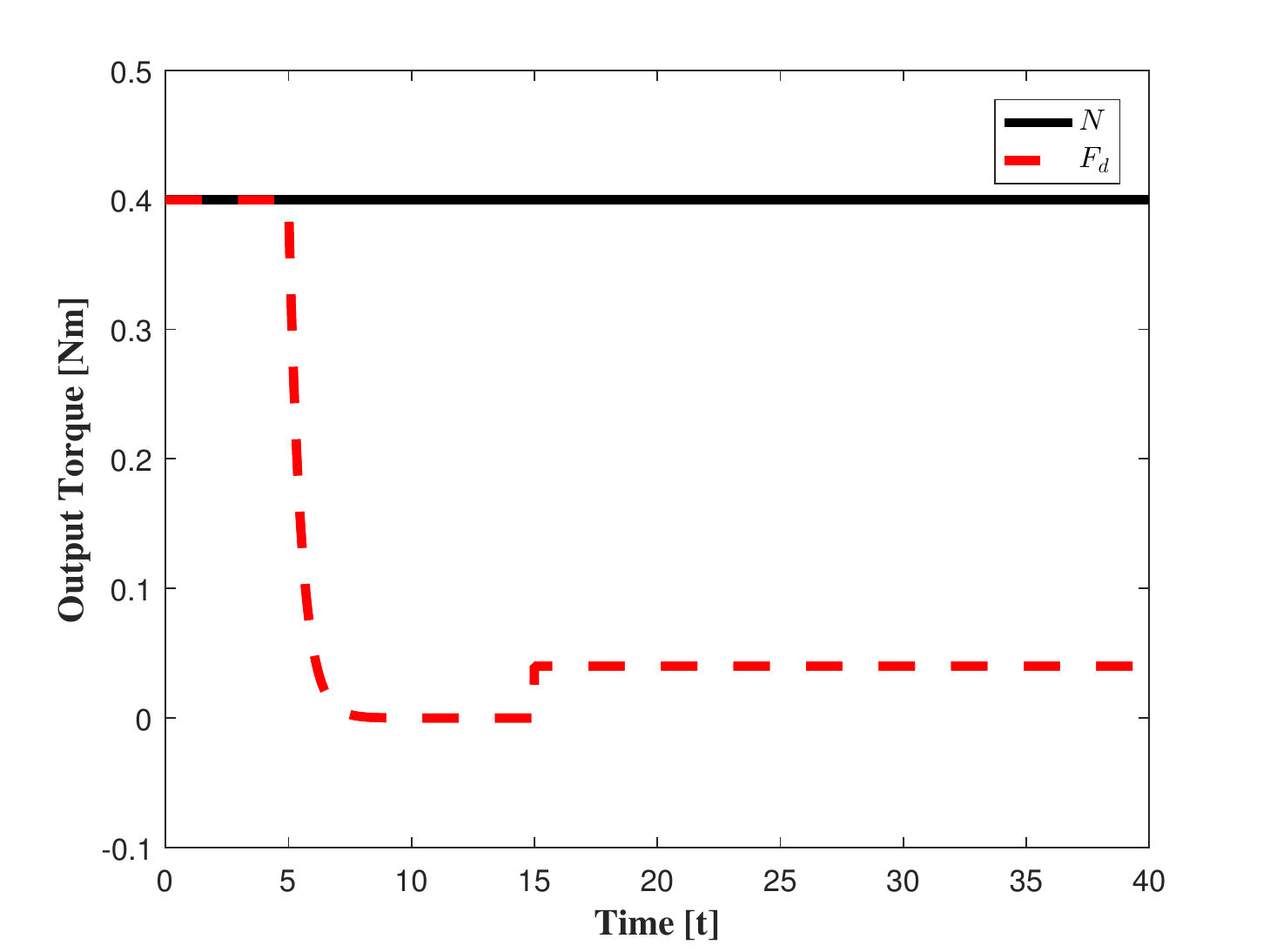}}
  \caption{Fault scenarios of RWs}
  \label{fig5}
\end{figure}

To demonstrate the EM-VSD performance under different fault conditions,
all the listed potential faults $F_a$ to $F_e$ in Section \ref{sec_SGCMG_F} are compared with the nominal condition. Taking the practical variation of possible fault and failure into consideration, an exponential function $u_{out} = \eta + (1-\eta) e^{- t_a (t - t_c)}$ is adopted to describe the dynamic characteristic of the multiplicative fault and this exponential function is also used in the SGCMG-actuated spacecraft simulations. In this equation, $\eta$ represents the effectiveness factor of the motor after the occurrence of fault or failure, $t_a$ represents the time constant for the fault or failure, and $t_c$ is the time instant that fault or failure happens. In the simulation, the time constant $t_a$ is chosen to be $2$ and $1$ for fault and failure, respectively.
In the nominal condition $N$, the output torque is set as 0.4 Nm.
The actuator is assumed to lose its effectiveness at $t=5 s$ and the bias offset is added at $t=15 s$. As shown in Fig.\ref{fig5}, the effectiveness factor $\eta$ is chosen as $0.75$ and the offset is chosen as $0.04$, which is $10\%$ of the nominal command. These parameters are also used in the RW-actuated spacecraft simulation. For the working condition $F_e$, it is similar to part of the Fig. \ref{fig5c} and Fig. \ref{fig5d} when the offset is added and the demonstration is omitted here.

%

\subsection{Attitude Control Results of RW-Atuated Spacecraft} \label{RW_Sim}
\ychf{Before we move forward to the demonstrate the control results of RW-actuated and CMG-actuated spacecraft under actuator faults, the dynamics and kinematics of the system can be found in \cite{cao2016time} as:}
\begin{equation}\label{dyandkin}
\ychf{\left\{ \begin{matrix}
   \bm{\dot{q}}=\frac{1}{2}\left[ \begin{matrix}
   -\bm{q}_{v}^{T}  \\
   {{q}_{o}}{{\bm{I}}_{3}}+\bm{q}_{v}^{\times }  \\
\end{matrix} \right]\bm{\omega }\quad \quad \quad \quad \quad   \\
   \bm{J\dot{\omega }}+{{\bm{\omega }}^{\times }}\left( \bm{J\omega }+\bm{H} \right)=\bm{\tau}_c+\bm{d}  \\
\end{matrix} \right.}
\end{equation}
\ychf{where $\bm{q}=[q_0,\ \bm{q}_v^T]^T$ is the unit quaternion, $\bm{\omega}$ is the angular velocity of the spacecraft, $\bm{J} = \bm{J}^T$ is moment of inertia matrix, $\bm{H}$ is the angular momentum of the actuators and $\bm{d}$ is the external disturbance. $\bm{\tau}_c=- \dot{\bm{H}}$ is the output of the actuators. }

To have a better demonstration the influence of the RW fault/failure to the attitude control system, we do not consider RW redundancy in the simulation. \ychf{Then the installation matrix of the RWs is identity.  Considering the fault model of the RW given by \eqref{22}, the output of the RW cluster can be given by: }
\begin{equation}\label{rwfault}
  \bm{\tau} = \bm{E}_{\eta} \bm{u}+\bm{E}_o
\end{equation}
\ychf{where $\bm{u}$ is the control command calculated by the controller, $\bm{E}_{\eta} = \text{diag}[\eta_1, \ \eta_2, \ \eta_3]$ is the effectiveness matrix and $\bm{E}_o = \text{diag}[u_{o1}, \ u_{o2}, \ u_{o3}]$ is the offset caused by the RW faults.}

\begin{figure}[!b]
  \centering
  \subfloat[Euler angle]{
    \label{fig7a}
    \includegraphics[scale=0.42]{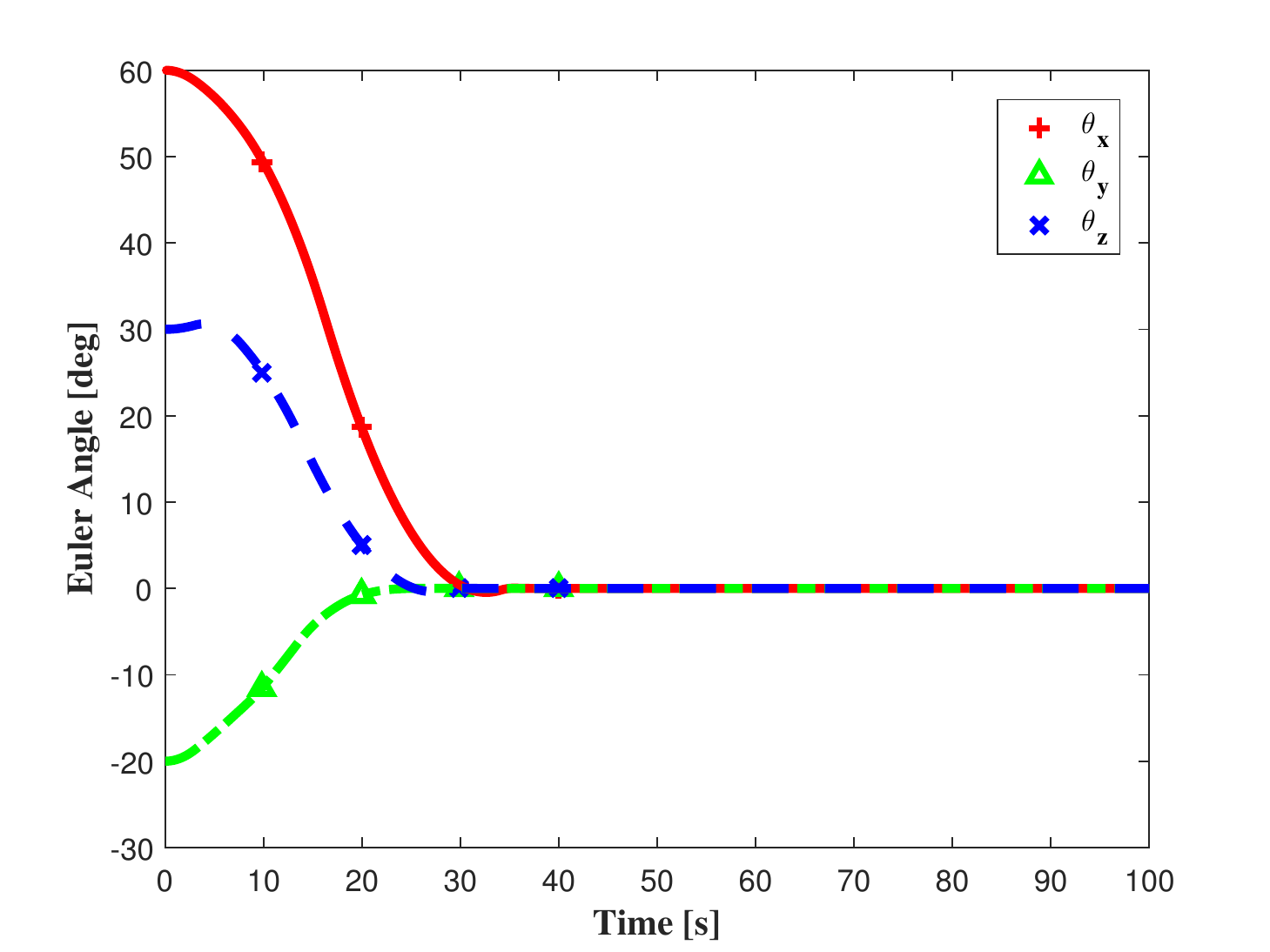}}
  \subfloat[Angular velocity]{
    \label{fig7b}
    \includegraphics[scale=0.42]{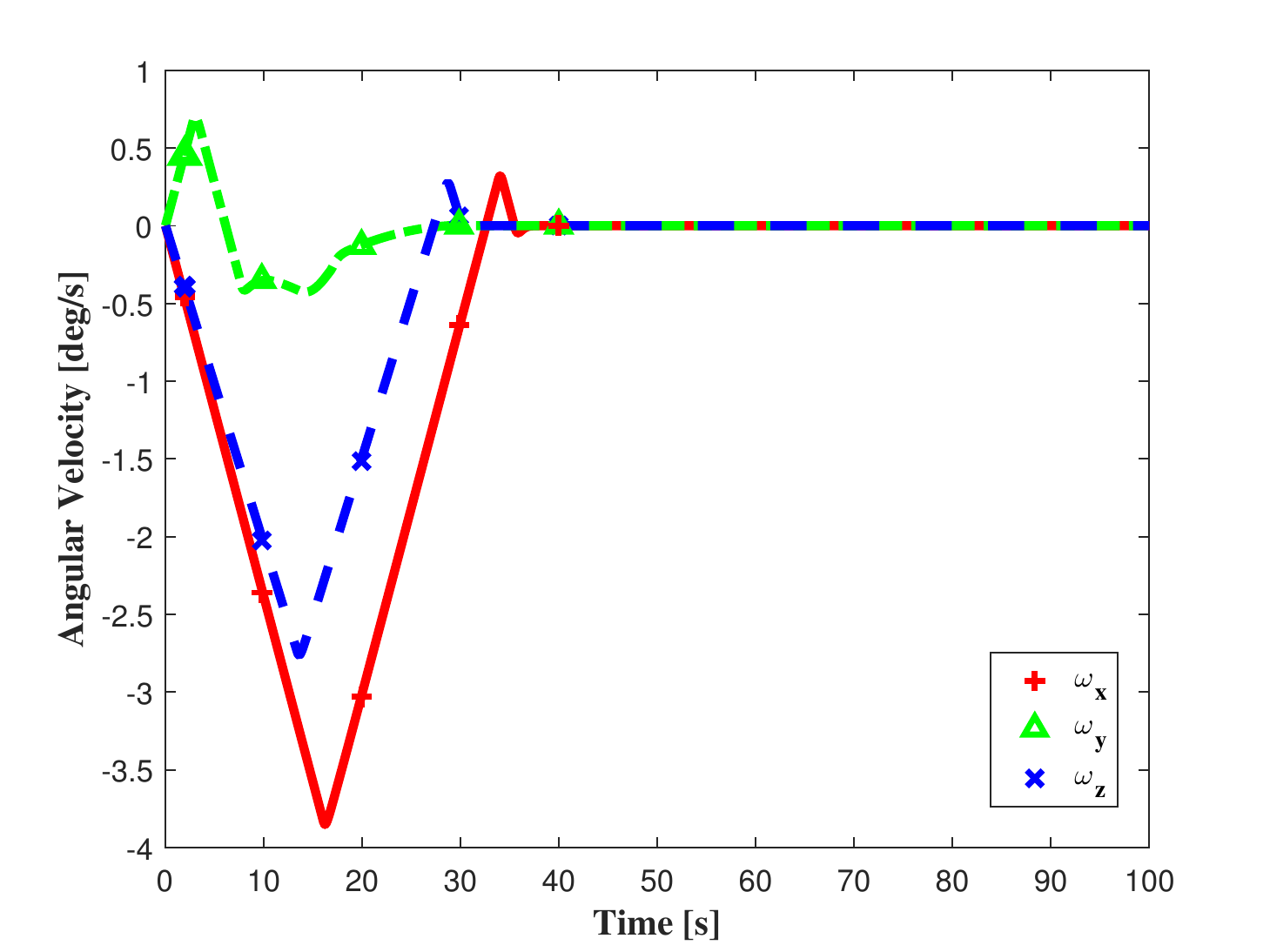}}

\subfloat[Output control torque]{
    \label{fig7c}
    \includegraphics[scale=0.42]{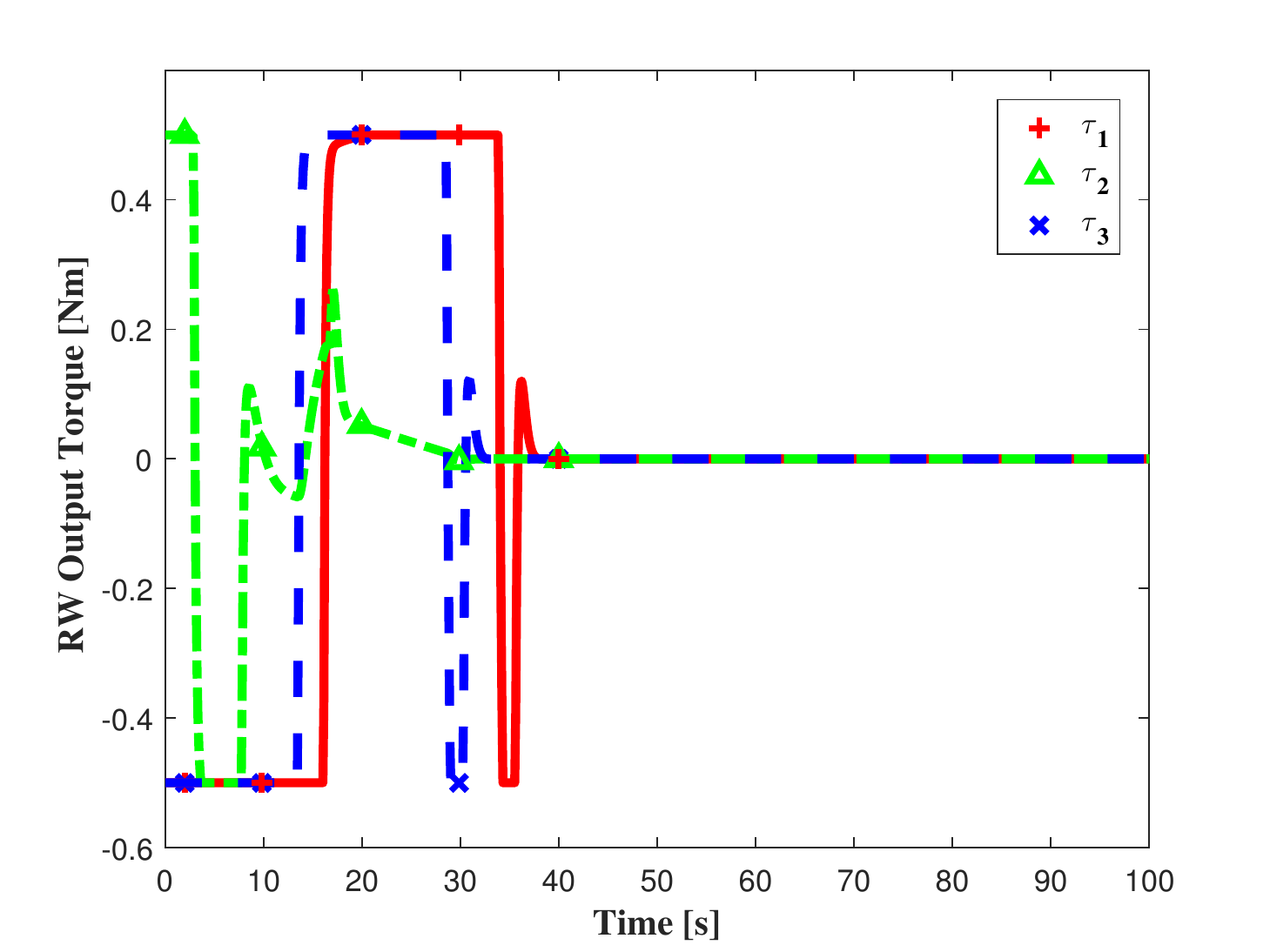}}
  \subfloat[Angular momentum]{
    \label{fig7d}
    \includegraphics[scale=0.42]{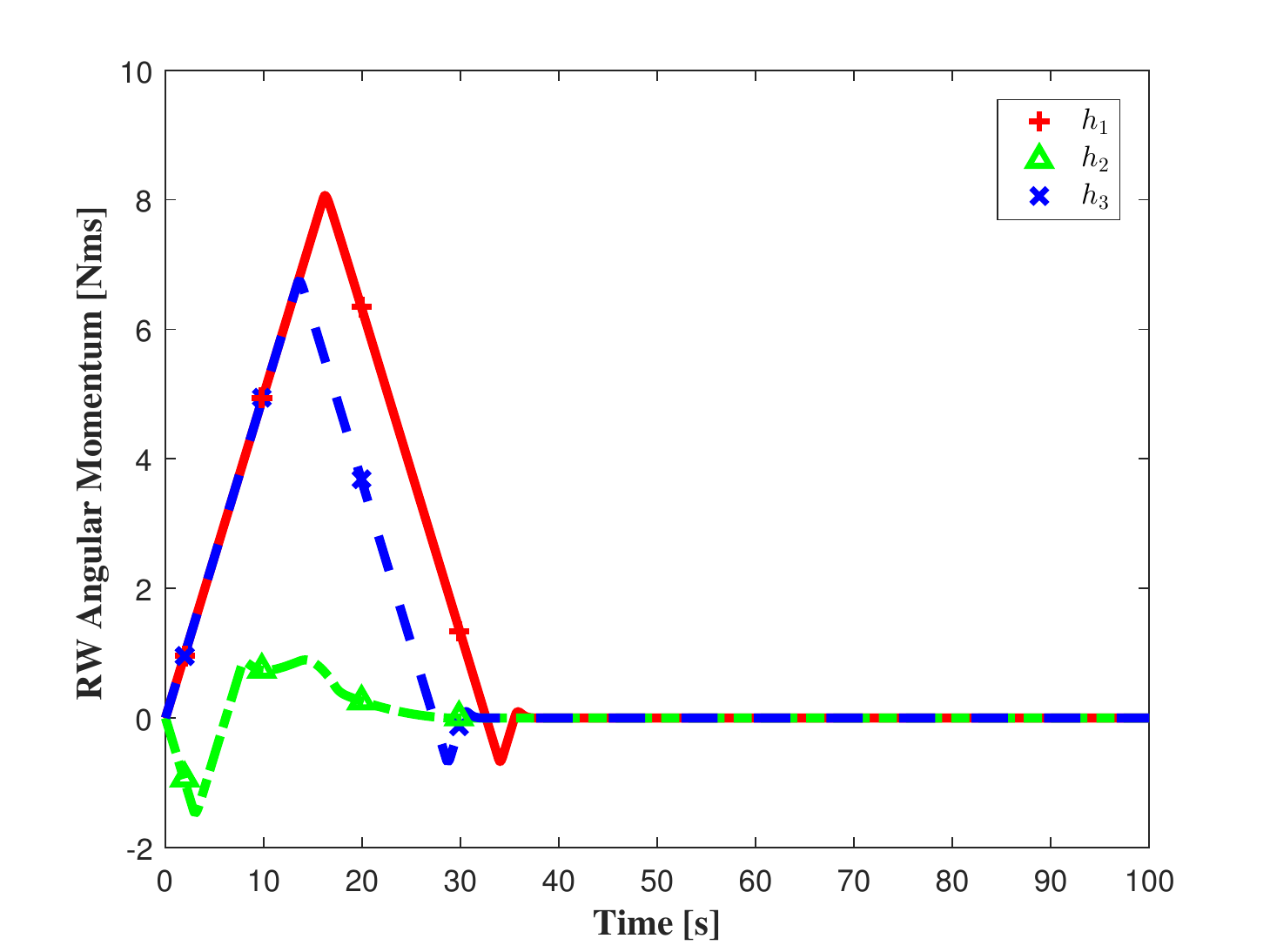}}
  \caption{RW-actuated attitude control result under nominal condition}
  \label{fig7}
\end{figure}

\begin{figure}[!t]
  \centering
  \subfloat[Euler angle]{
    \label{fig8a}
    \includegraphics[scale=0.42]{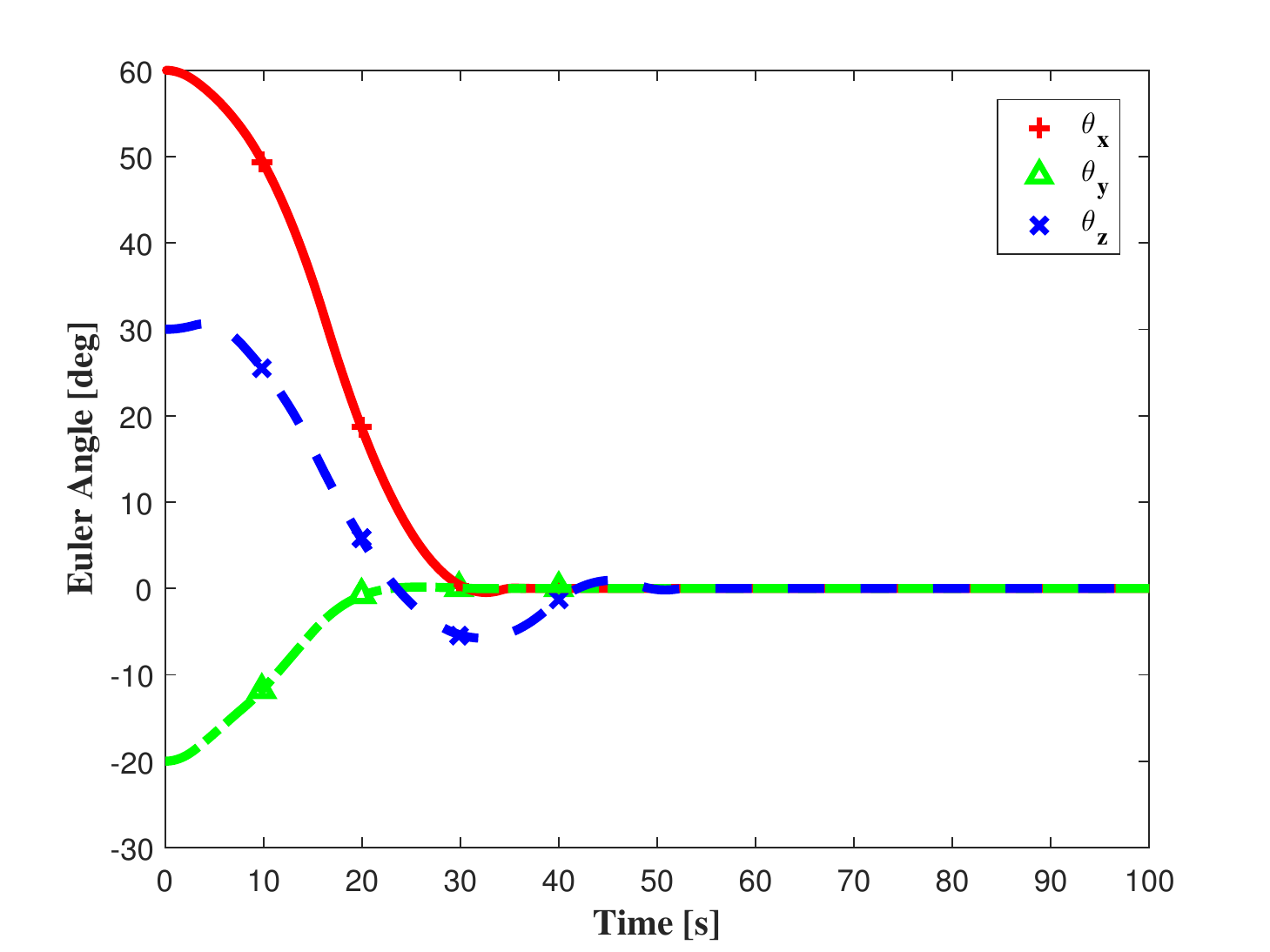}}
  \subfloat[Angular velocity]{
    \label{fig8b}
    \includegraphics[scale=0.42]{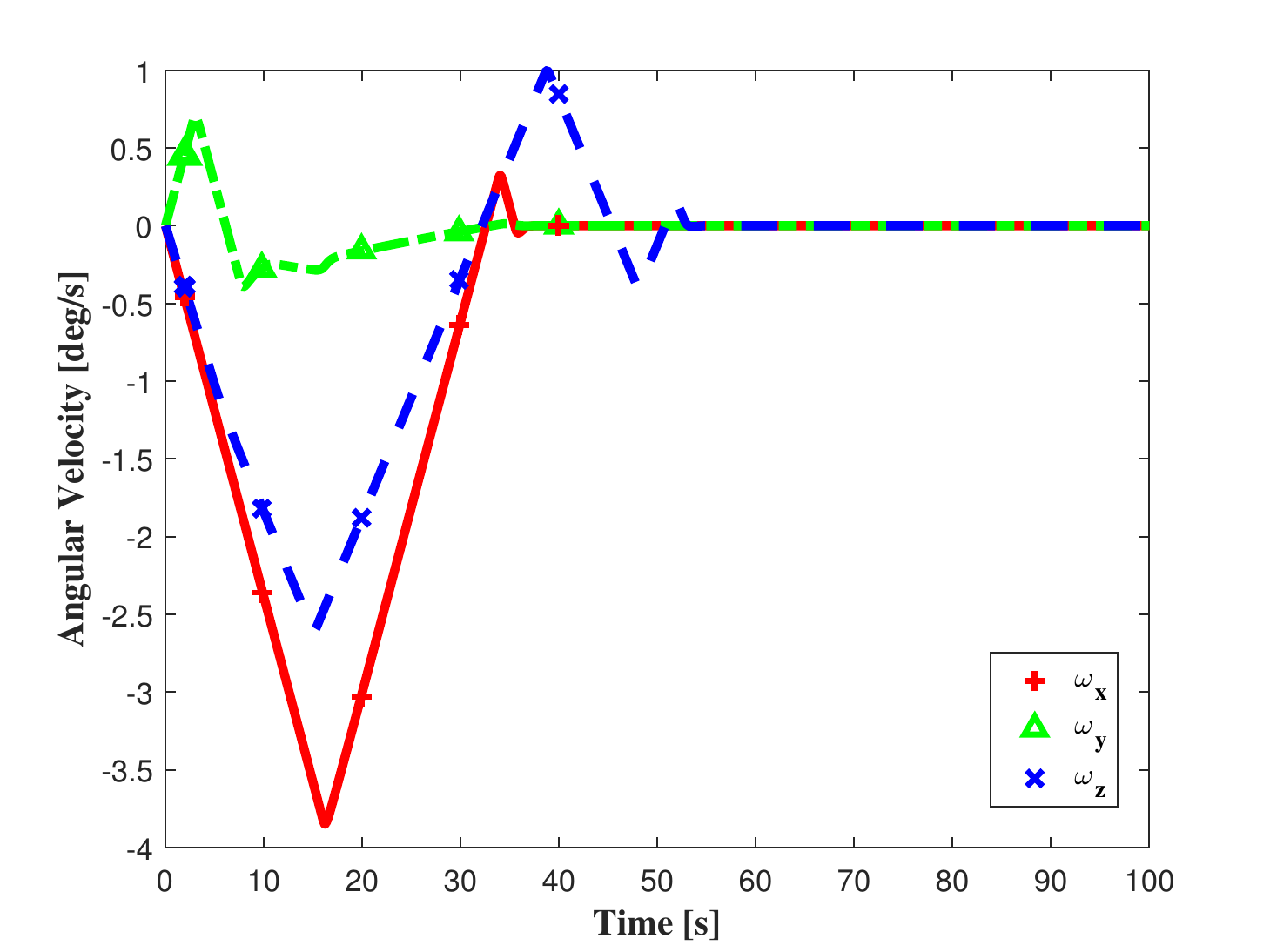}}

\subfloat[Output control torque]{
    \label{fig8c}
    \includegraphics[scale=0.42]{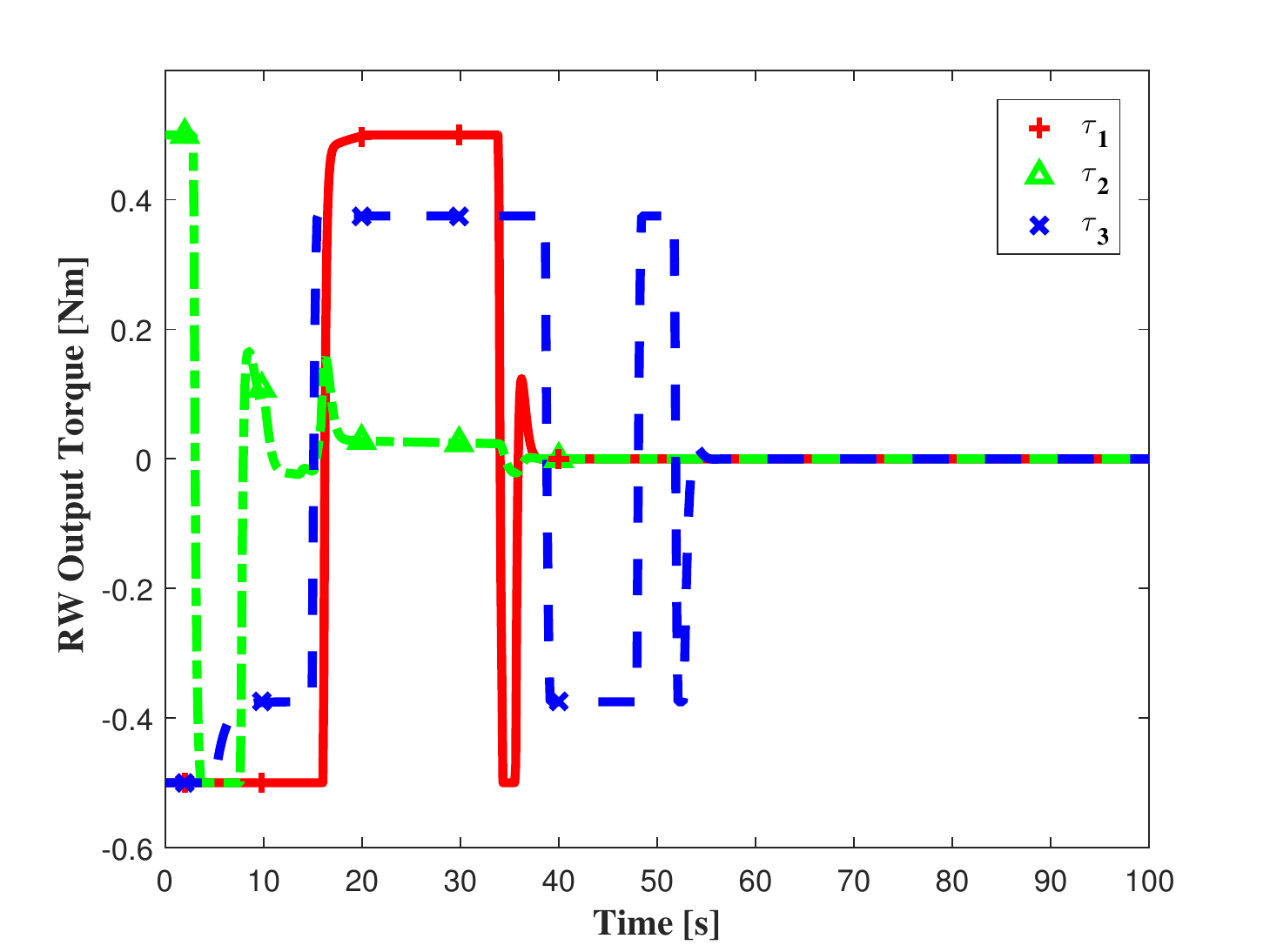}}
  \subfloat[Angular momentum]{
    \label{fig8d}
    \includegraphics[scale=0.42]{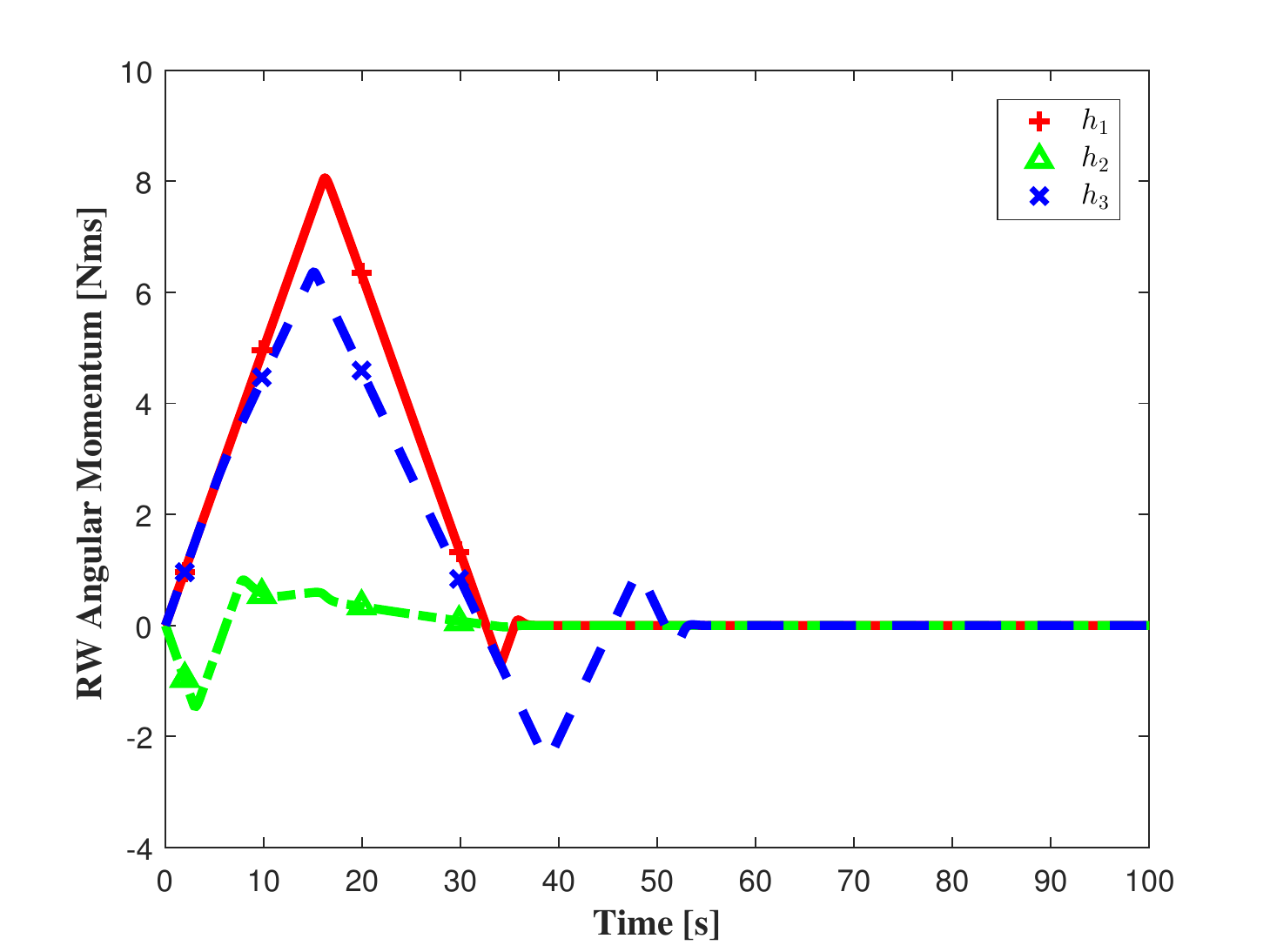}}
  \caption{RW-actuated attitude control result under $F_a$}
  \label{fig8}
\end{figure}

\begin{figure}[!t]
  \centering
  \subfloat[Euler angle]{
    \label{fig9a}
    \includegraphics[scale=0.42]{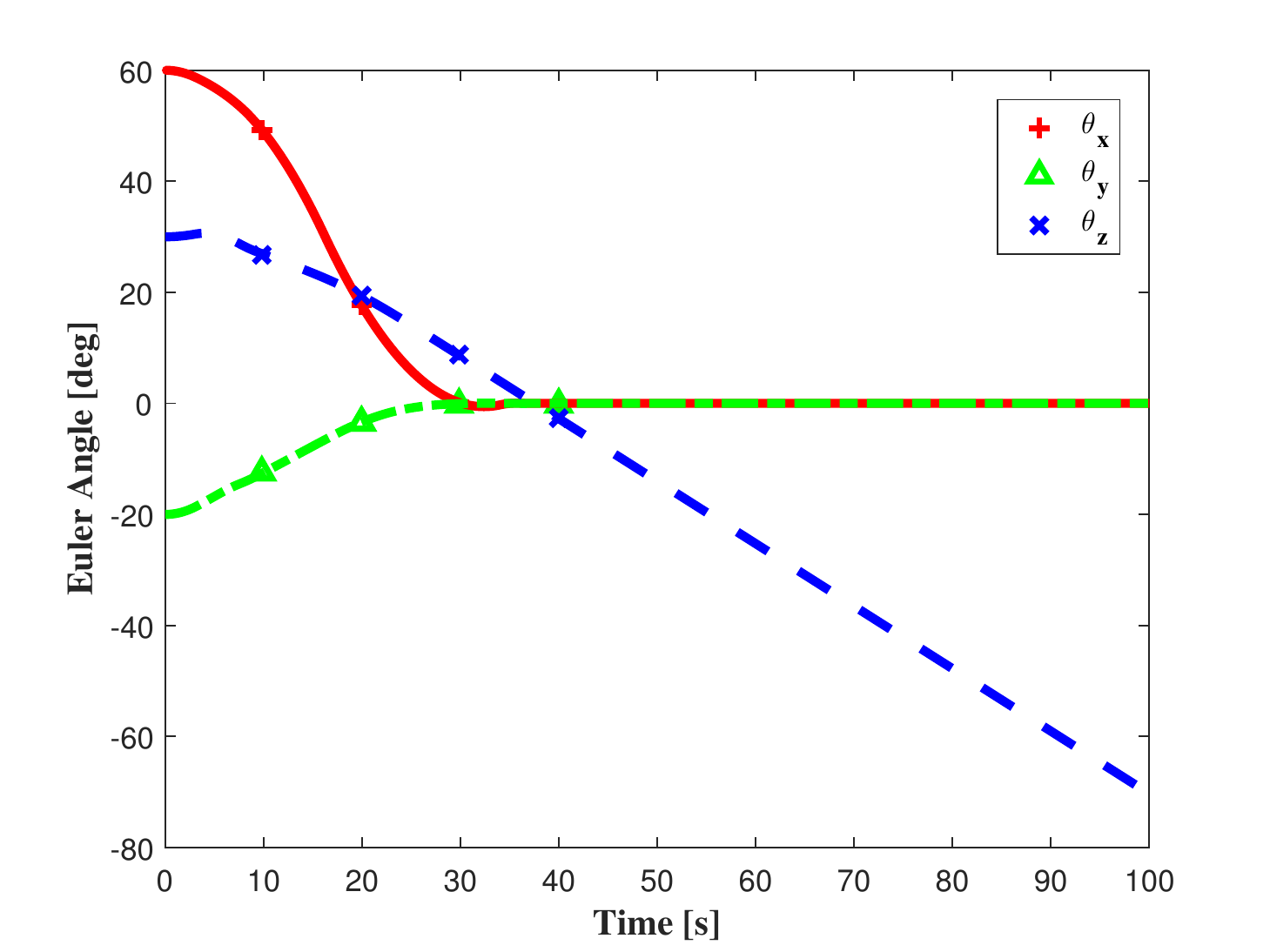}}
  \subfloat[Angular velocity]{
    \label{fig9b}
    \includegraphics[scale=0.42]{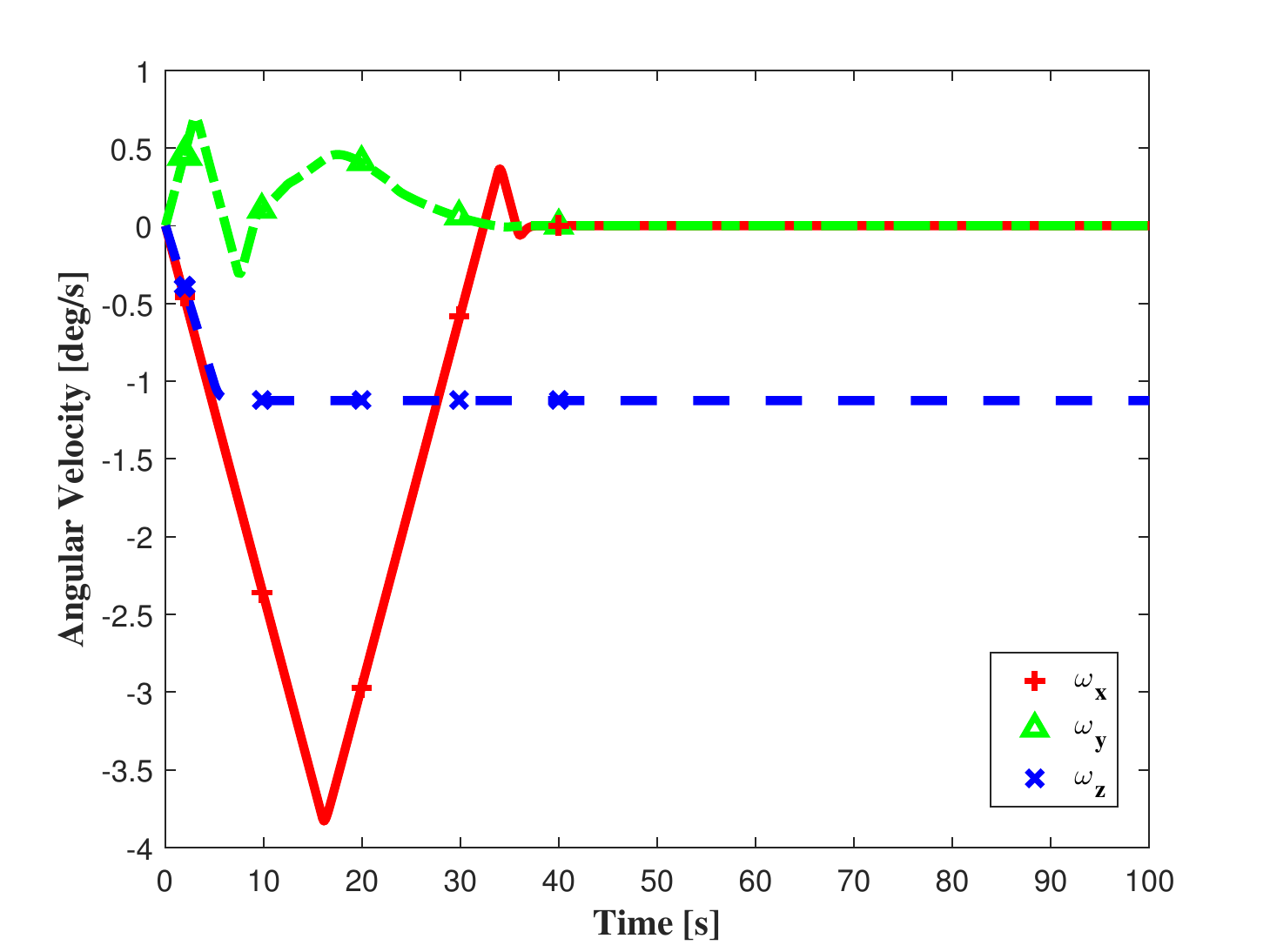}}

\subfloat[Output control torque]{
    \label{fig9c}
    \includegraphics[scale=0.42]{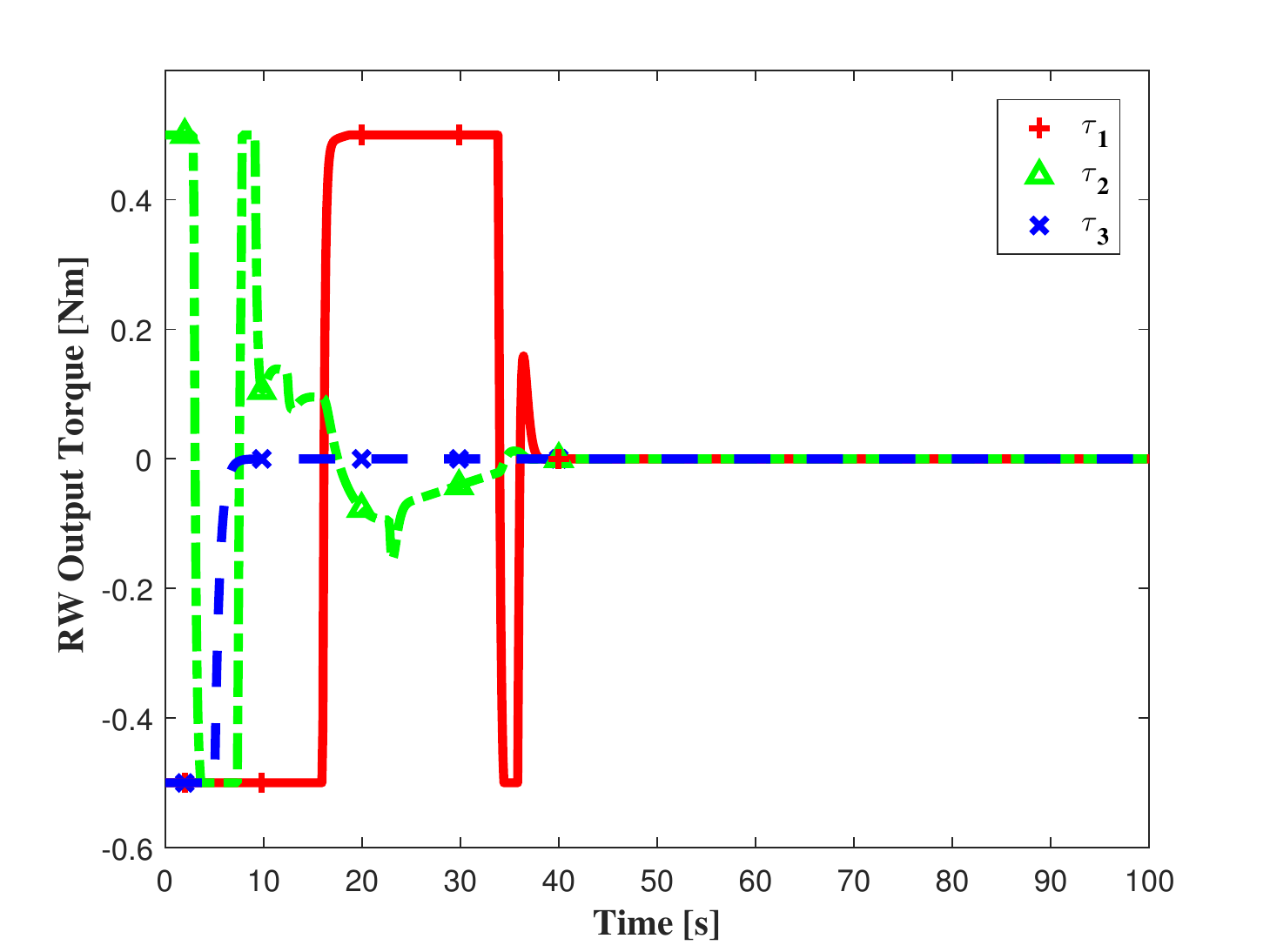}}
  \subfloat[Angular momentum]{
    \label{fig9d}
    \includegraphics[scale=0.42]{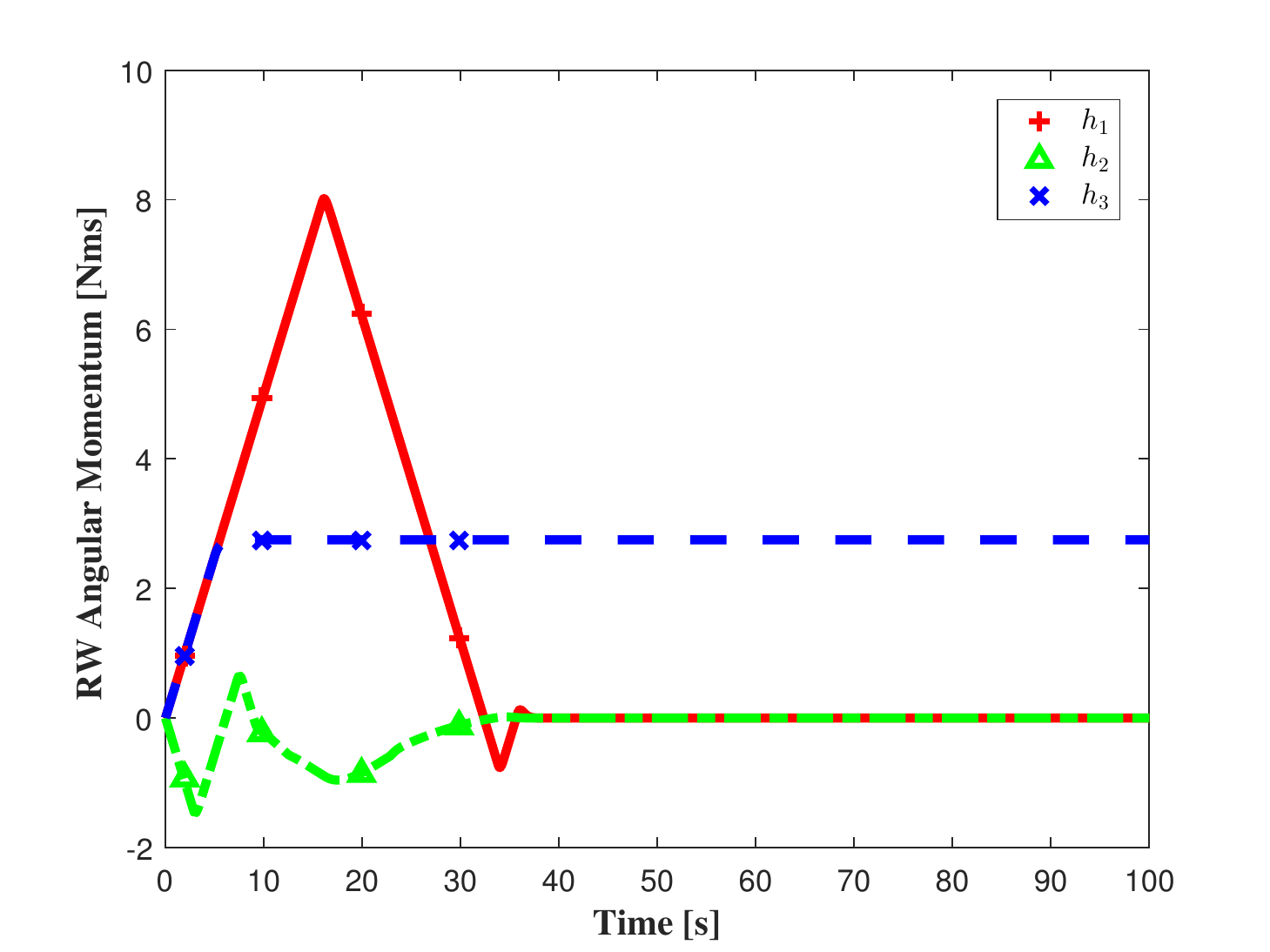}}
  \caption{RW-actuated attitude control result under $F_b$}
  \label{fig9}
\end{figure}

\begin{figure}[!t]
  \centering
  \subfloat[Euler angle]{
    \label{fig10a}
    \includegraphics[scale=0.42]{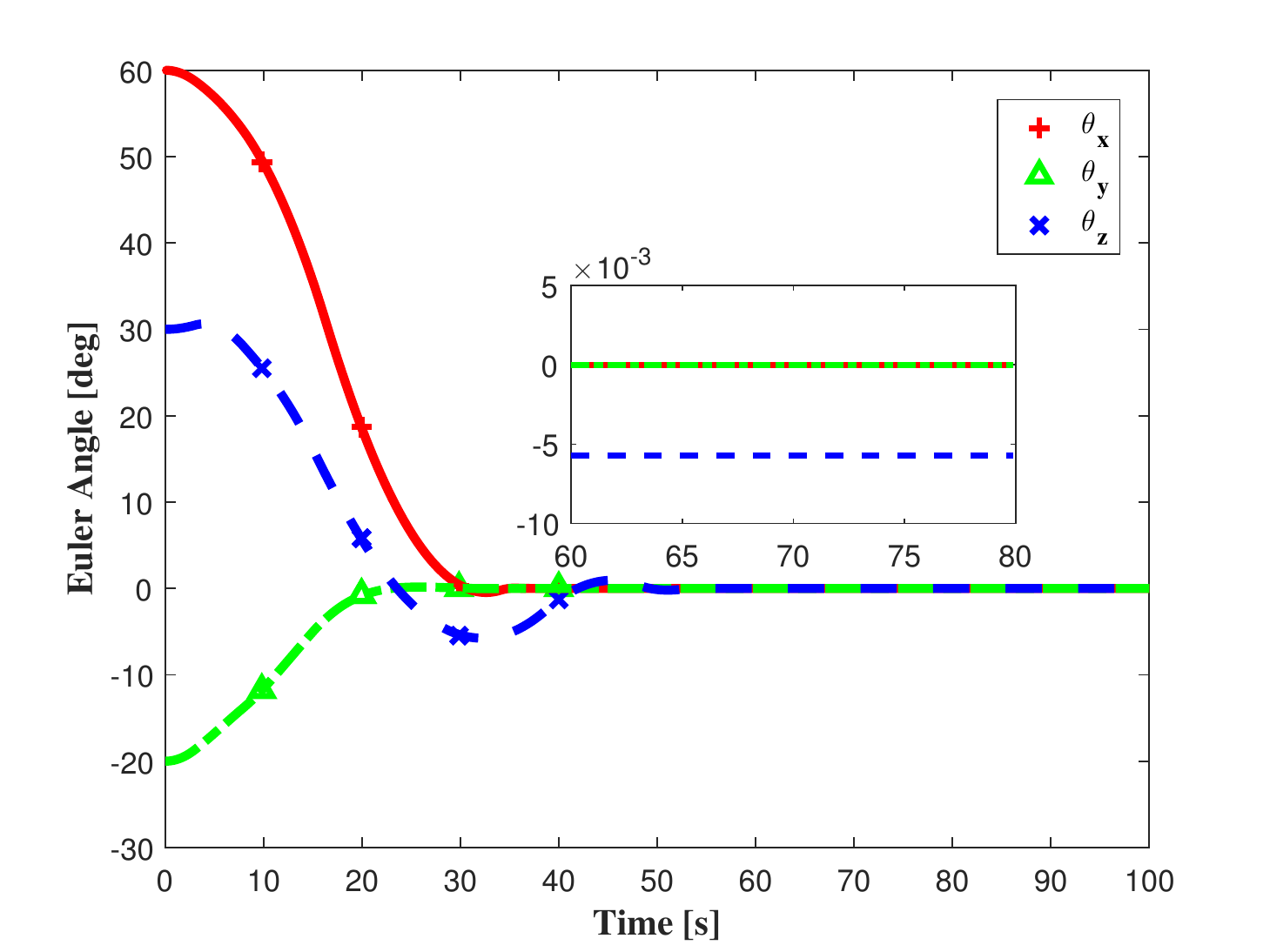}}
  \subfloat[Angular velocity]{
    \label{fig10b}
    \includegraphics[scale=0.42]{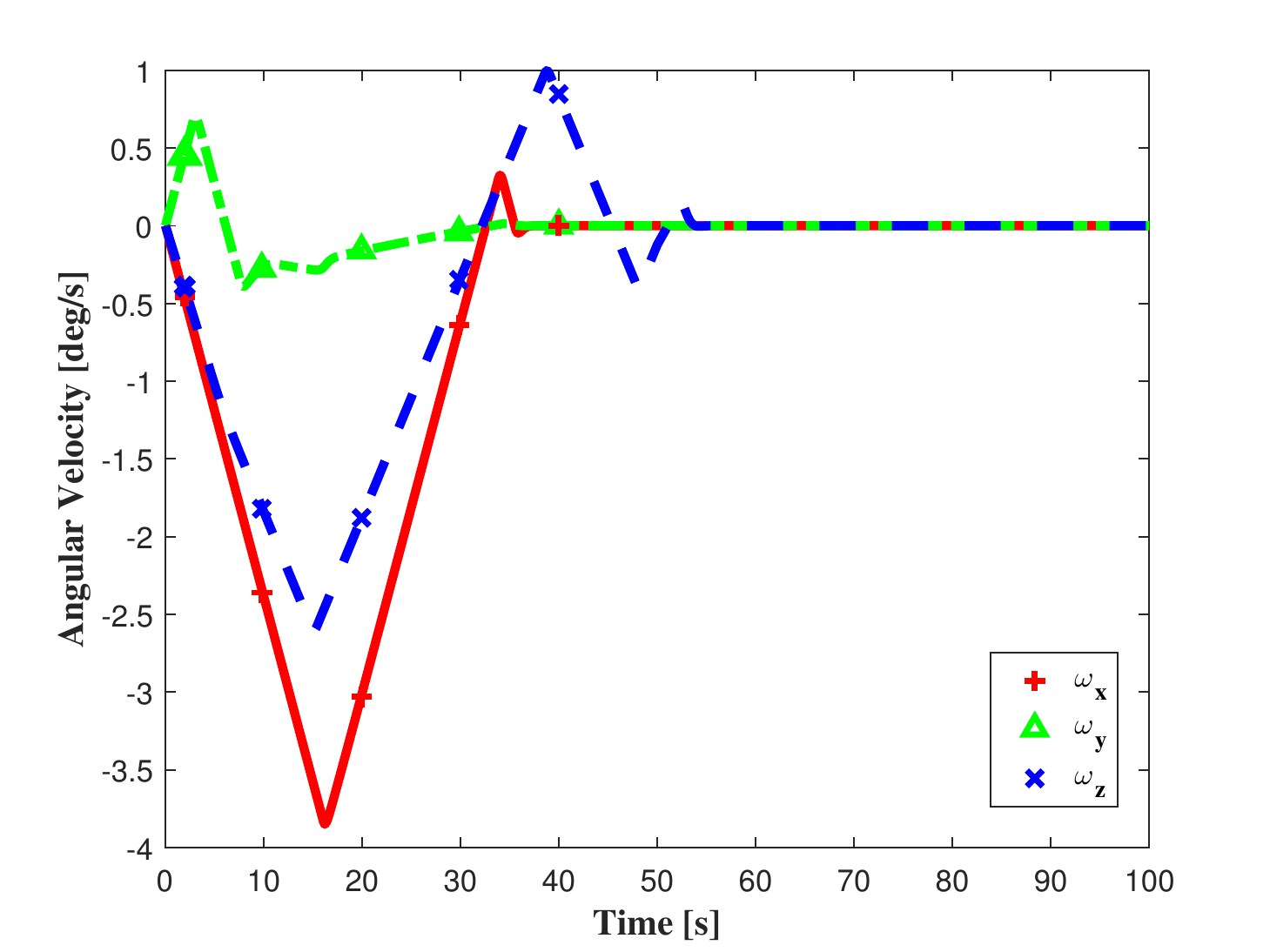}}

\subfloat[Output control torque]{
    \label{fig10c}
    \includegraphics[scale=0.42]{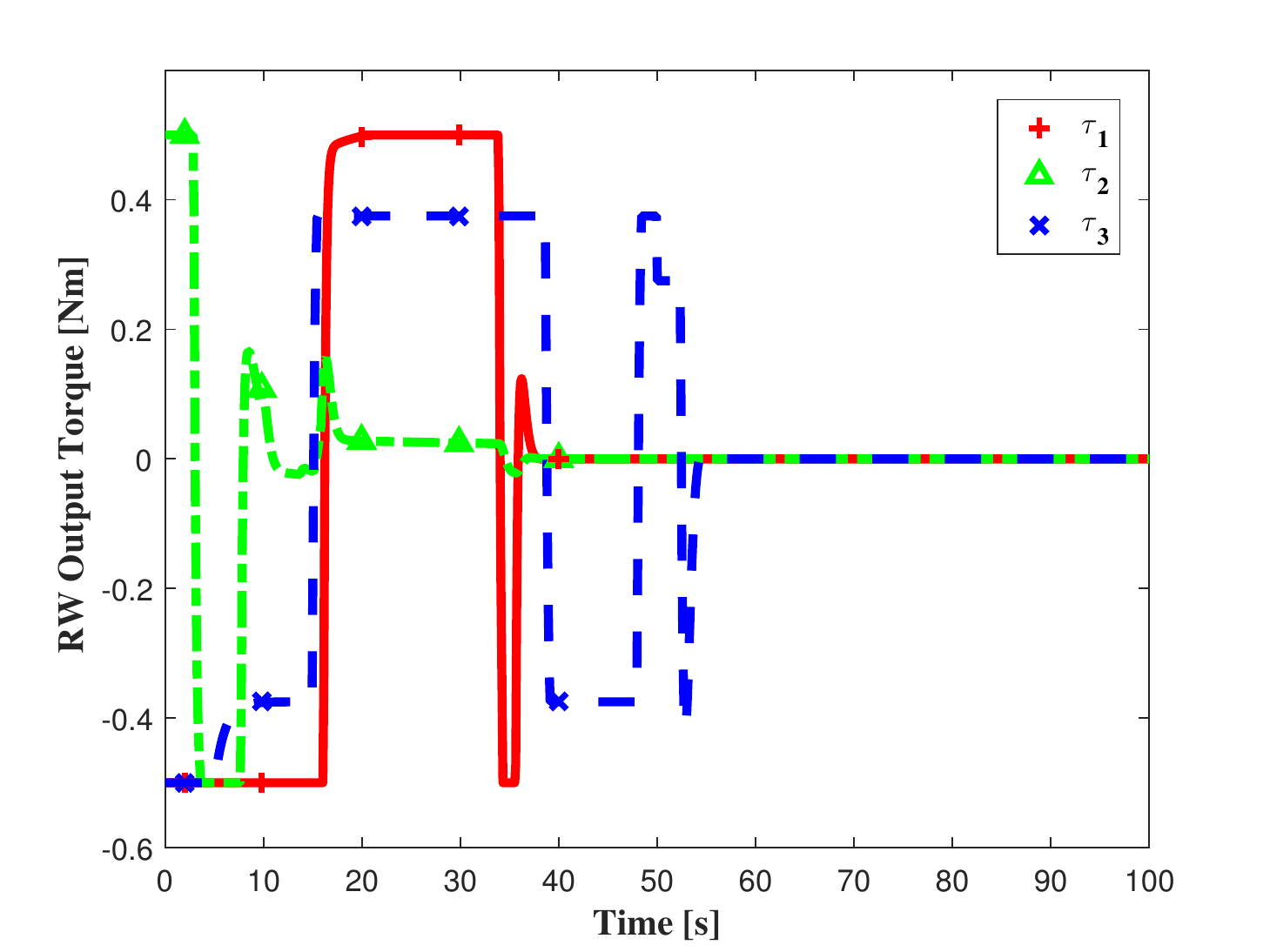}}
  \subfloat[Angular momentum]{
    \label{fig10d}
    \includegraphics[scale=0.42]{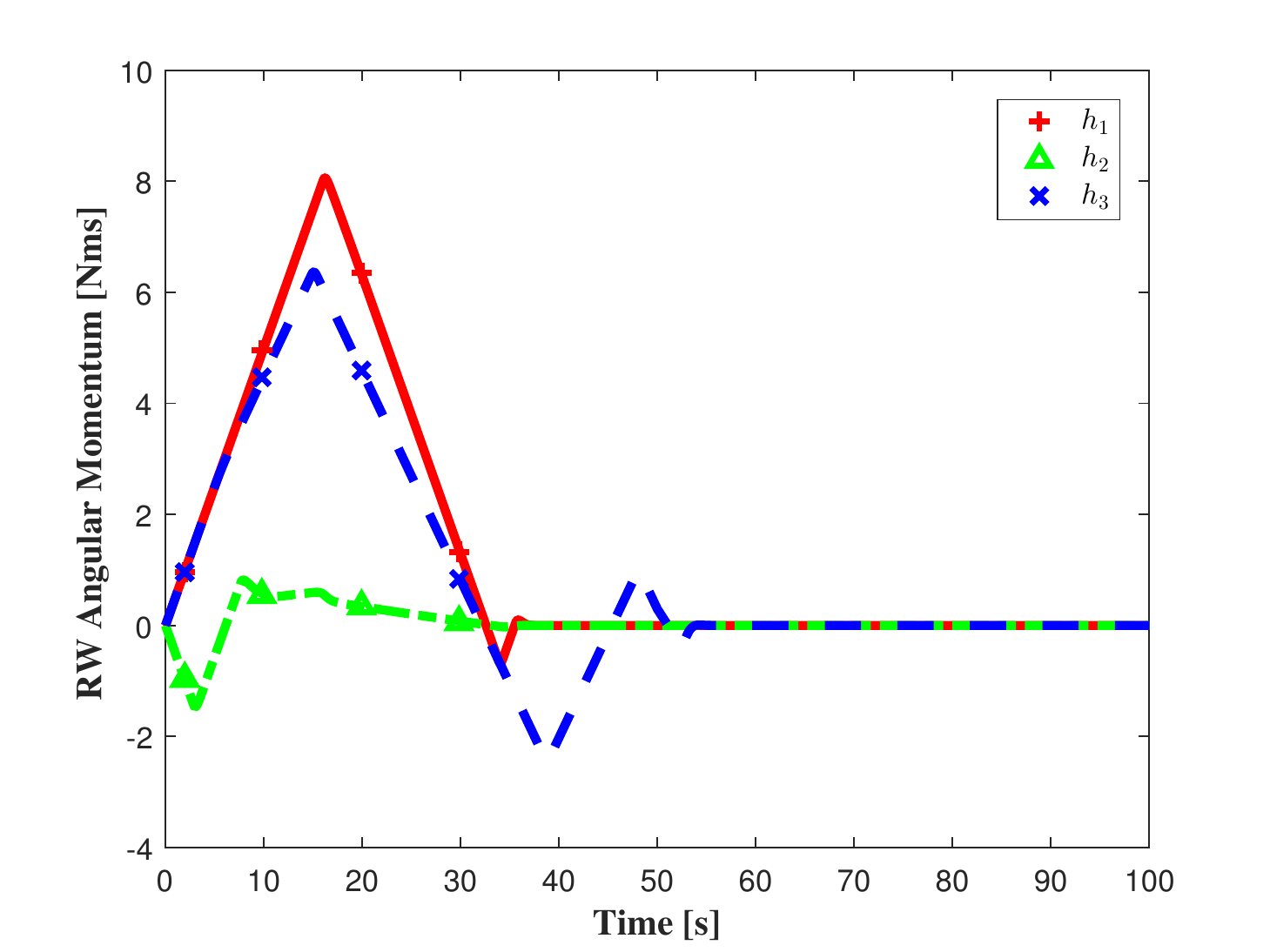}}
  \caption{RW-actuated attitude control result under $F_c$}
  \label{fig10}
\end{figure}

\begin{figure}[!h]
  \centering
  \subfloat[Euler angle]{
    \label{fig11a}
    \includegraphics[scale=0.42]{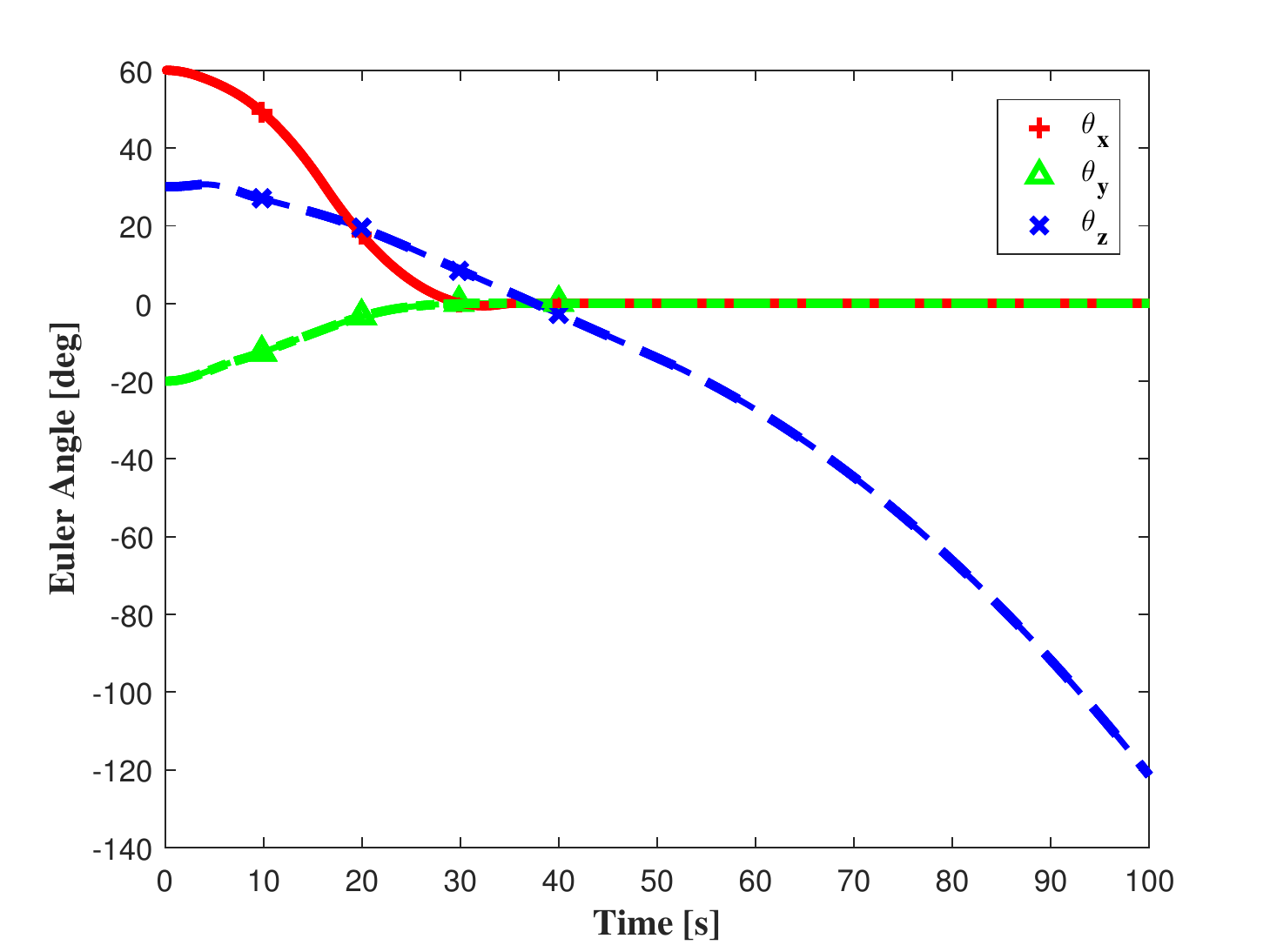}}
  \subfloat[Angular velocity]{
    \label{fig11b}
    \includegraphics[scale=0.42]{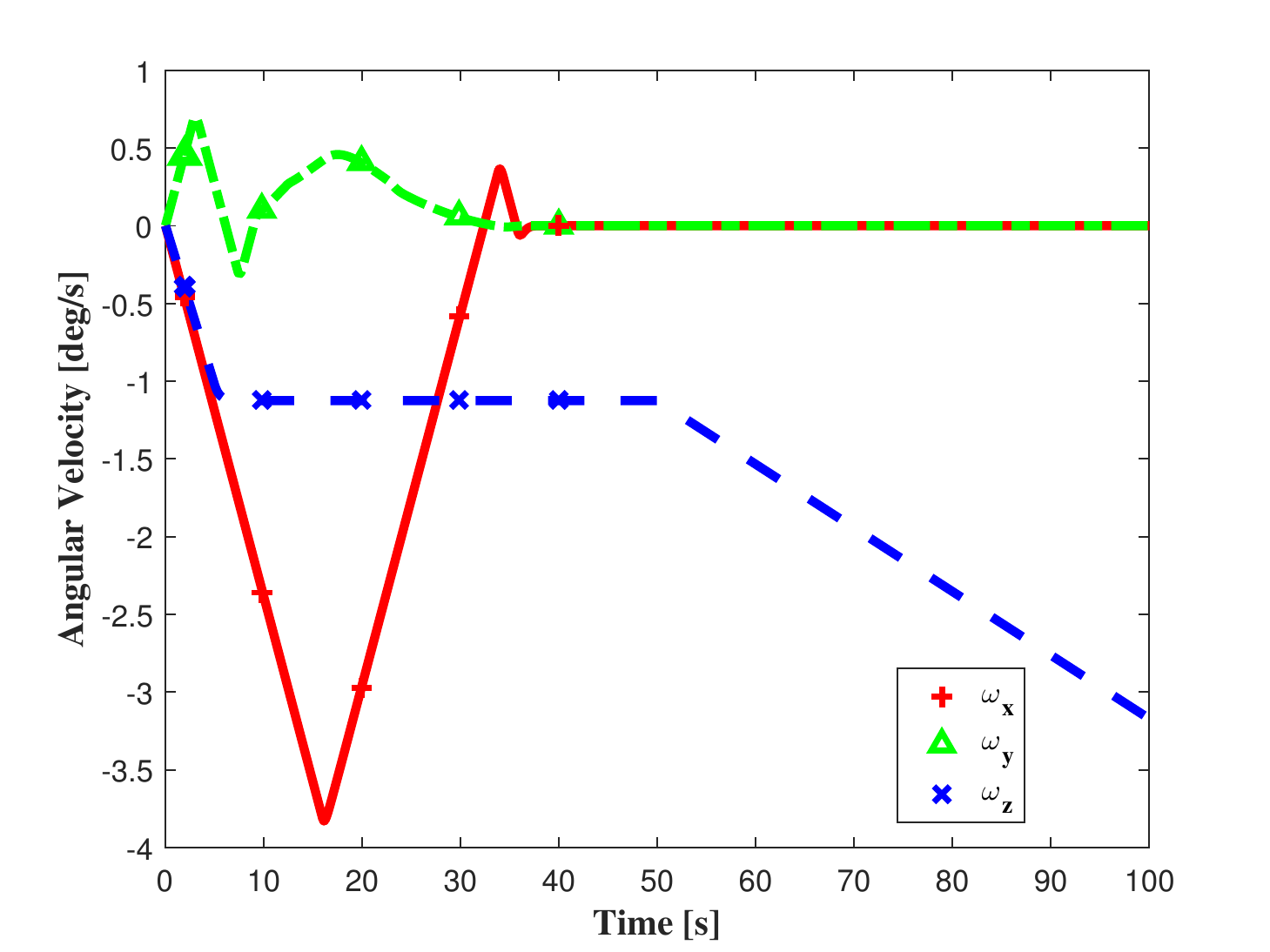}}

\subfloat[Output control torque]{
    \label{fig11c}
    \includegraphics[scale=0.42]{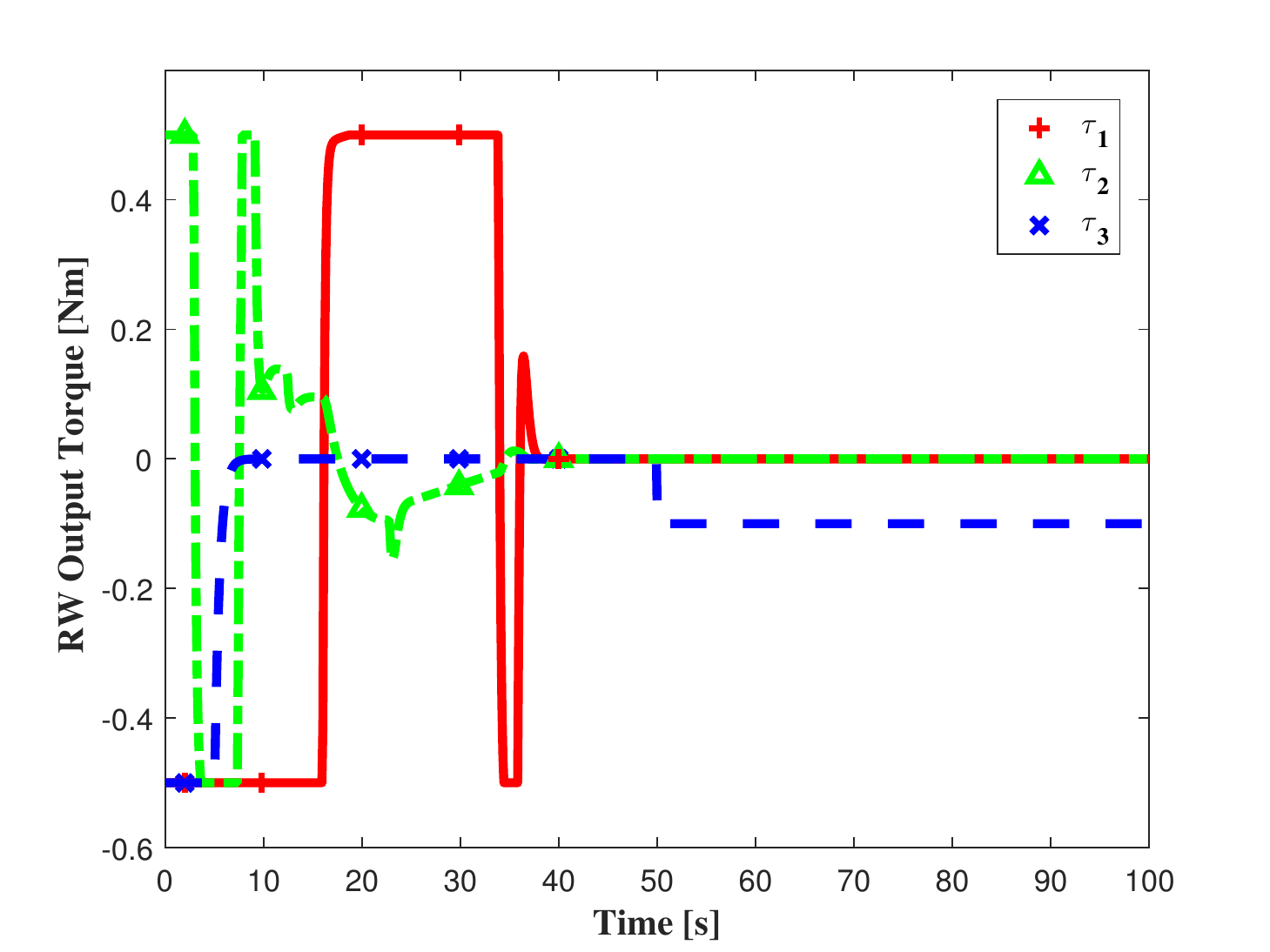}}
  \subfloat[Angular momentum]{
    \label{fig11d}
    \includegraphics[scale=0.42]{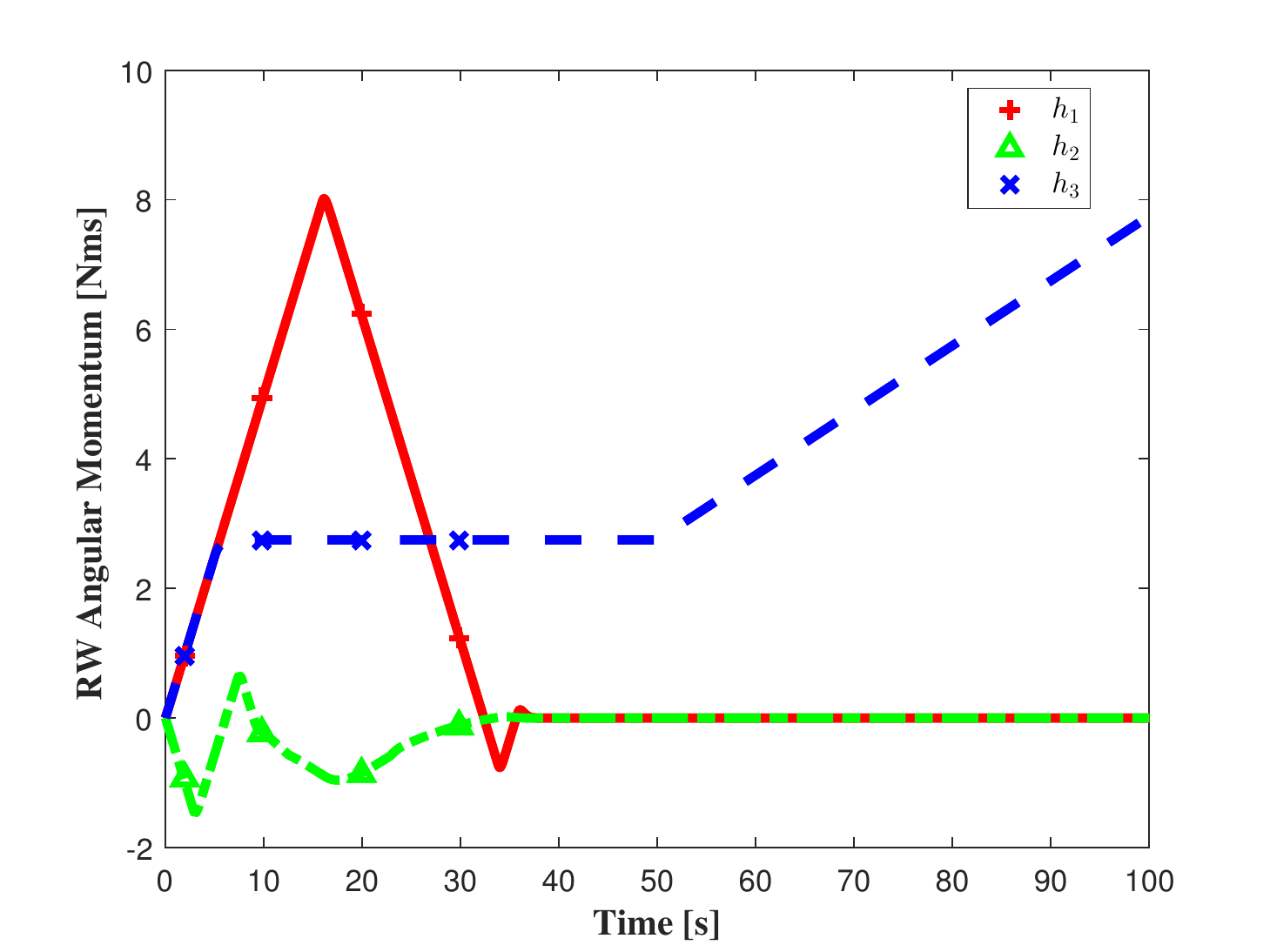}}
  \caption{RW-actuated attitude control result under $F_d$}
  \label{fig11}
\end{figure}

The objective of the attitude control of this simulation is to stabilize the attitude and angular velocity such that the spacecraft can change its orientation from the initial value to the target.
The initial Euler angle is set to be $60$ deg, $-20$ deg and $30$ degree and the initial angular velocity is zero. The attitude controller in the simulation is the widely used cascade PD control in \cite{wie2002rapid}, which is in the form of:
\begin{equation}\label{CS}
{\bm{u}} =  - {\bm{J}}\left\{ {\left. {2k\mathop {sat}\limits_{{L_i}} \left( {{\bm{q}}_e} \right) + c{\bm{\omega }}} \right\}} \right. ,
\end{equation}
with
\begin{equation}\label{Li}
{L_i} = (c/2k)\min \left\{ {\left. {\sqrt {4{a_i}\left| {{q_{ei}}} \right|} ,{{\left| {{\omega _i}} \right|}_{\max }}} \right\}} \right. ,
\end{equation}
where $\bm{J}$ represents the moment of inertia of the satellite, $\bm{q_e}$ is the quaternion-error vector, $\bm{\omega }$ is the spacecraft angular velocity, $c$ and $k$ are natural frequency and damping ratio related parameters. The parameters $a_i=u_{\max}/J_{ii}$ and ${\left| {{\omega _i}} \right|}_{\max }$ are the maximum acceleration and angular velocity along the $i$th axis,  ${\left| {{\omega _i}} \right|}_{\max }$ is set to be $4 \text{ deg}/\text{s}$, and the PD gains are $k=9.54$ and $c=5.5$.

The simulation results of the RW-actuated spacecraft under different fault scenarios including the fault-free situation are shown in Fig. \ref{fig7} to \ref{fig11}. In Fig. \ref{fig7}, the trajectories of Euler angle, angular velocity, control torque and angular momentum of RWs in the nominal condition are presented. It can be seen from Fig. \ref{fig7a} to \ref{fig7d} that the required maneuver is completed in $40$ s. The scenario that the wheel along Z axis partially loses its effectiveness at $5$ s (the fault $F_a$) is demonstrated in Fig. \ref{fig8}. Compared to Fig. \ref{fig7}, the whole system is controllable, but the stabilization is achieved in a longer period. It should be mentioned from Fig. \ref{fig8c} that the maximum output torque drops to $\eta$ times of the command gradually after $5$ s. Fig. \ref{fig9} shows the control result when the third wheel totally fails at $5$ s (the failure $F_b$). It is clear that the axis about which the failed RW is installed becomes uncontrollable and the Euler angle diverges as shown in Fig. \ref{fig9a}. Fig. \ref{fig9b} states the angular velocity in Z axis keeps at a constant after the wheel fails. This is because the speed of the third RW does not change after failure happens as shown in Fig. \ref{fig9d}. Fig. \ref{fig10} shows the control result in the condition that the third wheel partially loses its effectiveness at $5$ s and experiences the additive fault in $50$ s (the fault $F_c$). Comparing with Fig. \ref{fig8}, the transient process is almost the same, but a larger steady-state error is observed in Fig. \ref{fig10a}. The abrupt bias after $50$ s is clearly observed in Fig. \ref{fig10c}. As shown in Fig. \ref{fig11c}, the wheel suffers from failure at $5s$ and offset at $50$ s (the failure $F_d$). With the influence of bias, the angular velocity of the spacecraft and the wheel speed diverges as depicted in Fig. \ref{fig11b} and Fig. \ref{fig11d}. Same as Fig. \ref{fig9}, the whole system is uncontrollable under the RW failure.

From the simulation results in Fig. \ref{fig7} to \ref{fig11}, we can obtain the following qualitative conclusions:
\begin{itemize}
  \item The RW failure has more serious consequences than the RW fault in attitude control, and will directly make the system uncontrollable and unstable when there is no RW redundancy.
  \item The additive fault is more serious than the multiplicative fault, and it can result in steady-state error;
  \item The partial loss of effectiveness fault can be regarded that the wheel is replaced by another one with a smaller control capacity. So, it does not affect the system's controllability.
\end{itemize}

\begin{figure}[!h]
 \centering
\includegraphics[scale=1.2]{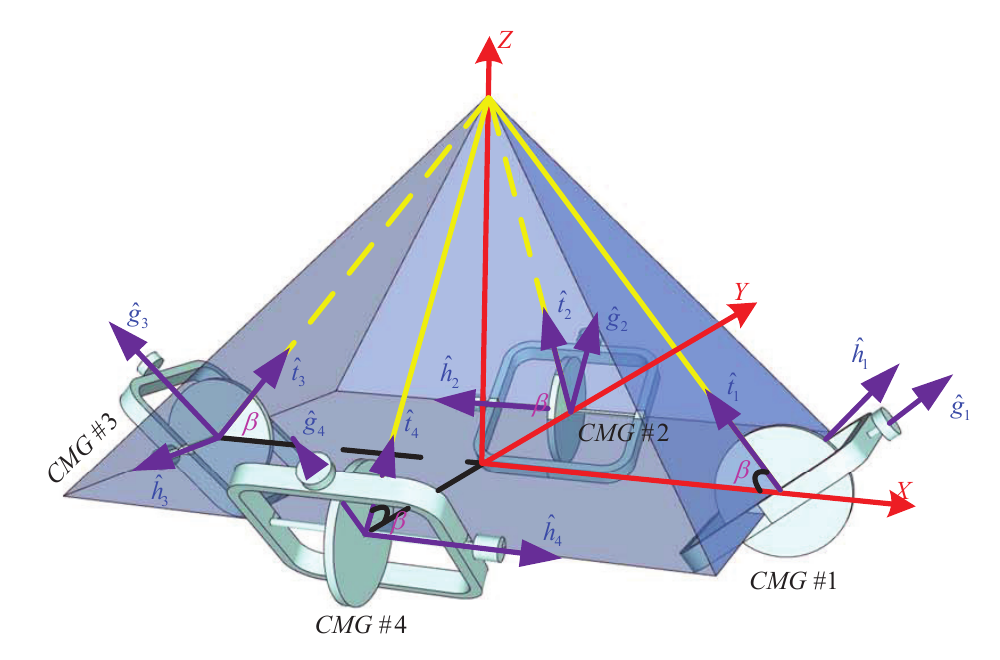}
\caption{Pyramid configuration of the SGCMGs}
  \label{pyramid}
\end{figure}

\begin{figure}[!b]
  \centering
  \subfloat[Euler angle]{
    \label{fig12a}
    \includegraphics[scale=0.42]{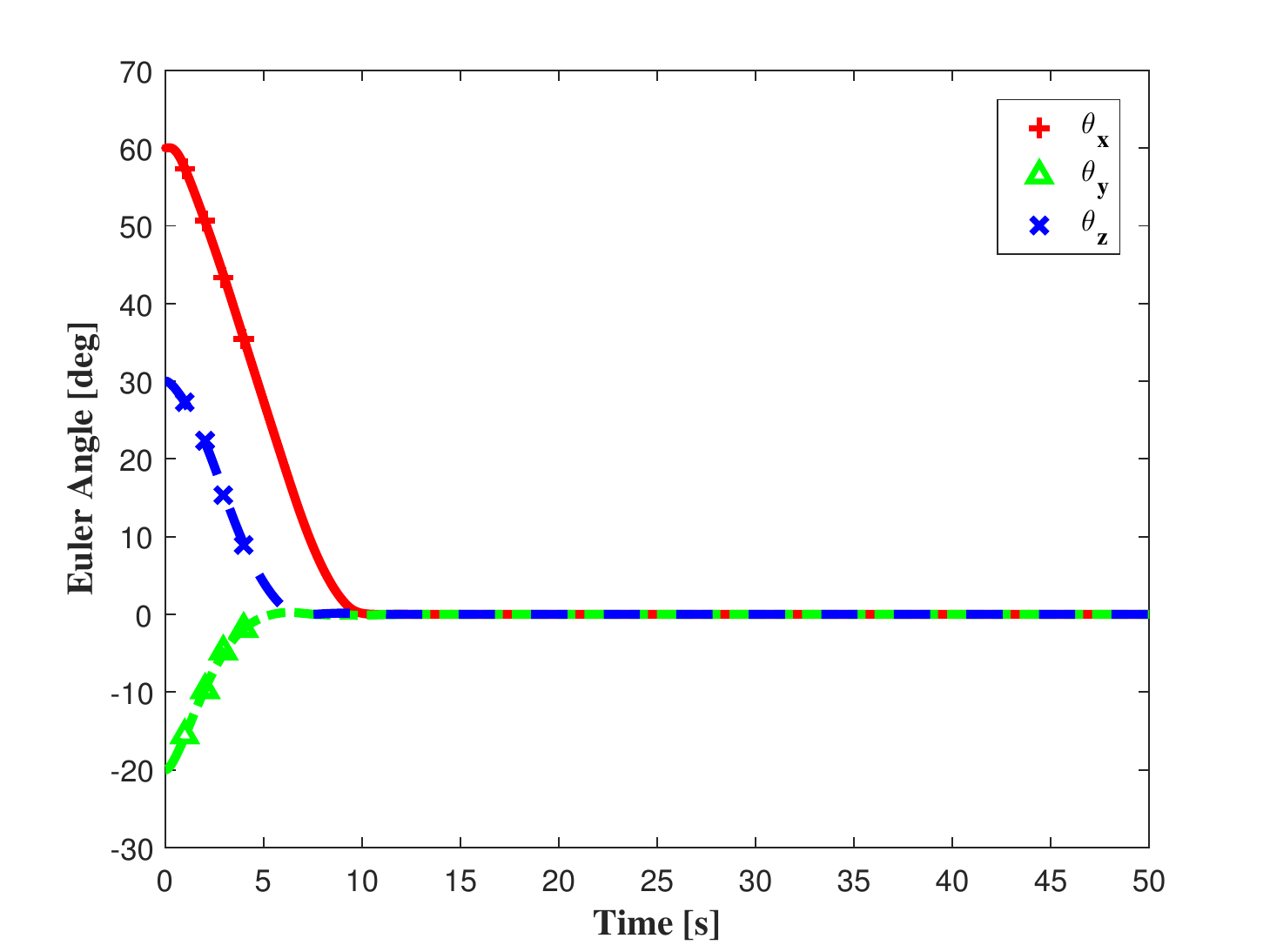}}
  \subfloat[Angular velocity]{
    \label{fig12b}
    \includegraphics[scale=0.42]{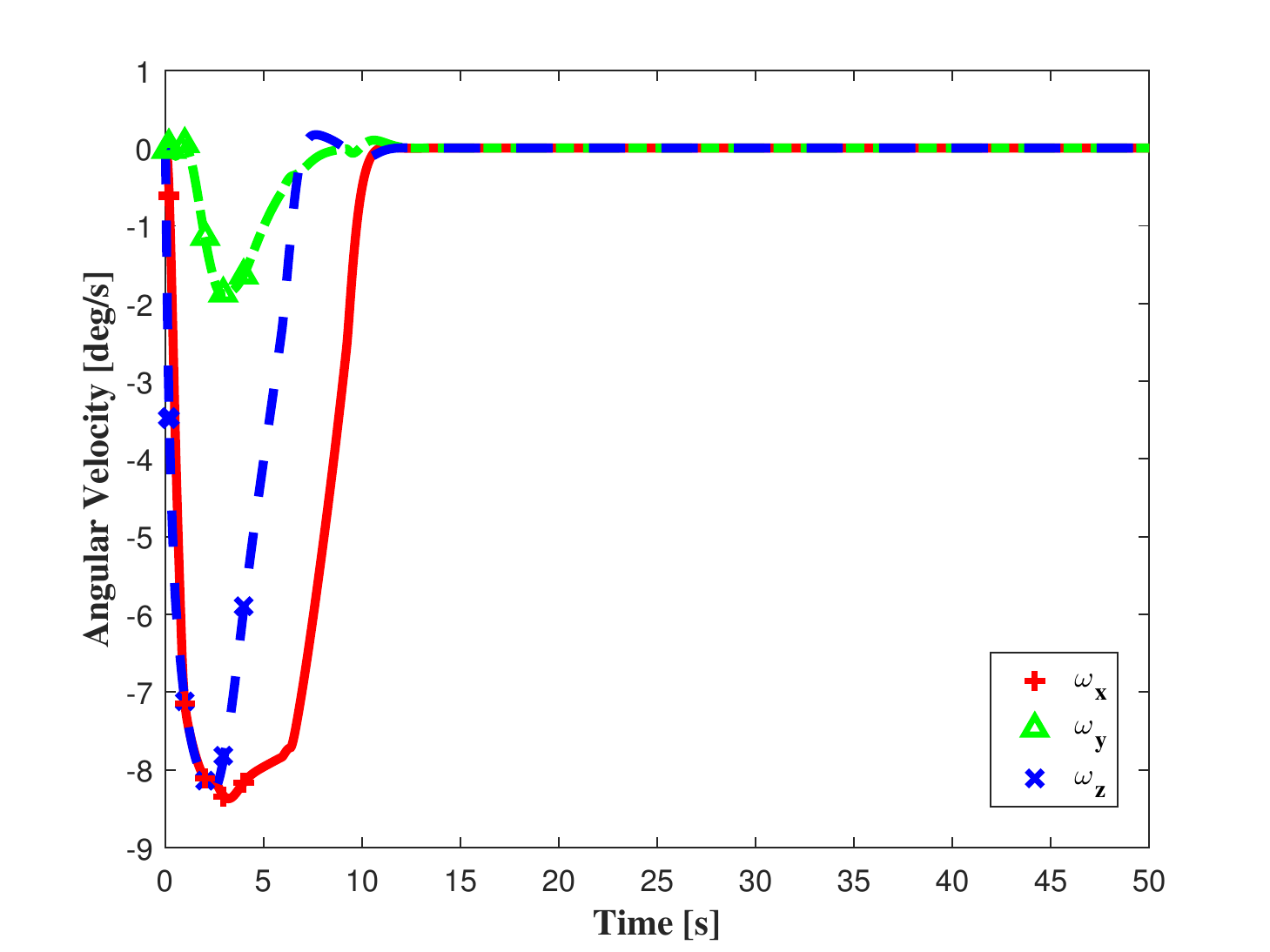}}

 \subfloat[Gimbal angle]{
    \label{fig12c}
    \includegraphics[scale=0.42]{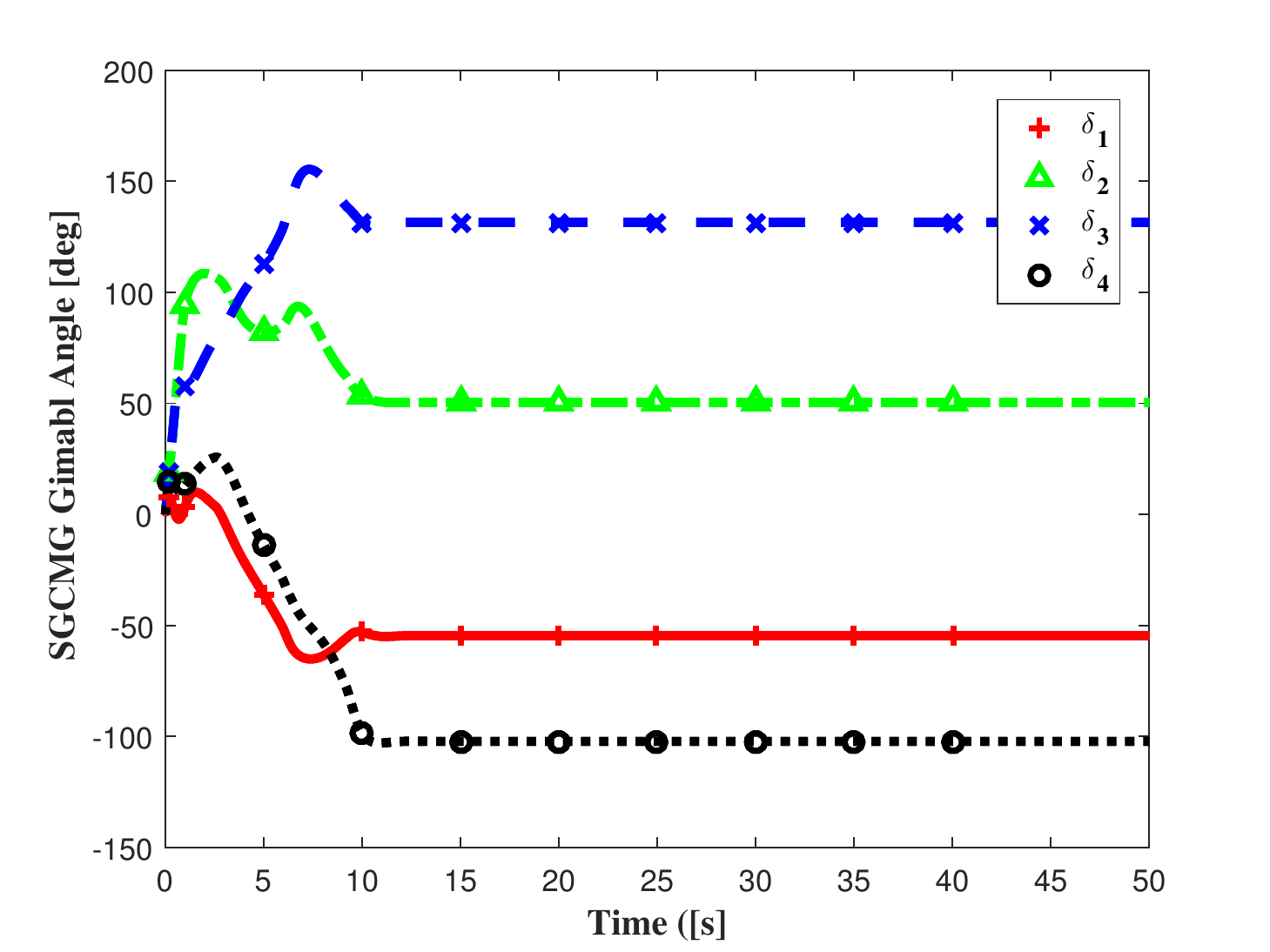}}
  \subfloat[Gimbal rate]{
    \label{fig12d}
    \includegraphics[scale=0.42]{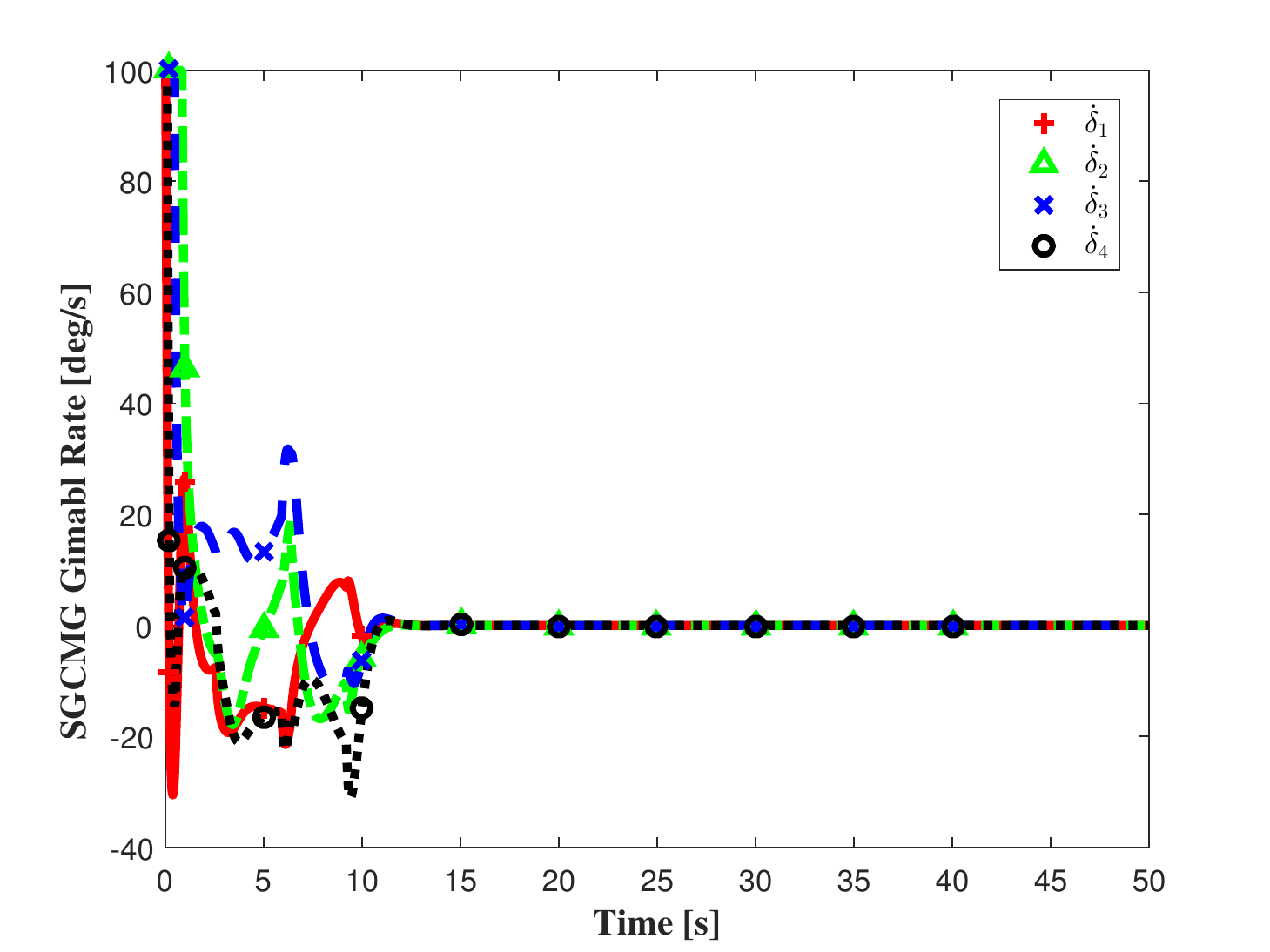}}

\subfloat[Output control torque]{
    \label{fig12e}
    \includegraphics[scale=0.42]{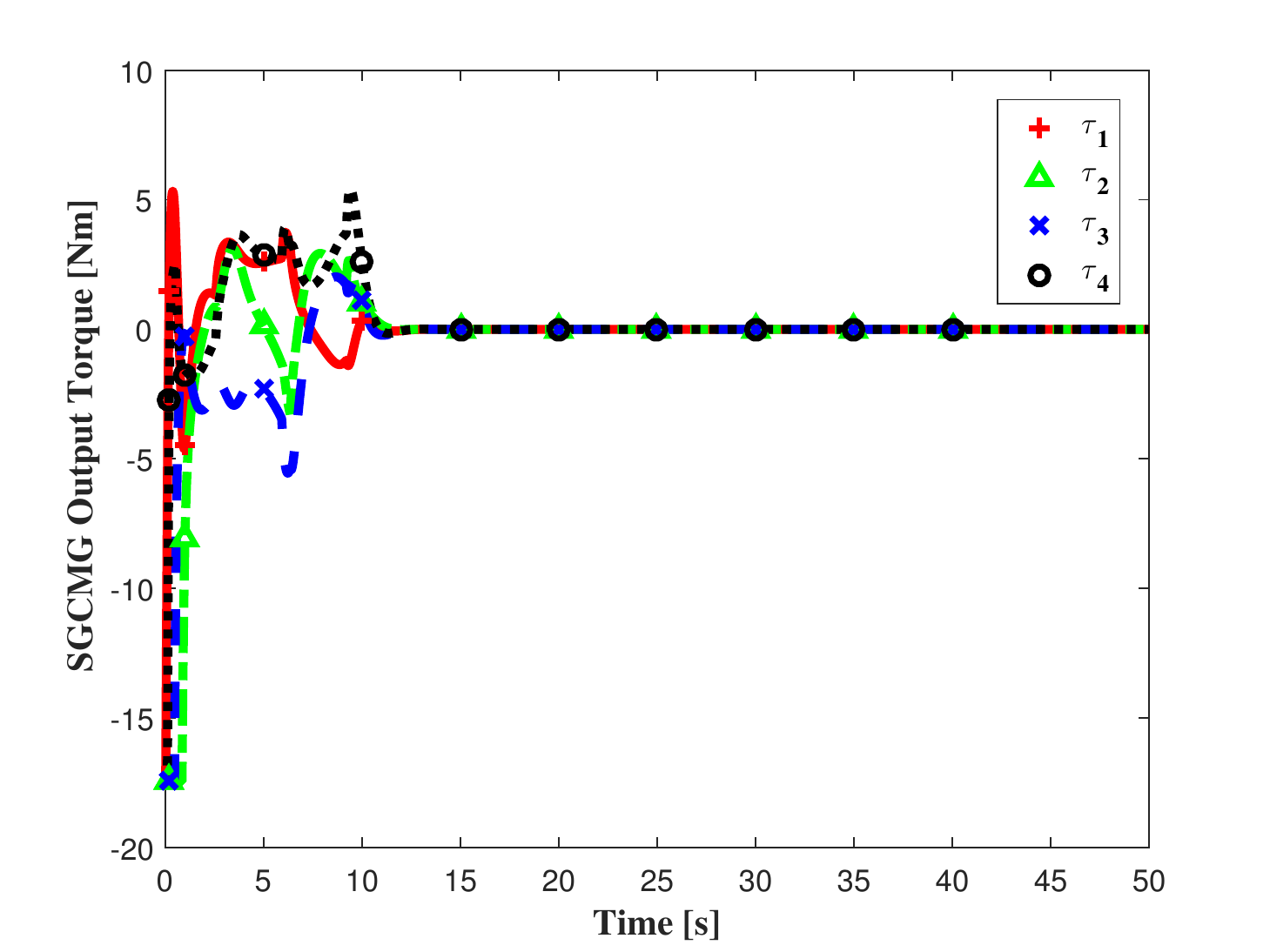}}
  \subfloat[Singularity measurement]{
    \label{fig12f}
    \includegraphics[scale=0.42]{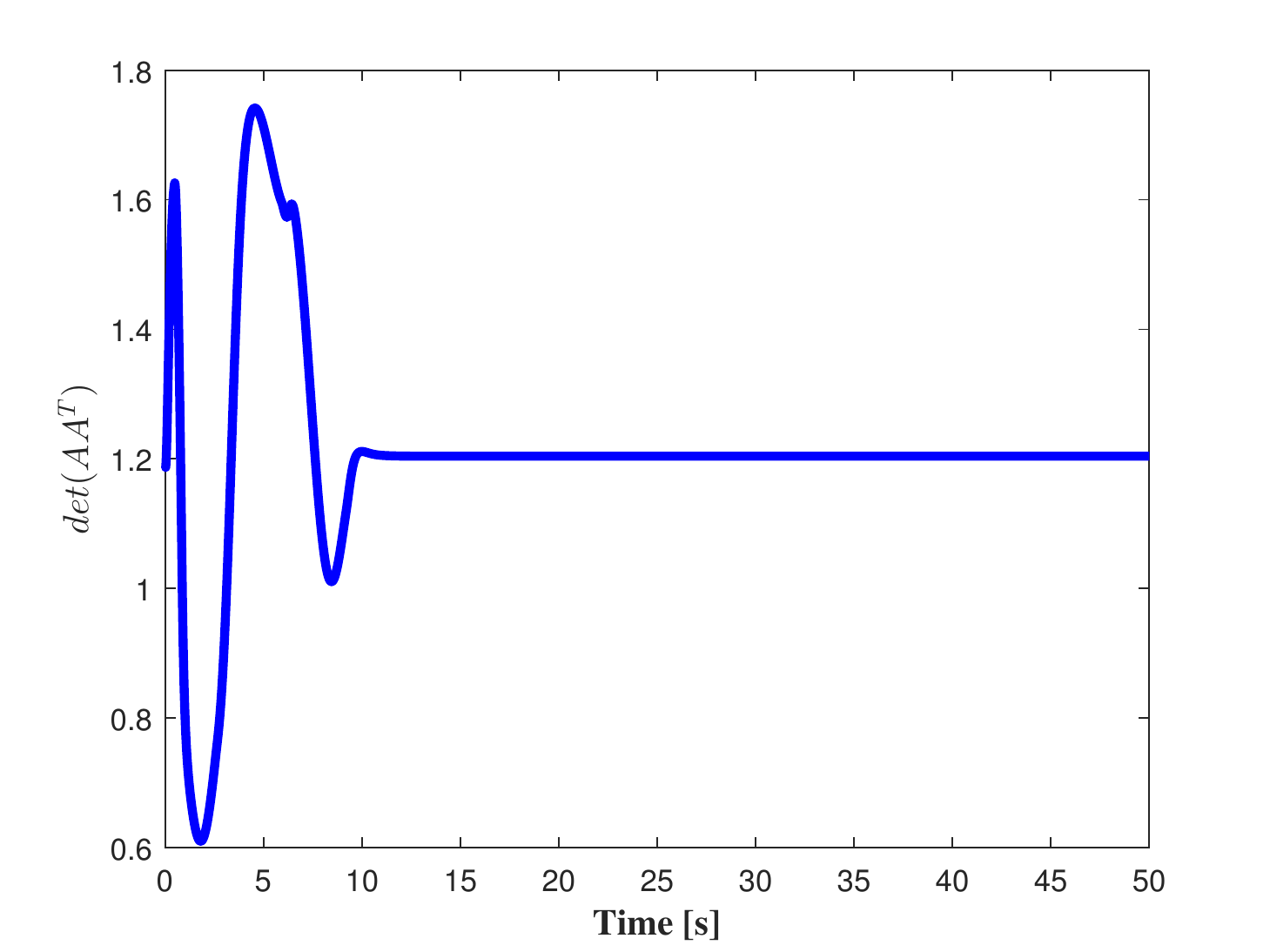}}
  \caption{SGCMG-actuated attitude control result under nominal condition}
  \label{fig12}
\end{figure}

\subsection{Attitude Control Results of SGCMG-Actuated Spacecraft}
Simulation of SGCMG-actuated spacecraft system is conducted in this Section. A pyramid configuration of four SGCMGs as shown in Fig. \ref{pyramid} is adopted to control the spacecraft. The attitude controller, the initial states and target are the same as they are in Section \ref{RW_Sim}. \ychf{The generalized singular robust inverse method is used to steer the gimbal as in \cite{wie2002rapid}. Thus the gimbal rate command can be calculated as follows:}
\ychf{\begin{equation}\label{ratecom}
 \bm{\dot{\delta }}_c=\frac{1}{{{h}_{0}}}{{\bm{A}}^{\#}}\bm{u} \quad \quad \text{and}\quad \quad {{\bm{A}}^{\#}}={{\bm{A}}^{T}}{{\left[ \bm{A}{{\bm{A}}^{T}}+\lambda \bm{E} \right]}^{-1}}
\end{equation}}
\ychf{where $\bm{A}$ is the Jacobian matrix, $\lambda =0.01\exp \left[ -10\det \left( \mathbf{A}{{\mathbf{A}}^{T}} \right) \right]$, and the matrix $\bm{E}$ is expressed as:}
\ychf{\[\mathbf{E}=\left[ \begin{matrix}
   1 & {{\varepsilon }_{3}} & {{\varepsilon }_{2}}  \\
   {{\varepsilon }_{3}} & 1 & {{\varepsilon }_{1}}  \\
   {{\varepsilon }_{2}} & {{\varepsilon }_{1}} & 1  \\
\end{matrix} \right]>0\]}
\ychf{with ${{\varepsilon }_{i}}=0.01\sin \left( 0.5\pi t+{{\phi }_{i}} \right)$, $\phi_1 =0$, $\phi_2 = \pi/2$ and $\phi_3 = \pi$. }

\ychf{Since we only consider the gimbal fault, then the actual gimbal rate can be given using the fault model described by \eqref{26} as:}
\ychf{
\begin{equation}\label{gimfault}
\bm{\dot{\delta }}={{\bm{E}}_{\eta^g}}{{\bm{\dot{\delta }}}_{c}}+{{\bm{E}}_{ga}}
\end{equation}
where ${\bm{E}}_{\eta^g} = \text{diag}[\eta_1^g, \ \eta_2^g, \ \eta_3^g, \ \eta_4^g]$ is the effectiveness matrix of the gimbal loop and ${\bm{E}}_{ga} = \text{diag}[\dot{\delta}_{o_1}, \ \dot{\delta}_{o_2}, \ \dot{\delta}_{o_3}, \ \dot{\delta}_{o_4}]$ is the additive bias of the gimbal loop caused by faults.}

\begin{figure}[!t]
  \centering
  \subfloat[Euler angle]{
    \label{fig13a}
    \includegraphics[scale=0.42]{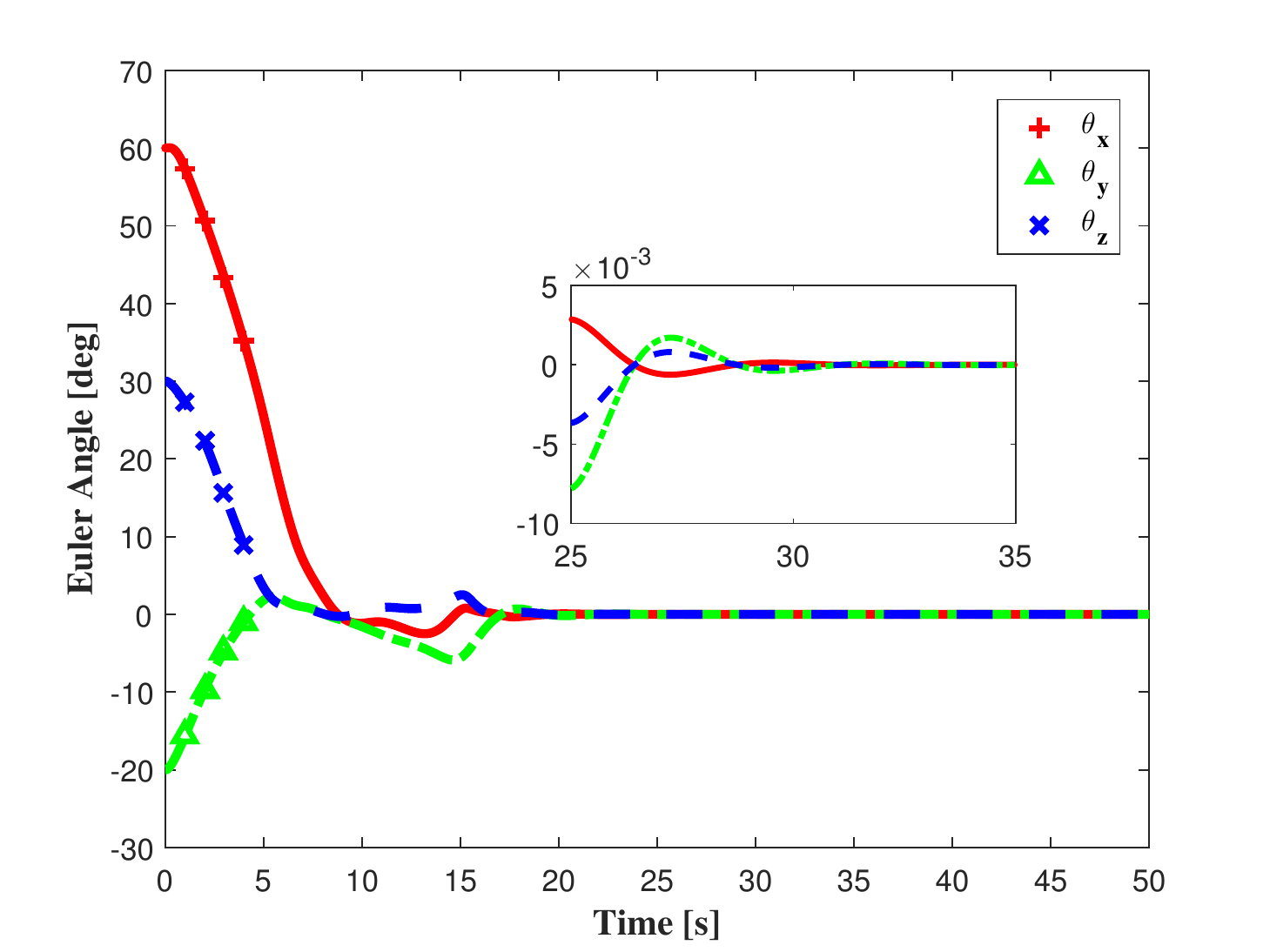}}
  \subfloat[Rotor angular momentum]{
    \label{fig13b}
    \includegraphics[scale=0.42]{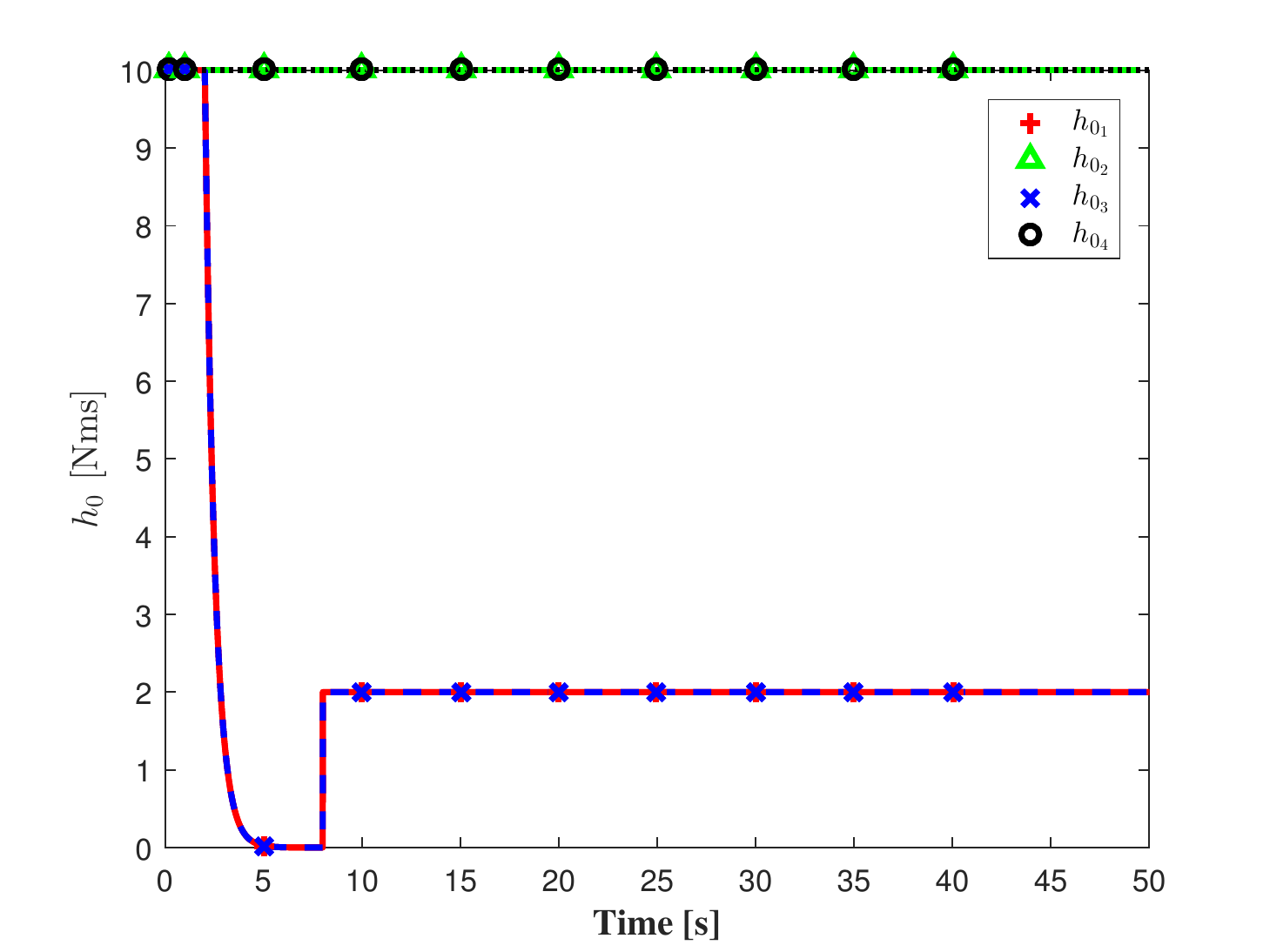}}

 \subfloat[Gimbal angle]{
    \label{fig13c}
    \includegraphics[scale=0.42]{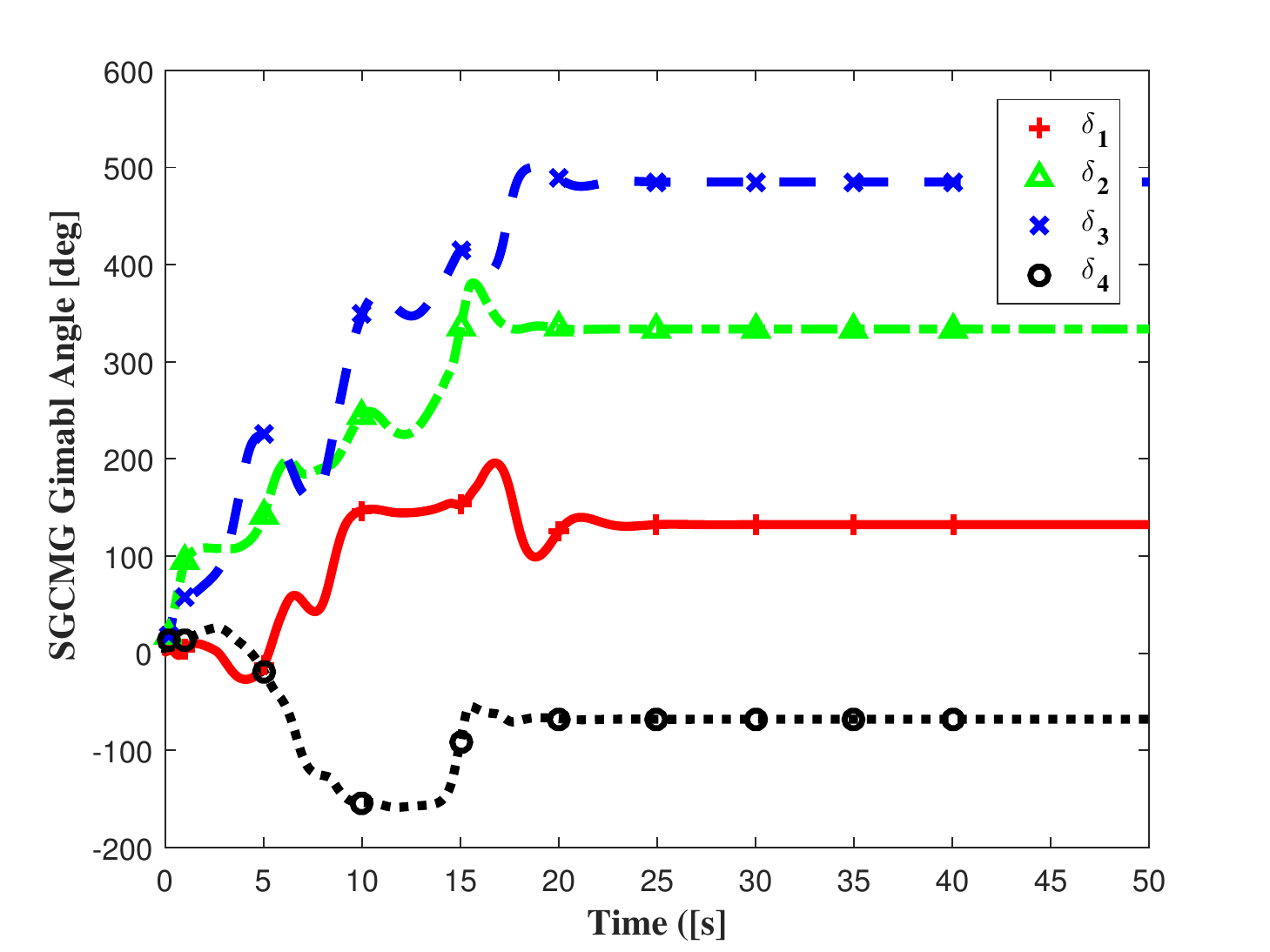}}
  \subfloat[Output control torque]{
    \label{fig13d}
    \includegraphics[scale=0.42]{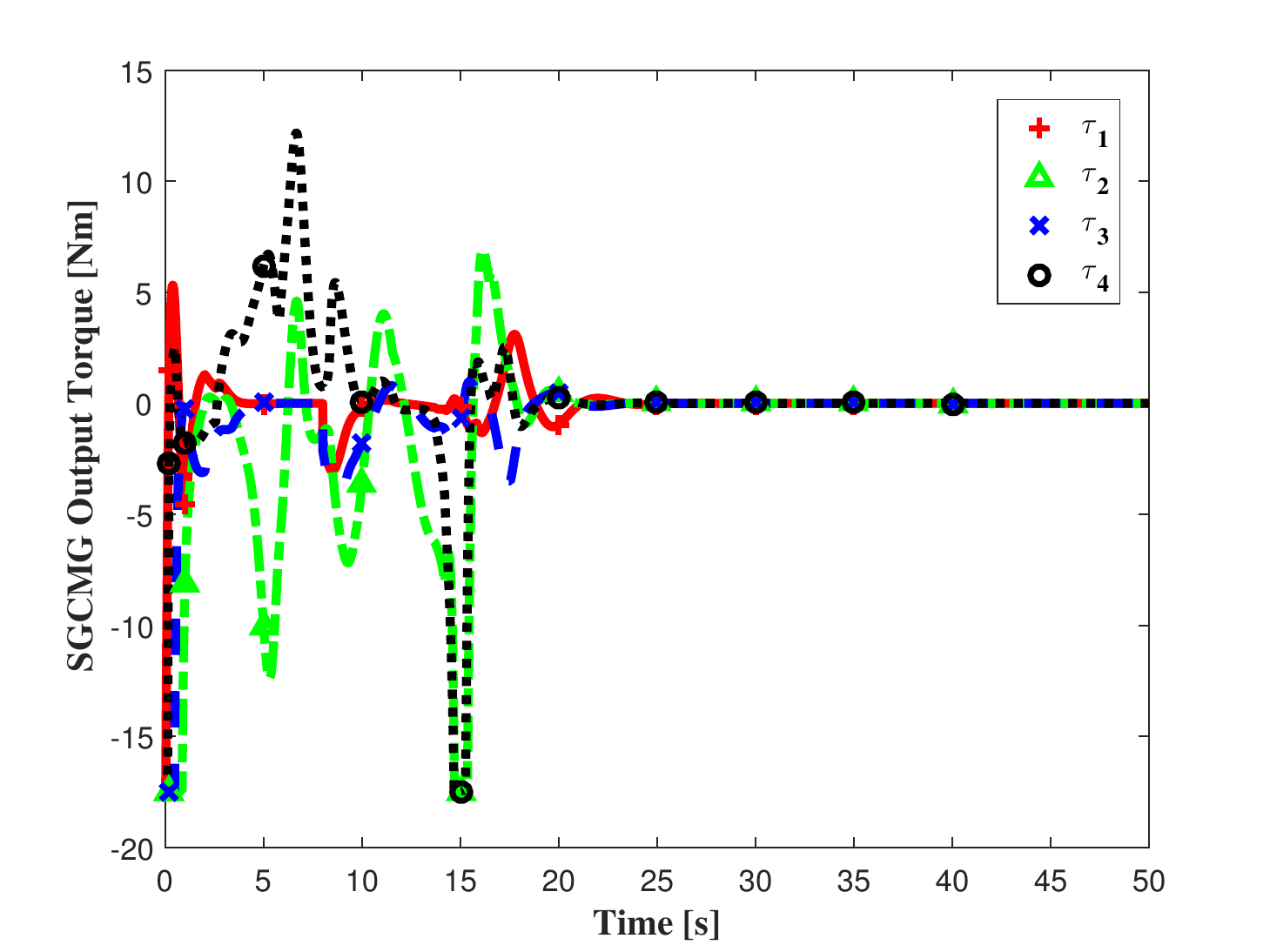}}
  \caption{SGCMG-actuated attitude control result in the presence of only rotor fault}
  \label{fig13}
\end{figure}

\begin{figure}[!h]
  \centering
  \subfloat[Euler angle]{
    \label{fig14a}
    \includegraphics[scale=0.42]{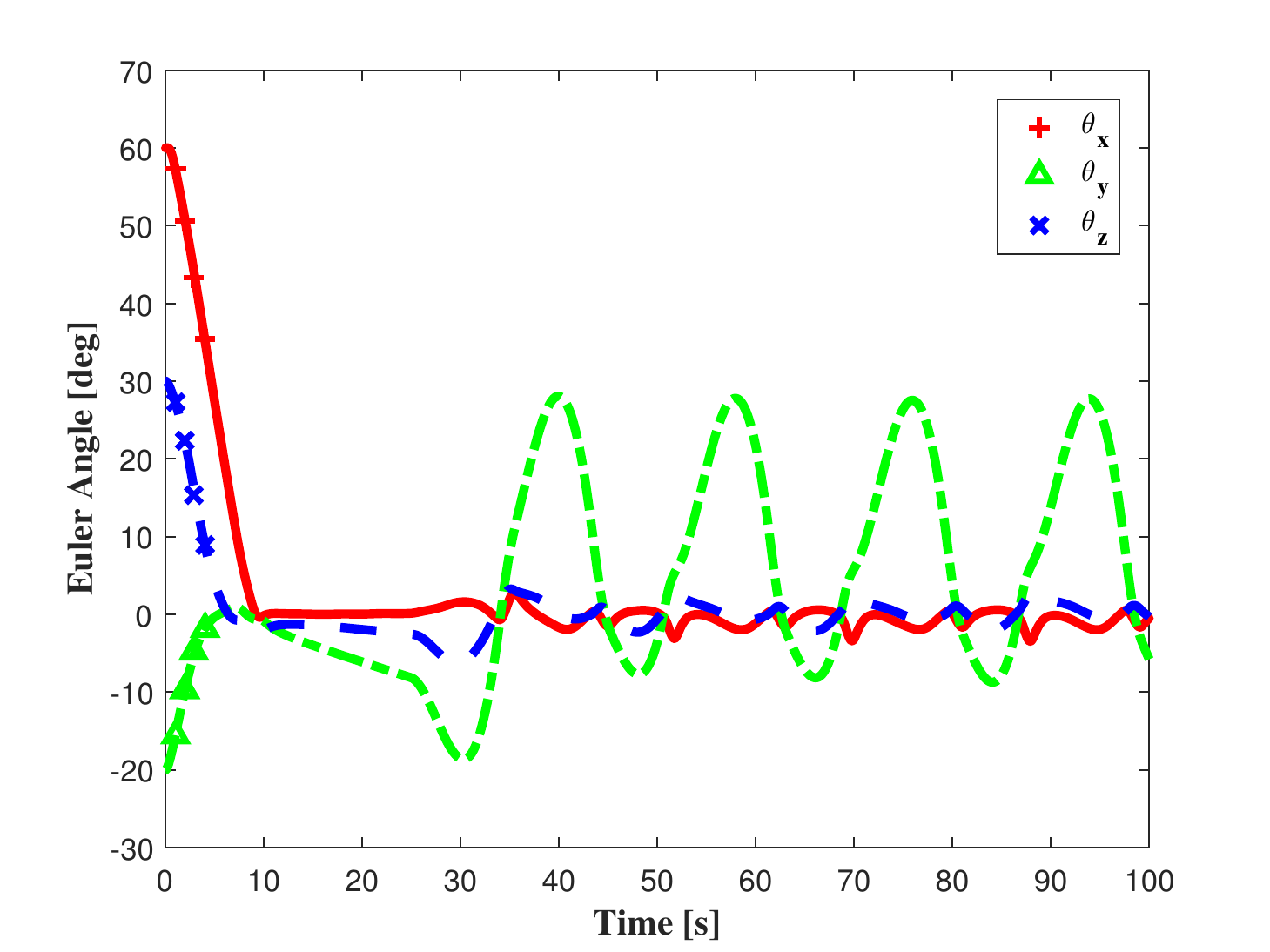}}
  \subfloat[Gimbal angle]{
    \label{fig14b}
    \includegraphics[scale=0.42]{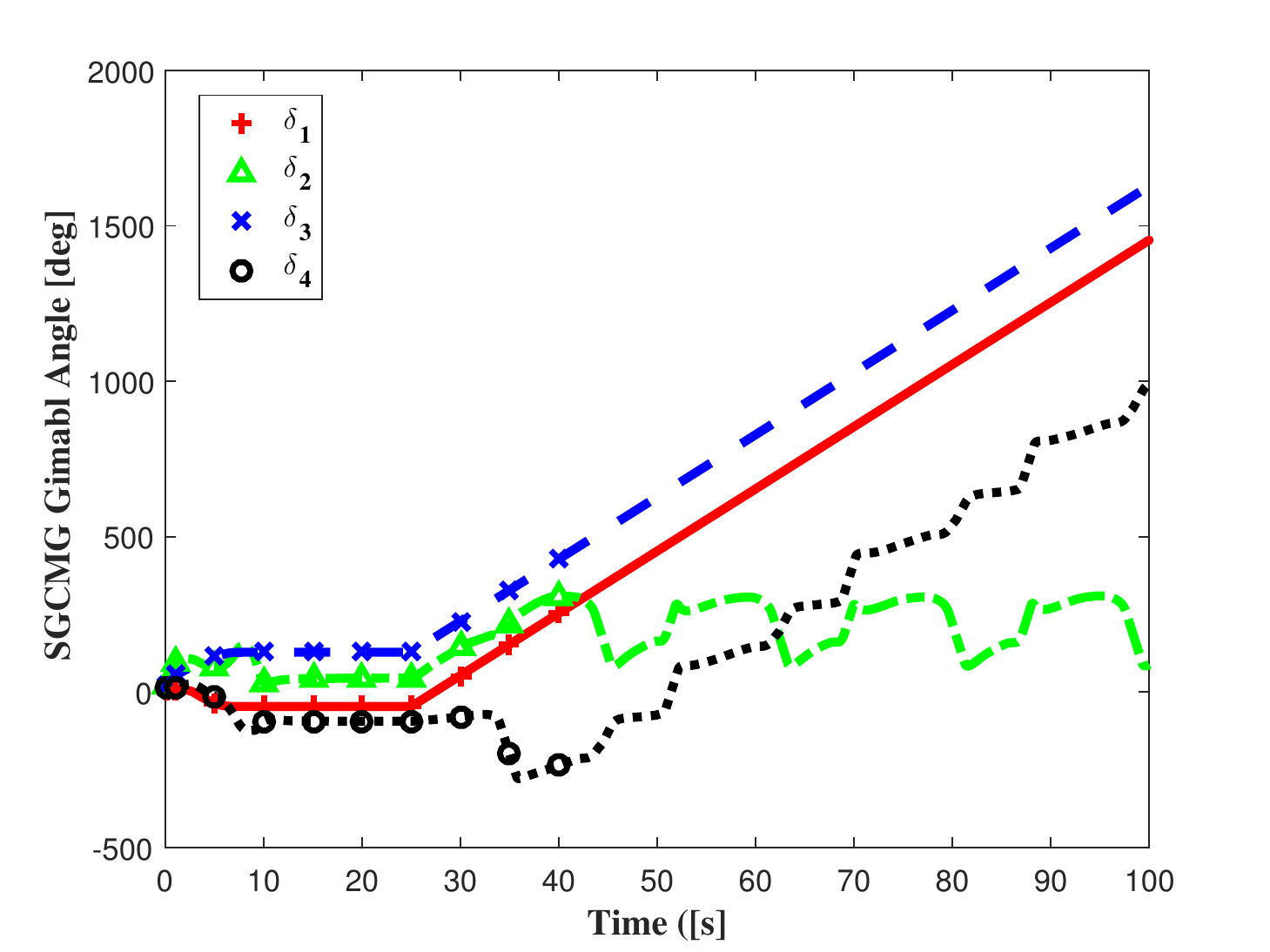}}

 \subfloat[Gimbal rate]{
    \label{fig14c}
    \includegraphics[scale=0.42]{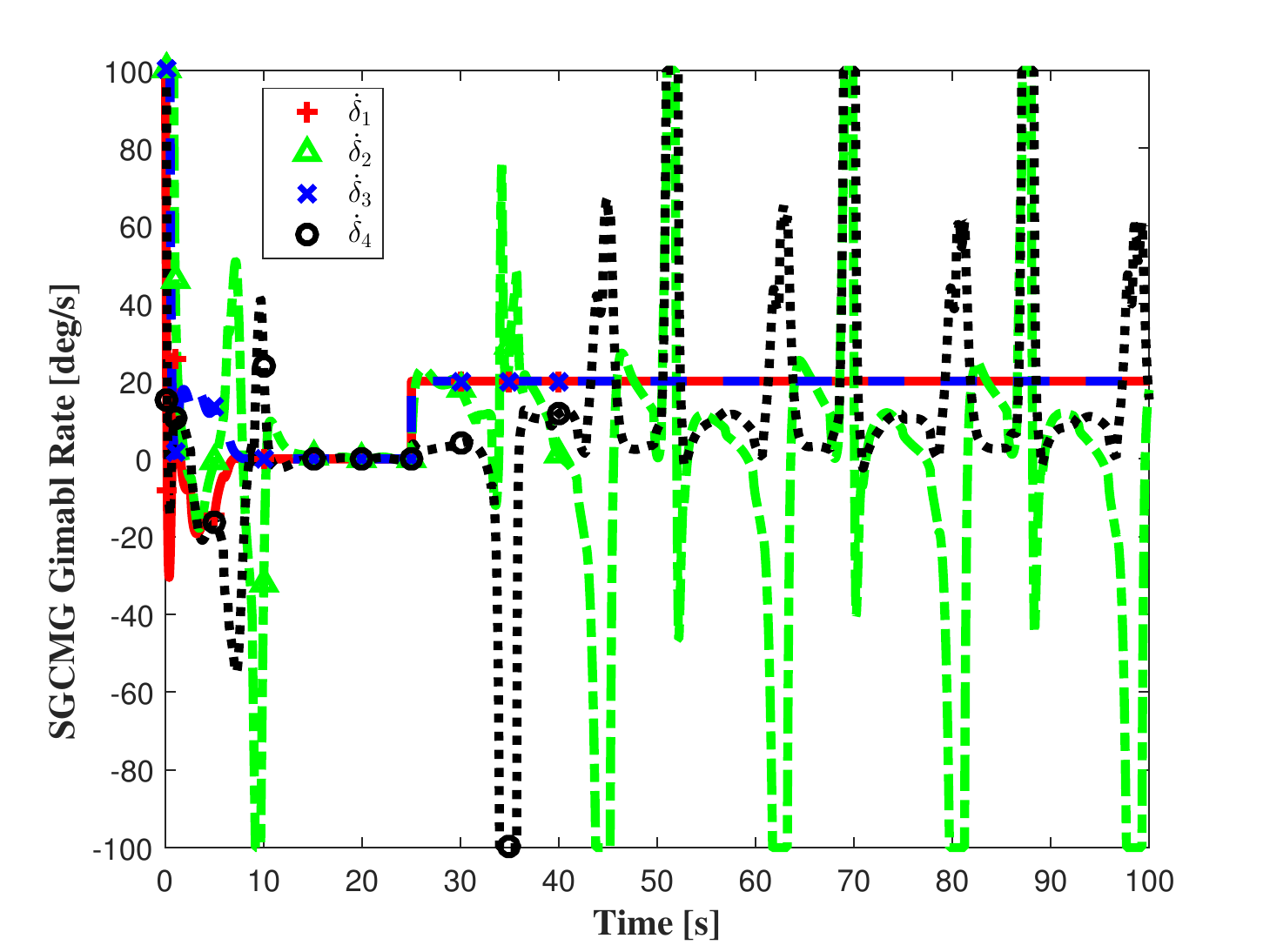}}
  \subfloat[Output control torque]{
    \label{fig14d}
    \includegraphics[scale=0.42]{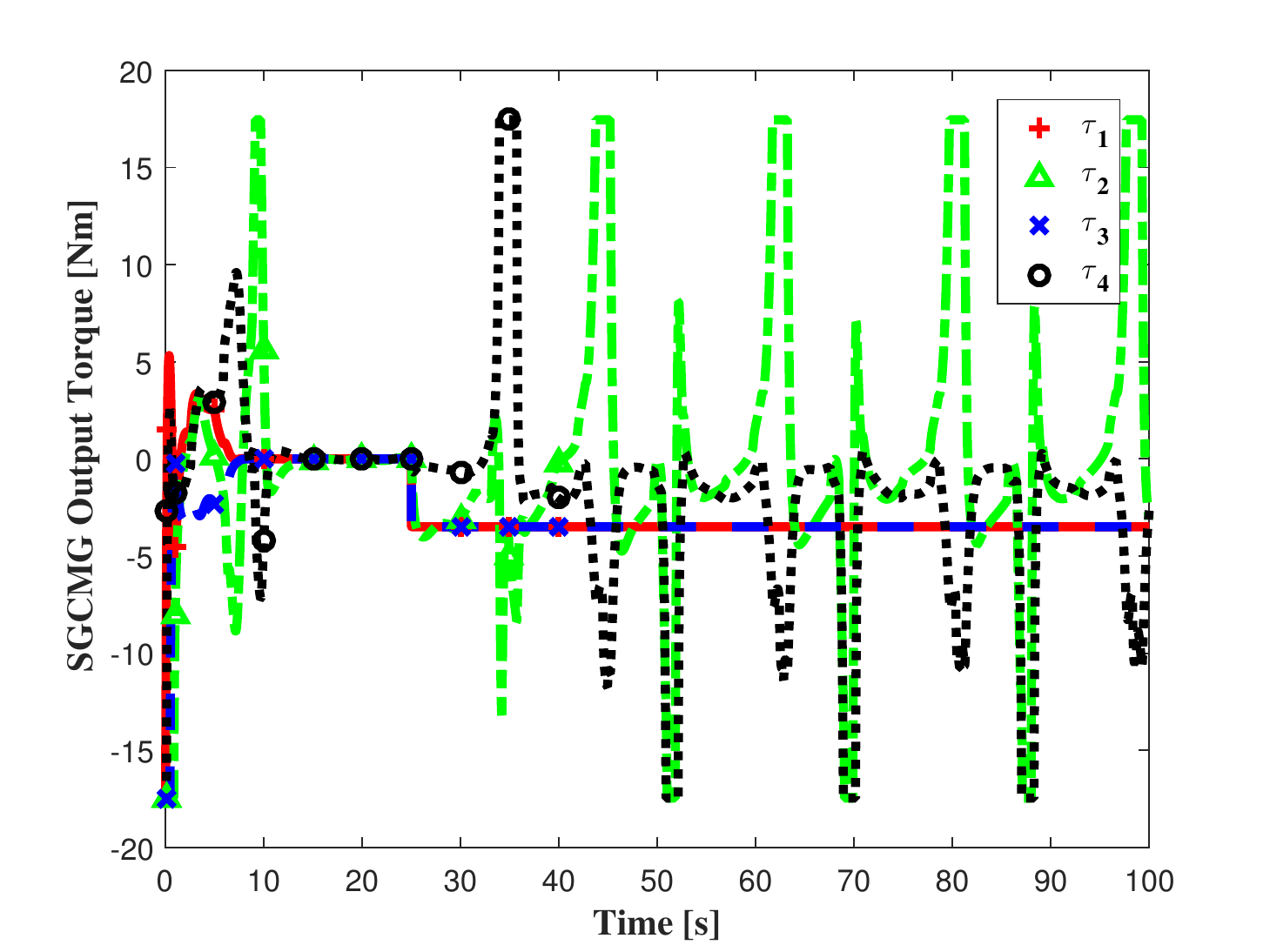}}
  \caption{SGCMG-actuated attitude control result in the presence of only gimbal fault}
  \label{fig14}
\end{figure}

The nominal angular momentum of the rotor is set as $10$ Nm and the maximum gimbal angular velocity is set as $100 \ \text{deg}/\text{s}$.
The initial gimbal angle is $[0, \, 0 , \, 0 ,\, 0]^T$ deg.
 Considering the one degree of redundancy in the pyramid configuration, the SGCMG pair in the $x-z$ plane, i.e. $1$st and $3$rd SGCMGs are assumed to be faulty simultaneously.
 To avoid fuzzy, duplication and repetition, we only implement some typical scenarios in Table \ref{tab1} in the simulation, which include 1-3 rotors failure with offset in Fig. \ref{fig13}, 1-3 gimbal failure with offset in Fig. \ref{fig14} and both the rotor and gimbals are failure in Fig. \ref{fig15}. In the following simulations, the effectiveness factors and bias of the rotor control loop are set as $0.5$ and $2 \ \text{Nms}$, while $0.75$ and $20 \ \text{deg}/\text{s}$ for the gimbal control loop.

\begin{figure}[!h]
  \centering
  \subfloat[Euler angle]{
    \label{fig15a}
    \includegraphics[scale=0.42]{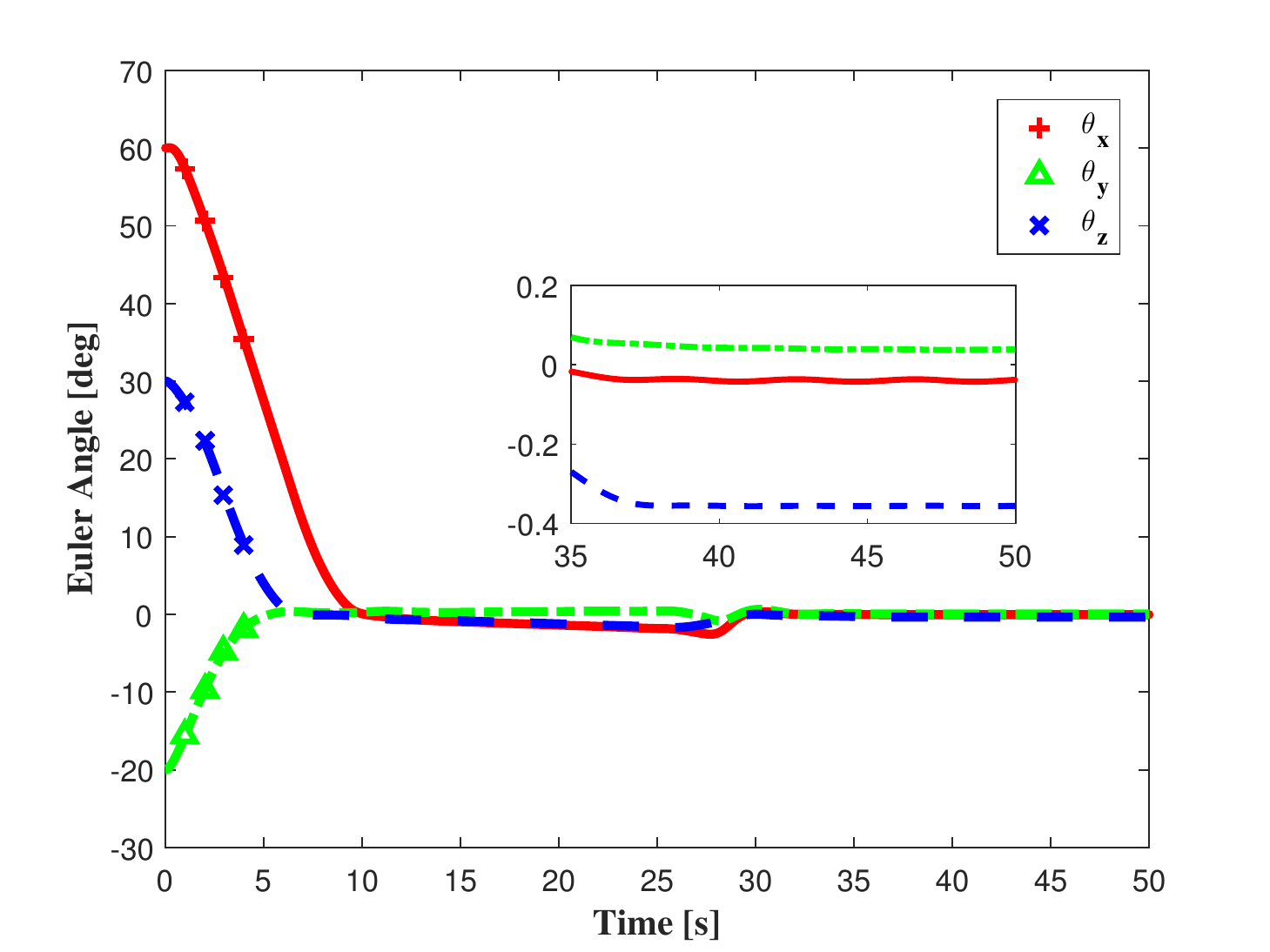}}
  \subfloat[Angular velocity]{
    \label{fig15b}
    \includegraphics[scale=0.42]{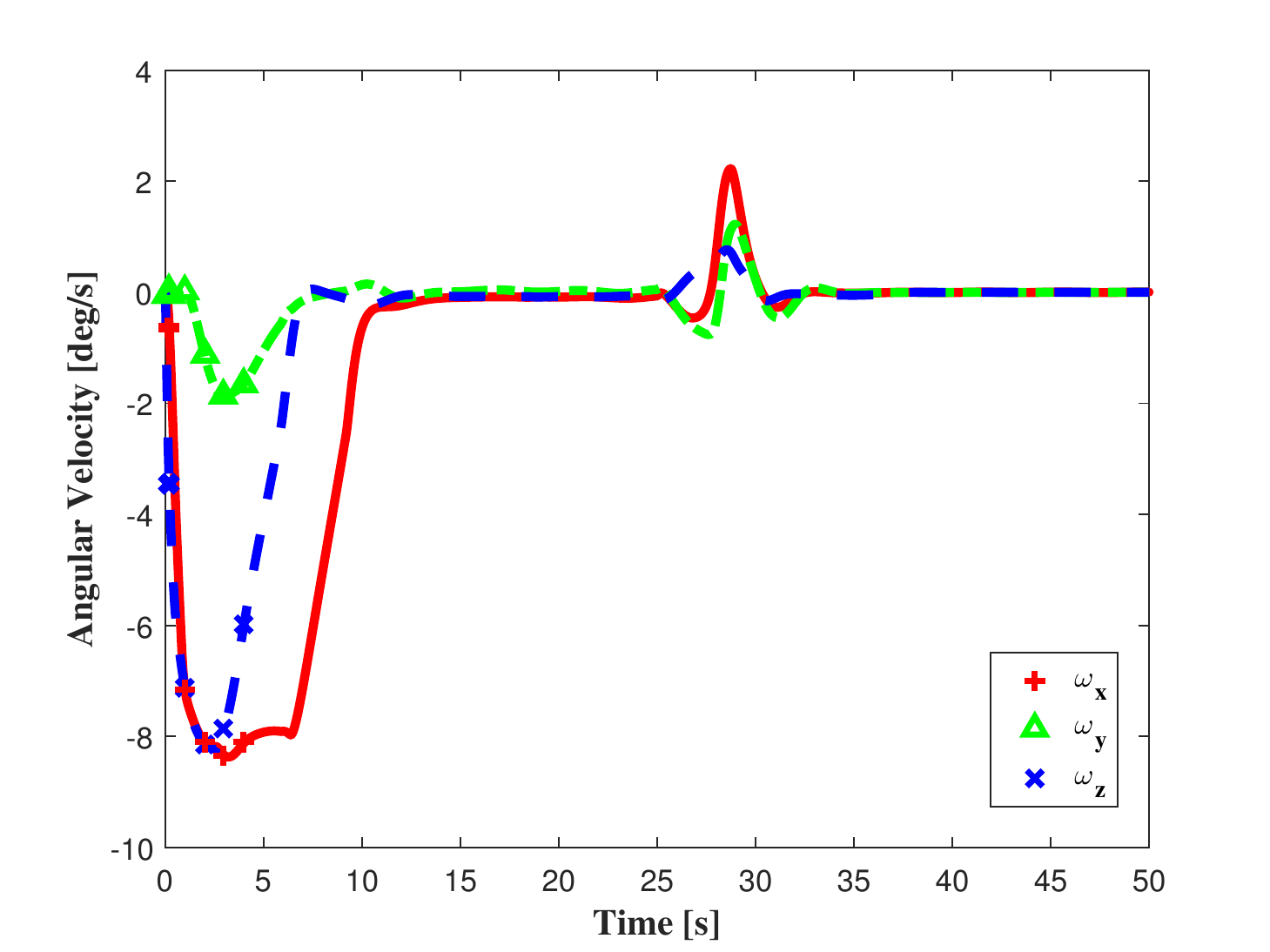}}

 \subfloat[Gimbal angle]{
    \label{fig15c}
    \includegraphics[scale=0.42]{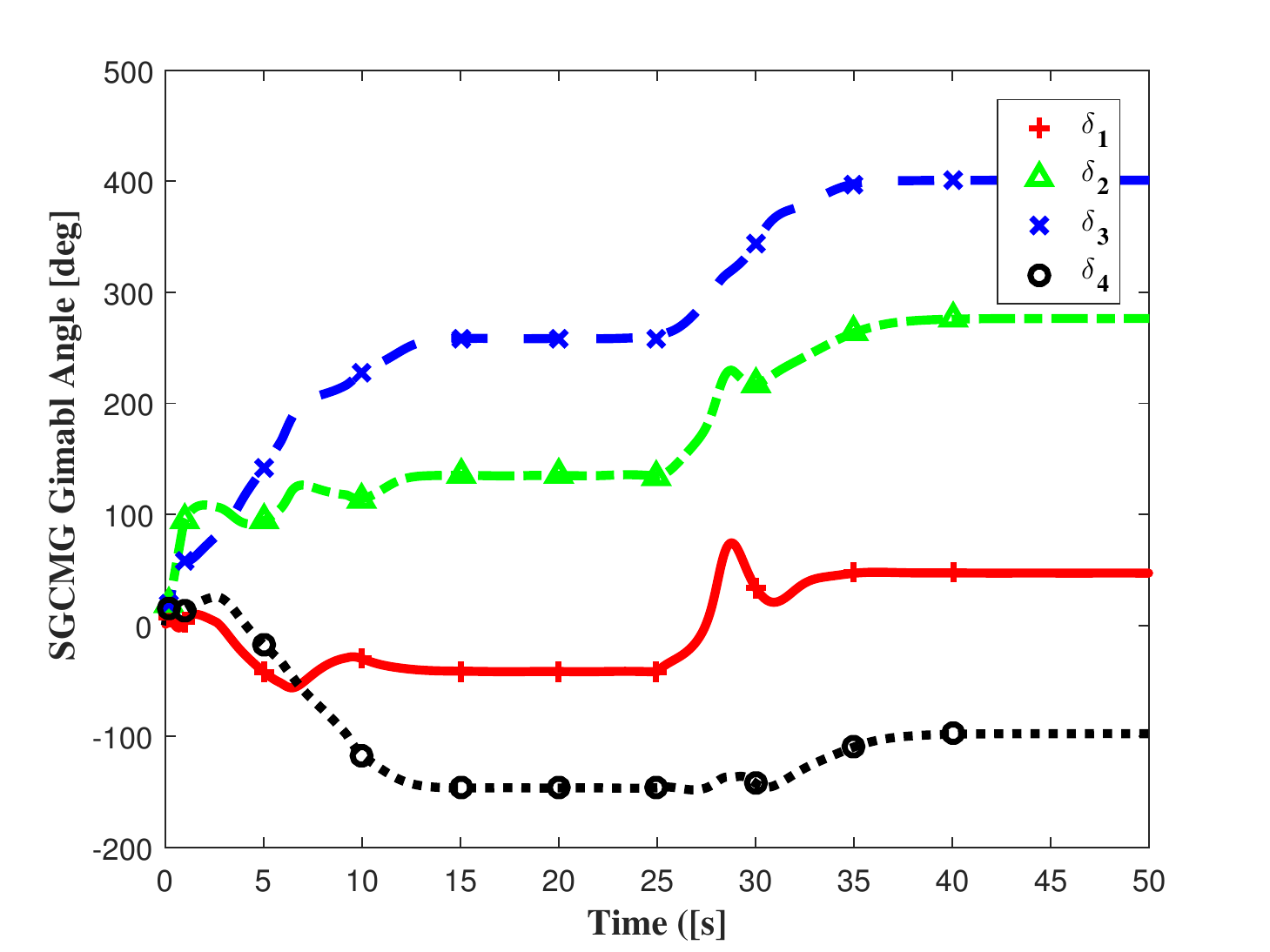}}
  \subfloat[Gimbal rate]{
    \label{fig15d}
    \includegraphics[scale=0.42]{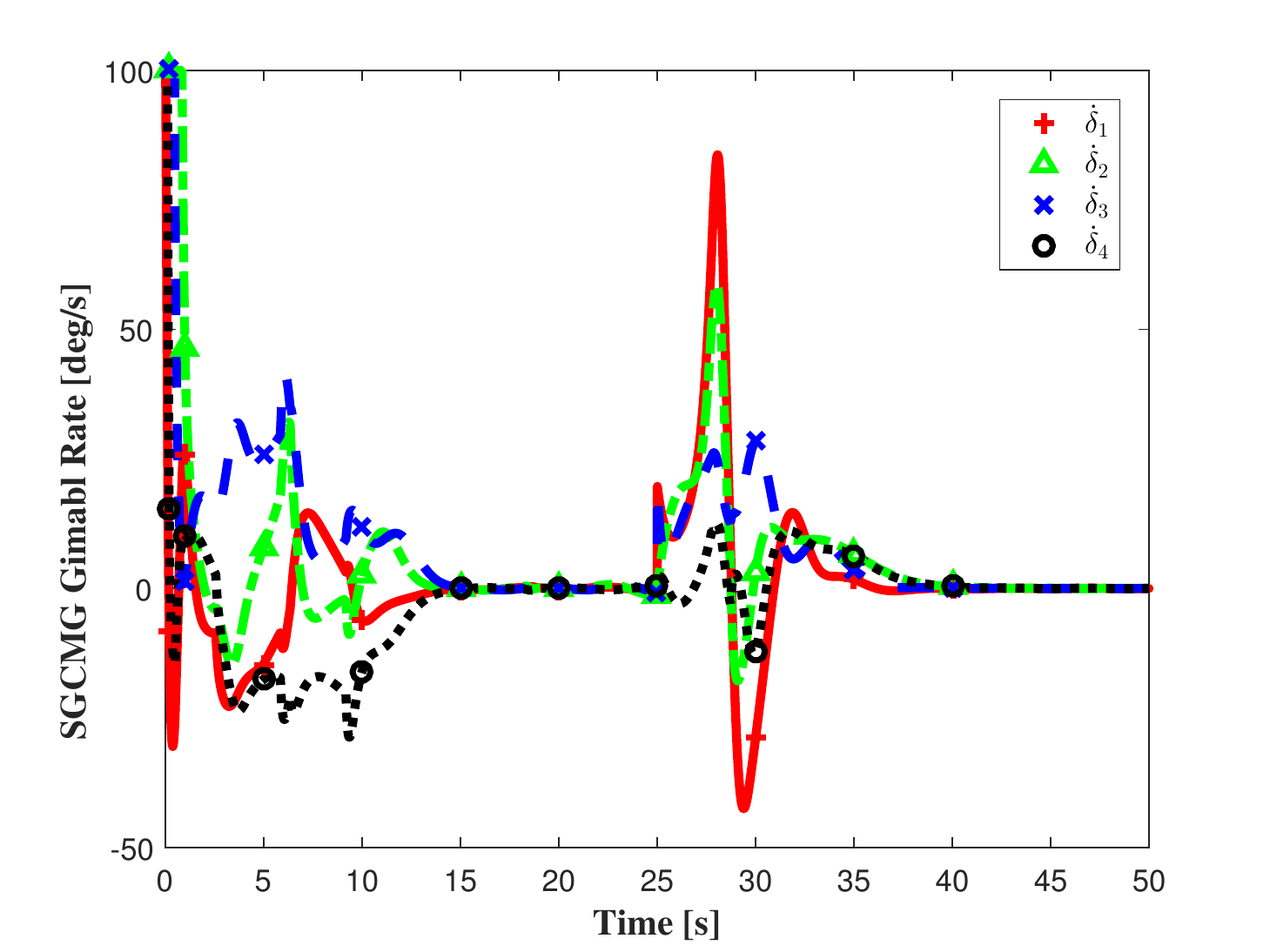}}

\subfloat[Rotor angular momentum]{
    \label{fig15e}
    \includegraphics[scale=0.42]{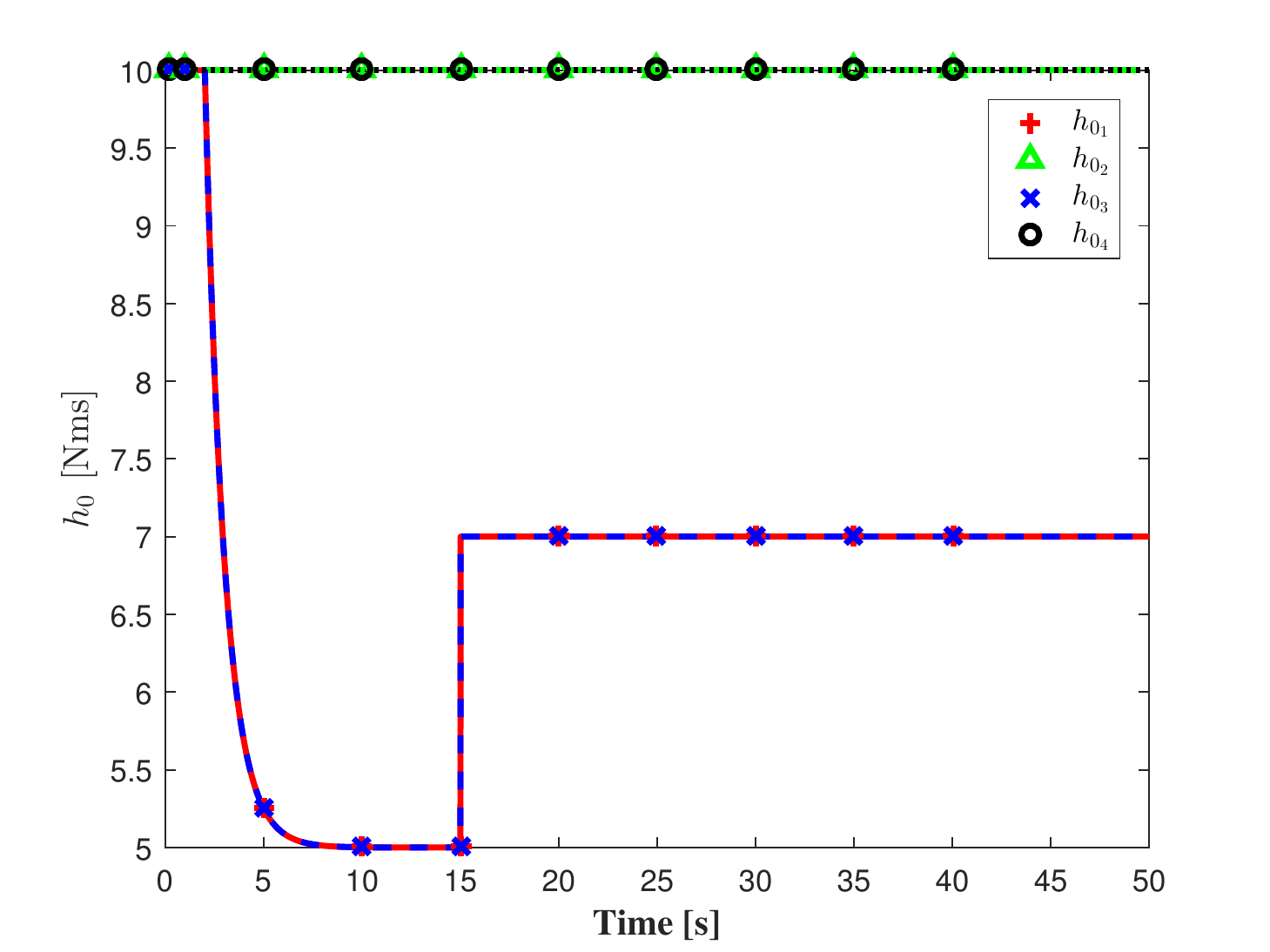}}
  \subfloat[Output control torque]{
    \label{fig15f}
    \includegraphics[scale=0.42]{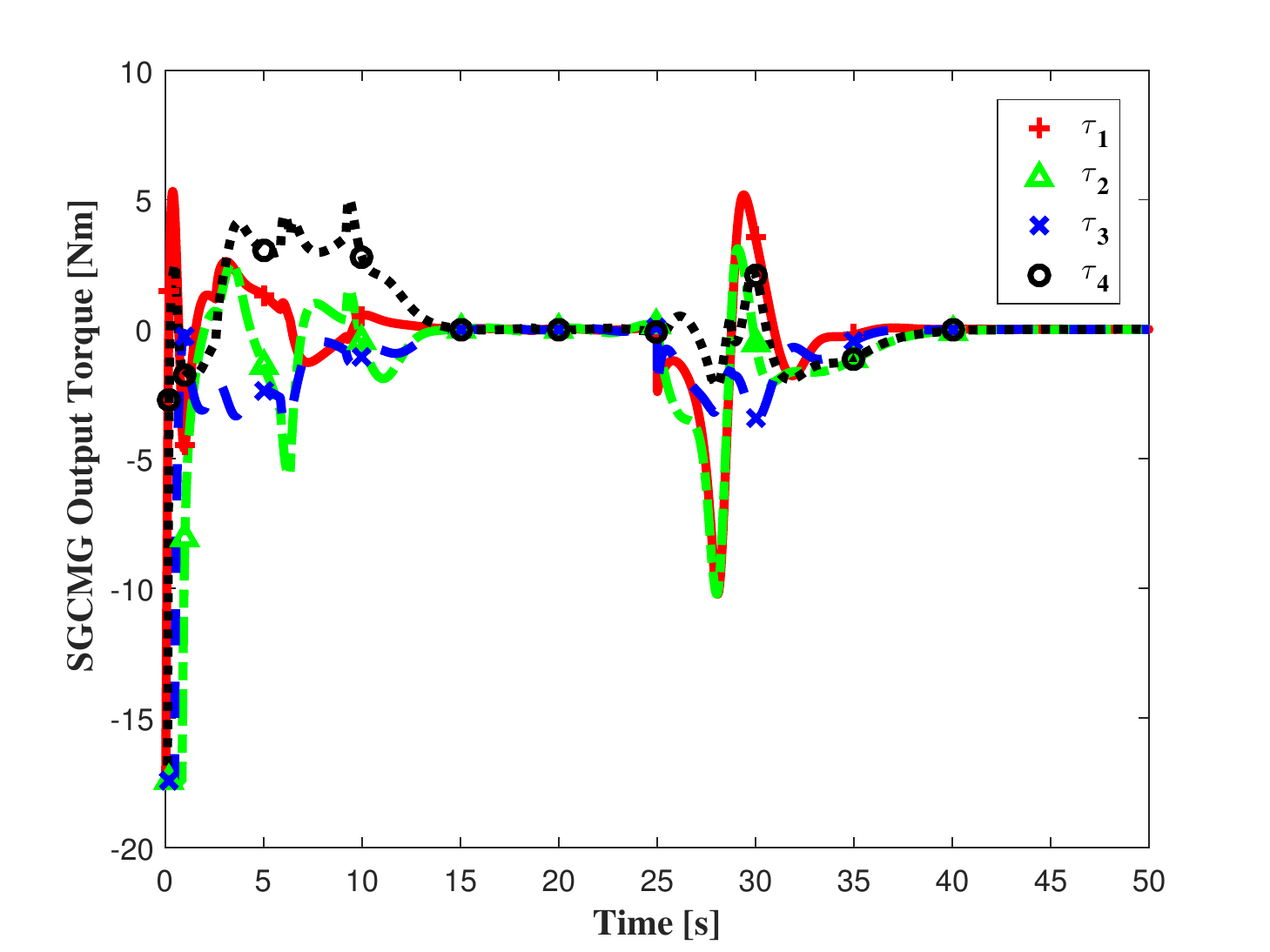}}
  \caption{SGCMG-actuated attitude control result in the presence of rotor fault and gimbal fault}
  \label{fig15}
\end{figure}

Fig. \ref{fig12} shows the attitude control results under nominal condition. It can be seen from Fig. \ref{fig12} that the maneuver is completed in about $10$ s. The maximum output torque is the product of rotor's maximum angular momentum and gimbal's maximum angular speed. During the whole process, there is no singularity. Fig. \ref{fig13} presents the Euler angle trajectory, rotor's angular momentum, gimbal angle and output torque in the presence of rotor failures. Fig. \ref{fig13b} shows that the rotor fails at $2$ s and the angular momentum drops to zero rapidly. Consequently, the output torque also becomes zero along with the variation of rotor's angular momentum, which is shown in Fig. \ref{fig13d}. When this pair of CMGs totally fails, the system will  turn to be uncontrollable. When the bias is added at $8$ s, there is a jump in the output torque. After the bias, the system goes to be stable state in about $20$ s in Fig. \ref{fig13a}. Fig. \ref{fig14} shows the control result when the EM-VSD system of gimbal fails into failure. When the gimbal fails into failure, the system becomes unstable until the additive fault occurs. After the additive fault, the system approaches to
stable state as shown in Fig. \ref{fig14a}. Fig. \ref{fig15} shows the simulation results when both the variable speed drive system of rotor and gimbal experience fault. We can see an obvious steady-state error in Fig. \ref{fig15a} and output jump due to the fault.

Parallel to the simulation results of RW-actuated system, we can obtain the similar qualitative conclusions:
\begin{itemize}
  \item The gimbal fault is more serious than the rotor fault when the rotor's angular momentum is greater than zero;
  \item The angular momentum of rotor is the amplification factor of the SGCMG. The fault in rotor can be regarded as the replacement of another SGCMG with a smaller control capacity.
  \item The additive fault may result in steady-state error.
\end{itemize}

{\section{Fault-tolerant control strategies}
Various fault-tolerant controllers (FTCs) have been developed, as summarized in the review of \cite{fekih2014fault} and \cite{yin2016review}. Generally speaking, the existing FTCs can be divided into the passive FTCs, active FTCs and the hybrid of passive and active FTCs. A comparative study between the active and passive approaches is given in \cite{jiang2012fault}.
For these FTCs, the redundancy of the system, including but not limited to the hardware redundancy, is the key factor and the significant difference lies in how these redundancies are used. For the passive FTCs, a single  fixed controller is designed among all admissible solution sets within the overlapped region considering both the nominal conditions and design basis faults. This kind of method emphasises on robustness for all identified cases rather than achieving optimal performance. Thus the passive FTCs are more conservative comparing with active FTCs. The active FTCs contain the fault detection and diagnosis (FDD) scheme, reconfigurable controller, and the decision making or redundancy management scheme. The performance of active FTCs depends highly on the FDD results.
More results can be found in \cite{zhang2008bibliographical} and \cite{yu2015survey}.

In this paper, we focus on how to incorporate the proposed fault model in fault-tolerant strategies design to accommodate SGCMG gimbal fault. Different from the previous section where the control torque generated by the SGCMGs expressed as $\bm{\tau}_c = -h_0 \bm{A} \dot{\bm{\delta}}$, the internal torque $\bm{\tau}= -h_0 \bm{A} \dot{\bm{\delta}} - \bm{\omega}^{\times}\bm{H}$ is adopted as the output of the SGCMG cluster, which makes the SGCMG to be a replaceable unit.  Then the dynamic of
spacecraft can be expressed as $\bm{J}\dot{\bm{\omega}}+\bm{\omega}^{\times}{\bm{J}\bm{\omega}}=\bm{\tau}+\bm{d} $, which is independent with the actuators. The schematic diagram of two potential fault-tolerant strategies to accommodate SGCMG gimbal fault are demonstrated in Fig. \ref{fig15}.

\begin{figure}[!h]
  \centering
  \subfloat[Additive equivalent strategy]{
    \label{fig15a}
    \includegraphics[scale=0.6]{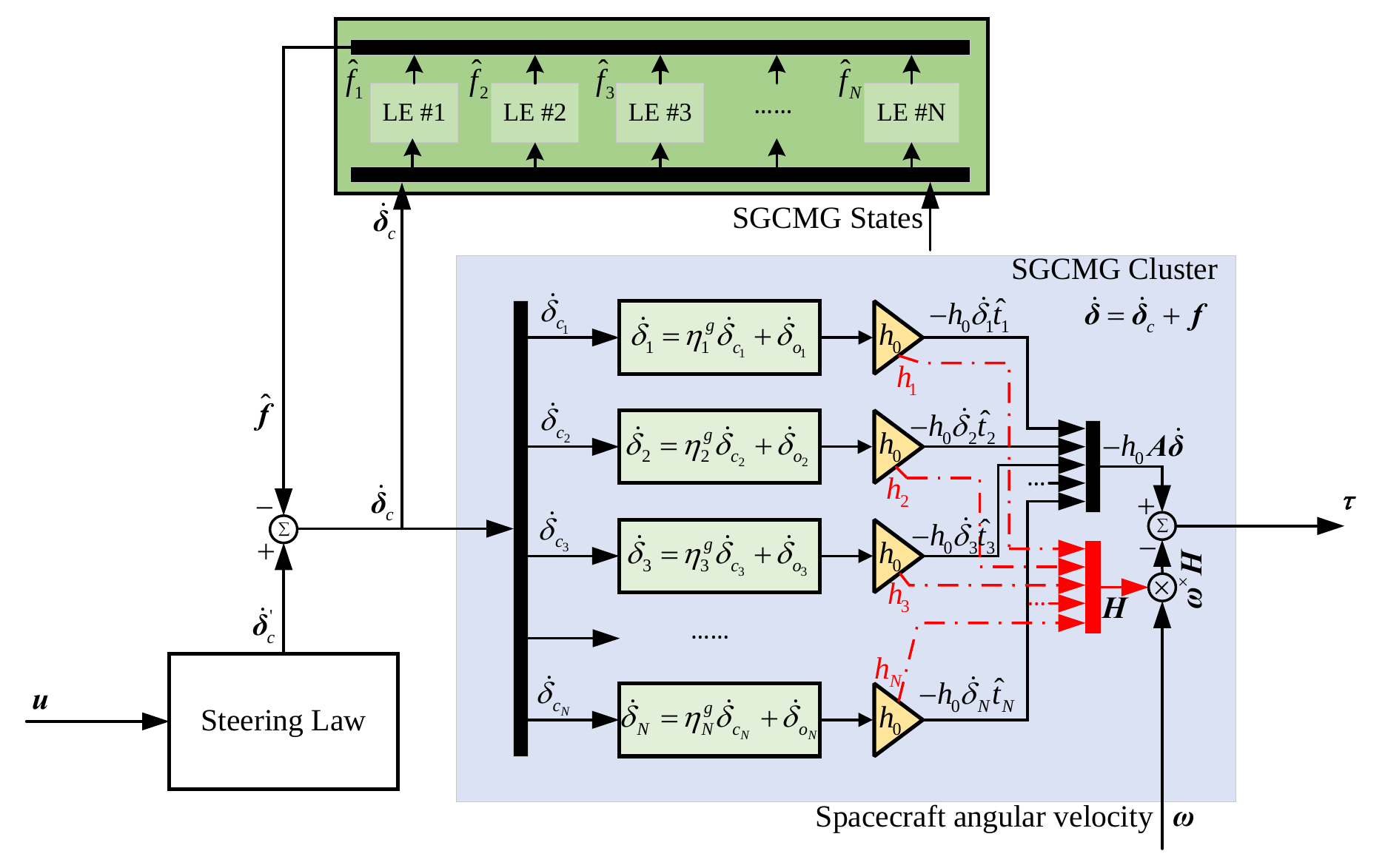}}

  \subfloat[Multiplicative equivalent strategy]{
    \label{fig15b}
    \includegraphics[scale=0.6]{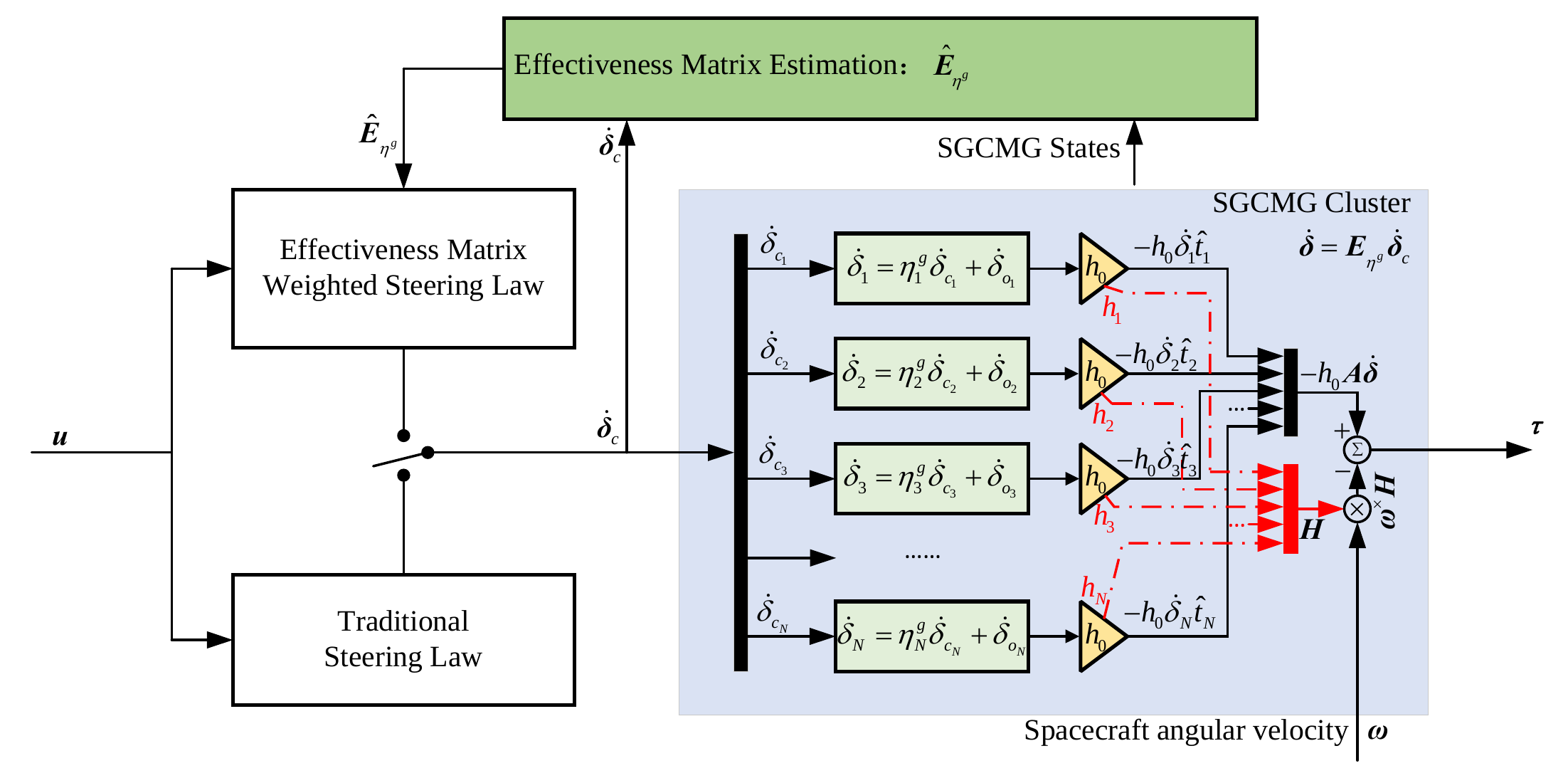}}
   \caption{\ychf{Potential fault-tolerant control strategies to handle gimbal fault} }
  \label{fig15}
\end{figure}

In Fig. \ref{fig15a}, an additive equivalent strategy is adopted and the real gimbal rate $\dot{\bm{\delta}}$ is expressed by the gimbal rate command $\dot{\bm{\delta}}_c$ and a lumped bias $\bm{f}$, i.e. $\dot{\bm{\delta}} = \dot{\bm{\delta}}_c+\bm{f}$. To estimate the lumped bias, local estimators (LE) can be designed to estimate each component of $\bm{f}$ with respect to each SGCMG. Then the gimbal rate command after compensation becomes $\dot{\bm{\delta}}_c = \dot{\bm{\delta}}_c^{'}-\hat{\bm{f}}$. As a consequence, the actual gimbal rate is
$\dot{\bm{\delta}}=\dot{\bm{\delta}}_c^{'}-\hat{\bm{f}}+{\bm{f}}\approx \bm{\delta}_c^{'}$. Thus the fault-tolerant target is achieved. Since the FDD scheme and the reconfiguration are not used, above additive equivalent fault-tolerant strategy can be regarded as a passive strategy.

 In Fig. \ref{fig15b}, the fault effect is described in a multiplicative way, i.e. $\dot{\bm{\delta}} =\bm{E}_{\eta^g} \dot{\bm{\delta}}_c$. To complete the fault-tolerant steering law design, some strategies are required to identify the fault and evaluate how serious the fault is, namely to estimate the effectiveness matrix ${\bm{E}}_{\eta^g}$. Using the estimated effectiveness matrix $\hat{\bm{E}}_{\eta^g}$, some new effectiveness matrix weighted steering law can be developed to minimize the use of the faulty SGCMG. Once the estimation is completed, the steering law is re-configured from the traditional one to the effectiveness weighted one. Then the control command is reallocated and the faulty SGCMGs can potentially be isolated. This strategy can be regarded as an active strategy. Together with the method shown in Fig. \ref{fig15a}, the gimbal fault of the SGCMGs can be handled. How to design the local estimator and estimate the equivalent effectiveness matrix will be our future works.}

\section{Conclusion}
This paper investigated the modeling of fault/failure in momentum exchange devices. A general fault model is established by regarding the momentum exchange devices as the cascade-connected EM-VSD systems. The potential faults of the EM-VSD system are identified, and the reason why these faults can be categorised into multiplicative fault and additive fault is given.
Based on the developed EM-VSD fault model, we further get the fault models of RW, SGCMG, SGCMG, and VSCMG.
Through simulations of spacecraft attitude control using RWs and SGCMGs as actuators, the potential faults in RW and SGCMG as well as their influences on the control performance are analyzed in detail.
Through simulation studies, we also obtain the qualitative conclusion that the additive fault has more serious influence than the multiplicative fault from the viewpoint of control accuracy.
This observation can be a guideline in developing fault-tolerant control system for the momentum exchange devices actuated spacecraft. By the end, how to incorporate the proposed fault model in fault-tolerant strategies design to accommodate the gimbal fault of the SGCMGs are demonstrated.

\section*{Acknowledgments}
\ychf{The authors would like to express their sincere gratitude to the
Editor-in-chief, the Associated Editor and anonymous reviewers whose constructive
comments have helped them to significantly improve both the
technical quality and presentation of this paper.}

\bibliographystyle{unsrt}
\bibliography{Franklin_R1}

\begin{thebibliography}{10}

\bibitem{tsiotras2001satellite}
P.~Tsiotras, H.~Shen, and C.~Hall.
\newblock Satellite attitude control and power tracking with energy/momentum
  wheels.
\newblock {\em Journal of Guidance, Control, and Dynamics}, 24(1):23--34, 2001.

\bibitem{shen2017robust}
Q.~Shen, D.~Wang, S.~Zhu, and E.~K. Poh.
\newblock Robust control allocation for spacecraft attitude tracking under
  actuator faults.
\newblock {\em IEEE Transactions on Control Systems Technology},
  25(3):1068--1075, 2017.

\bibitem{wie2002rapid}
B.~Wie, D.~Bailey, and C.~Heiberg.
\newblock Rapid multitarget acquisition and pointing control of agile
  spacecraft.
\newblock {\em Journal of Guidance, Control, and Dynamics}, 25(1):96--104,
  2002.

\bibitem{ahmed2002adaptive}
J.~Ahmed and D.~S. Bernstein.
\newblock Adaptive control of double-gimbal control-moment gyro with unbalanced
  rotor.
\newblock {\em Journal of guidance, control, and dynamics}, 25(1):105--115,
  2002.

\bibitem{wie2001singularity}
B.~Wie, D.~Bailey, and C.~Heiberg.
\newblock Singularity robust steering logic for redundant single-gimbal control
  moment gyros.
\newblock {\em Journal of Guidance, Control, and Dynamics}, 24(5):865--872,
  2001.

\bibitem{gleyzes2012pleiades}
M~Alain Gleyzes, Lionel Perret, and Philippe Kubik.
\newblock Pleiades system architecture and main performances.
\newblock {\em International Archives of the Photogrammetry, Remote Sensing and
  Spatial Information Sciences}, 39(1):537--542, 2012.

\bibitem{poli2010radiometric}
D~Poli, E~Angiuli, and F~Remondino.
\newblock Radiometric and geometric analysis of worldview-2 stereo scenes.
\newblock {\em International archives of photogrammetry and remote sensing and
  spatial information sciences}, 38(Part 1):15--18, 2010.

\bibitem{markley2014fundamentals}
F.~L. Markley and J.~L. Crassidis.
\newblock {\em Fundamentals of spacecraft attitude determination and control},
  volume~33.
\newblock Springer, 2014.

\bibitem{fekih2014fault}
A.~Fekih.
\newblock Fault diagnosis and fault tolerant control design for aerospace
  systems: A bibliographical review.
\newblock In {\em American Control Conference (ACC), 2014}, pages 1286--1291.
  IEEE, 2014.

\bibitem{jiang2012fault}
J.~Jiang and X.~Yu.
\newblock Fault-tolerant control systems: A comparative study between active
  and passive approaches.
\newblock {\em Annual Reviews in control}, 36(1):60--72, 2012.

\bibitem{murugesan1987fault}
S.~Murugesan and P.S. Goel.
\newblock Fault-tolerant spacecraft attitude control system.
\newblock {\em Sadhana}, 11(1-2):233--261, 1987.

\bibitem{shen2015finite}
Q.~Shen, D.~Wang, S.~Zhu, and E.~K. Poh.
\newblock Finite-time fault-tolerant attitude stabilization for spacecraft with
  actuator saturation.
\newblock {\em IEEE Transactions on Aerospace and Electronic Systems},
  51(3):2390--2405, 2015.

\bibitem{noumi2013fault}
A.~Noumi and M.~Takahashi.
\newblock Fault-tolerant attitude control systems for satellite equipped with
  control moment gyros.
\newblock In {\em AIAA Guidance, Navigation, and Control (GNC) Conference},
  page 5119, 2013.

\bibitem{zhang2017fault}
F.~Zhang, L.~Jin, and S.~Xu.
\newblock Fault tolerant attitude control for spacecraft with sgcmgs under
  actuator partial failure and actuator saturation.
\newblock {\em Acta Astronautica}, 132:303--311, 2017.

\bibitem{bialke98high}
B.~Bialke.
\newblock High fidelity mathematical modeling of reaction wheel performance.
\newblock In {\em 1998 Annual American Astronautical Socitety Rocky Mountain
  Guidance and Control Conference on Advances in the Astronautical Sciences},
  pages 483--496, 1998.

\bibitem{rahimi2017fault}
A.~Rahimi, K.~D. Kumar, and H.~Alighanbari.
\newblock Fault estimation of satellite reaction wheels using covariance based
  adaptive unscented kalman filter.
\newblock {\em Acta Astronautica}, 134:159--169, 2017.

\bibitem{choi2015fault}
Y.C. Choi, J.~H. Son, and H.~S. Ahn.
\newblock Fault detection and isolation for a small cmg-based satellite: A
  fuzzy q-learning approach.
\newblock {\em Aerospace Science and Technology}, 47:340--355, 2015.

\bibitem{book:931874}
H.~A. Toliyat, S.~Nandi, S.~Choi, and H.~Meshgin-Kelk.
\newblock {\em Electric Machines: Modeling, Condition Monitoring, and Fault
  Diagnosis}.
\newblock CRC Press, 2013.

\bibitem{kastha1994investigation}
D.~Kastha and B.~K. Bose.
\newblock Investigation of fault modes of voltage-fed inverter system for
  induction motor drive.
\newblock {\em IEEE Transactions on Industry Applications}, 30(4):1028--1038,
  1994.

\bibitem{nandi2005condition}
S.~Nandi, H.~A. Toliyat, and X.~Li.
\newblock Condition monitoring and fault diagnosis of electrical motors
  \text{--- }a review.
\newblock {\em IEEE transactions on energy conversion}, 20(4):719--729, 2005.

\bibitem{campos2008fault}
D.~U. Campos-Delgado, D.~R. Espinoza-Trejo, and E.~Palacios.
\newblock Fault-tolerant control in variable speed drives: a survey.
\newblock {\em IET Electric Power Applications}, 2(2):121--134, 2008.

\bibitem{rajagopalan2006detection}
S.~Rajagopalan, J.~M. Aller, J.~A. Restrepo, T.~G. Habetler, and R.~G. Harley.
\newblock Detection of rotor faults in brushless dc motors operating under
  nonstationary conditions.
\newblock {\em IEEE Transactions on Industry Applications}, 42(6):1464--1477,
  2006.

\bibitem{moseler2000application}
O.~Moseler and R.~Isermann.
\newblock Application of model-based fault detection to a brushless dc motor.
\newblock {\em IEEE Transactions on industrial electronics}, 47(5):1015--1020,
  2000.

\bibitem{rajagopalan2006thesis}
S.~Rajagopalan.
\newblock {\em Detection of rotor and load faults in brushless DC motors
  operating under stationary and non-stationary conditions}.
\newblock PhD thesis, Georgia Institute of Technology, 2006.

\bibitem{chen2012robust}
J.~Chen and R.~J. Patton.
\newblock {\em Robust model-based fault diagnosis for dynamic systems},
  volume~3.

\bibitem{joksimovic2000detection}
G.~M. Joksimovic and J.~Penman.
\newblock The detection of inter-turn short circuits in the stator windings of
  operating motors.
\newblock {\em IEEE Transactions on Industrial electronics}, 47(5):1078--1084,
  2000.

\bibitem{betin1999closed}
F.~Betin, M.~Deloizy, and C.~Goeldel.
\newblock Closed loop control of a stepping motor drivecomparison between pid
  control, self tuning regulationand fuzzy logic control.
\newblock {\em EPE Journal}, 8(1-2):33--39, 1999.

\bibitem{morar2003stepper}
A.~Morar.
\newblock Stepper motor model for dynamic simulation.
\newblock {\em Acta Electrotehnica}, 44(2):117--122, 2003.

\bibitem{collacott2012mechanical}
R.~Collacott.
\newblock {\em Mechanical fault diagnosis and condition monitoring}.
\newblock Springer Science \& Business Media, 2012.

\bibitem{pal1991direct}
S.~K. Pal.
\newblock Direct drive high energy permanent magnet brush and brushless dc
  motors for robotic applications.
\newblock In {\em IEE Colloquium on Robot Actuators}, pages 12--1. IET, 1991.

\bibitem{pal1991design}
S.~K. Pal.
\newblock Design criteria for brushless dc motors with hollow rotor of samarium
  cobalt for applications above 25000 rpm in vacuum.
\newblock In {\em Fifth International Conference on Electrical Machines and
  Drives}, pages 115--120. IET, 1991.

\bibitem{fisher1991design}
R.~Fisher.
\newblock Design and cost considerations for permanent magnet dc motor
  applications.
\newblock {\em Power Conversion \& Intelligent Motion}, 17(7):18--24, 1991.

\bibitem{motor1985report}
Motor Reliability~Working Group et~al.
\newblock Report of large motor reliability survey of industrial and commercial
  installations, part i.
\newblock {\em IEEE Trans. Industrial Applications}, 1(4):865--872, 1985.

\bibitem{isermann2011fault}
R.~Isermann.
\newblock {\em Fault-diagnosis applications: model-based condition monitoring:
  actuators, drives, machinery, plants, sensors, and fault-tolerant systems}.
\newblock Springer Science \& Business Media, 2011.

\bibitem{leve2015spacecraft}
F.~A. Leve, B.~J. Hamilton, and M.~A. Peck.
\newblock {\em Spacecraft momentum control systems}.
\newblock Springer, 2015.

\bibitem{jin2008fault}
J.~Jin, S.~Ko, and C.~K. Ryoo.
\newblock Fault tolerant control for satellites with four reaction wheels.
\newblock {\em Control Engineering Practice}, 16(10):1250--1258, 2008.

\bibitem{xiao2012adaptive}
B.~Xiao, Q.~Hu, and Y.~Zhang.
\newblock Adaptive sliding mode fault tolerant attitude tracking control for
  flexible spacecraft under actuator saturation.
\newblock {\em IEEE Transactions on Control Systems Technology},
  20(6):1605--1612, 2012.

\bibitem{sasaki2016fault}
T.~Sasaki and T.~Shimomura.
\newblock Fault-tolerant architecture of two parallel double-gimbal
  variable-speed control moment gyros.
\newblock In {\em AIAA Guidance, Navigation, and Control Conference},
  2016-0090.

\bibitem{schaub2000singularity}
H.~Schaub and J.~L. Junkins.
\newblock Singularity avoidance using null motion and variable-speed control
  moment gyros.
\newblock {\em Journal of Guidance, Control, and Dynamics}, 23(1):11--16, 2000.

\bibitem{yoon2004singularity}
H.~Yoon and P.~Tsiotras.
\newblock Singularity analysis of variable speed control moment gyros.
\newblock {\em Journal of Guidance, Control, and Dynamics}, 27(3):374--386,
  2004.

\bibitem{cui2013steering}
P.~Cui and J.~He.
\newblock Steering law for two parallel variable-speed double-gimbal control
  moment gyros.
\newblock {\em Journal of Guidance, Control, and Dynamics}, 37(1):350--359,
  2013.

\bibitem{tafazoli2009study}
M.~Tafazoli.
\newblock A study of on-orbit spacecraft failures.
\newblock {\em Acta Astronautica}, 64(2-3):195--205, 2009.

\bibitem{cao2016time}
X.~Cao, C.~Yue, M.~Liu, and B.~Wu.
\newblock Time efficient spacecraft maneuver using constrained torque
  distribution.
\newblock {\em Acta Astronautica}, 123:320--329, 2016.

\bibitem{yin2016review}
S.~Yin, B.~Xiao, S.~X. Ding, and D.~Zhou.
\newblock A review on recent development of spacecraft attitude fault tolerant
  control system.
\newblock {\em IEEE Transactions on Industrial Electronics}, 63(5):3311--3320,
  2016.

\bibitem{zhang2008bibliographical}
Y.~Zhang and J.~Jiang.
\newblock Bibliographical review on reconfigurable fault-tolerant control
  systems.
\newblock {\em Annual reviews in control}, 32(2):229--252, 2008.

\bibitem{yu2015survey}
X.~Yu and J.~Jiang.
\newblock A survey of fault-tolerant controllers based on safety-related
  issues.
\newblock {\em Annual Reviews in Control}, 39:46--57, 2015.

\end{thebibliography}

\end{document}